\documentstyle[prd,aps,epsf,array]{revtex}
\newcommand{\beqa}{\begin{eqnarray}}
\newcommand{\eeqa}{\end{eqnarray}}
\newcommand{\nn}{\nonumber}

\begin{document}

\draft
\title{
\vspace*{-2.cm}
\begin{flushright}
{\normalsize UTHEP-444}\\
{\normalsize UTCCP-P-106}\\
\end{flushright}
{\Large
Spectral function and excited states in lattice QCD
with\\ maximum entropy method
}
}
\author{\normalsize
  T.~Yamazaki\rlap,$^{1}$
  S.~Aoki\rlap,$^{1}$
  R.~Burkhalter\rlap,$^{1,2}$ 
  M.~Fukugita\rlap,$^{3}$
  S.~Hashimoto\rlap,$^{4}$
  N.~Ishizuka\rlap,$^{1,2}$
  Y.~Iwasaki\rlap,$^{1,2}$
  K.~Kanaya\rlap,$^{1}$ 
  T.~Kaneko\rlap,$^{4}$ 
  Y.~Kuramashi\rlap,$^{4}$
  M.~Okawa\rlap,$^{4}$ 
  Y.~Taniguchi\rlap,$^{1}$
  A.~Ukawa\rlap,$^{1,2}$ and
  T.~Yoshi\'e$^{1,2}$\\
(CP-PACS Collaboration)\\
}
\address{
$^{\rm 1}$Institute of Physics,
    University of Tsukuba, Tsukuba, Ibaraki 305-8571, Japan \\
$^{\rm 2}$Center for Computational Physics,
    University of Tsukuba, Tsukuba, Ibaraki 305-8577, Japan \\
$^{\rm 3}$Institute for Cosmic Ray Research,
    University of Tokyo, Kashiwa 277-8582, Japan \\
$^{\rm 4}$High Energy Accelerator Research Organization
    (KEK), Tsukuba, Ibaraki 305-0801, Japan
}
\date{\today}

\maketitle

\begin{abstract}
We apply the maximum entropy method
to extract the spectral functions 
for pseudoscalar and vector mesons
from hadron correlators 
previously calculated at four different lattice spacings in quenched 
QCD with the Wilson quark action.
We determine masses and decay constants for the ground and 
excited states of the pseudoscalar and vector channels
from position and area of peaks in the spectral functions.
We obtain the results,
$m_{\pi_1} = 660(590)$ MeV and $m_{\rho_1} = 1540(570)$ MeV
for the masses of the first excited state masses,
in the continuum limit of quenched QCD.
We also find unphysical states which have infinite mass 
in the continuum limit,
and argue that they are bound states of two doublers of the Wilson quark 
action.
If the interpretation is correct,
this is the first time that the state of doublers is identified in 
lattice QCD numerical simulations.
\end{abstract}

\pacs{11.15.Ha, 12.38.Gc}

\section{Introduction}

The spectral function of hadron correlation functions
contains information not only on the mass of the ground state but also
other quantities such as the masses for excited states, and decays and
scatterings of hadrons.
In lattice QCD simulations one can numerically obtain an euclidean time 
correlation function $D(\tau)$ of an operator $O(\tau)$, 
which is related to the spectral function $f(\omega)$ of this correlator
through
\beqa
D(\tau) & =& \langle 0\vert O(\tau) O^{\dagger}(0) \vert 0 \rangle \nn \\
& = & \int d \omega f(\omega) K(\omega ,\tau),
\eeqa
where 
$K(\tau ,\omega)$ is a kernel of the Laplace transformation given by
\[
K(\omega , \tau ) = e^{-\omega\tau} +e^{-\omega (T-\tau)}
\]
for $ 0 \le \tau \le T$ with the periodic boundary condition,
where $T$ is the lattice size in the euclidean time direction.
A typical form of $f(\omega)$ is
\beqa
f(\omega) &=& Z_0\delta (\omega - E_0) + 
\widetilde{f}(\omega ;\omega \ge 2m_{0}),
\eeqa
where $E_0$ is the energy of the ground state $ \vert E_0\rangle$
coupled to the operator $O$ and 
$Z_0 = \vert \langle 0 \vert O\vert E_0\rangle \vert^2$, 
and $\widetilde{f}(\omega)$ represents the continuous spectrum
which starts at $\omega = 2m_{0}$ for the 2-particle state.

In principle one can extract all the information for the states
which can couple to the operator $O$ from the spectral function $f(\omega)$.
In a usual analysis of lattice QCD simulations, however,
only the mass (or energy) of the ground state $E_0$ and its amplitude
$Z_0$ can be reliably extracted from the asymptotic behavior of 
the correlation functions at large euclidean times,
\[
D(\tau) \rightarrow Z_0 e^{-E_0 \tau}, \quad \tau\rightarrow\infty .
\]
Numerically it is unstable to extract masses of excited states
with a multi-exponential fit, so that a simultaneous fit to several
correlation functions which have the same intermediate states with different
amplitudes becomes necessary to stabilize the result.
Similar but more difficult problems appear in the calculation of the
decay amplitude\cite{MT,LL}.

If one could reconstruct $f(\omega)$ directly from the correlation
function $D(\tau)$ using data at all $\tau$,
all the difficulties mentioned above would be avoided.
Since the number of data for $D(\tau)$ with a discrete set of time $\tau$
is much smaller than the number of degree of freedom necessary for the
reconstruction of $f(\omega)$ in general, however,
the standard $\chi^{2}$ fit is ill-posed for this problem.
With some assumptions on the form of the spectral function
the $\chi^{2}$ fit may work, but this is essentially equivalent to
the multi-exponential or more complicated fit to the correlation function.

In condensed matter physics, the reconstruction of the spectral function 
in quantum Monte Carlo simulations has been attempted with
the maximum entropy method (MEM)\cite{jarr}. 
It has been also successfully applied for 
image reconstruction in astrophysics.
The most important assumption in MEM is that a probability for
spectral functions can be assigned for given data of $D(\tau)$.
Then MEM can numerically reconstruct the most probable spectral function,
using the Bayes's theorem in probability theory, 
without any strong constraints on its form.
Recently, this method has been tested in lattice QCD~\cite{for,nak} and 
first interesting results for the spectral function
have been obtained~\cite{hats:rev,diquark,Shi}.

In this paper,
we employ MEM to reconstruct the spectral functions of
pseudoscalar and vector mesons from the correlation functions
previously calculated on lattices with the spatial size about 3~fm 
at four different lattice spacings in quenched QCD
\cite{CPdata,pre}.
From the spectral functions
we extract masses and decay constants for excited states
as well as for the ground state.
While they agree with results obtained from
the exponential fits to correlation functions,
errors for excited state masses from the spectral function
are much smaller than those from the multi-exponential fit,
so that we can estimate masses for excited states in the continuum limit
with reasonable errors.
We also find evidence that some excited states are composed
of fermion doublers.

This paper is organized as follows. In Sec.~\ref{sec:MEM}, we summarize 
our implementation of MEM
and present results from tests
using mock-up data generated from a realistic spectral function.
Some details of the lattice QCD data and parameters used in our MEM analysis
are given in Sec.~\ref{sec:para}.
In Sec.~\ref{results}, we present our results for the spectral function,
which show excited state peaks as well as the ground state peak.
From the position and the area of these peaks
we extract masses and decay constants,
and compare them with those obtained directly
from correlation functions.
The continuum extrapolation is made for these quantities.
In Sec.~\ref{sec:doubler}, 
we argue that some peaks in the spectral functions correspond
to the state containing two doublers of the Wilson quarks.
Our conclusions are given in Sec~\ref{sec:conc}.
In the Appendix technical details of MEM are collected.

\section{Maximum Entropy Method (MEM)}
\label{sec:MEM}
\subsection{Implementation} 

The existence of a probability distribution for a spectral function
is a key assumption in the maximum entropy method.
Using this assumption
one can obtain the most probable spectral function for given lattice data $D$ 
and all prior knowledge $H$, such as $f(\omega) \ge 0$, by 
maximizing the conditional probability ${\mathrm P}[F|DH]$,
where 
${\mathrm P}[F|DH]$ is the probability of $F$ with
the condition that $D$ and $H$ are given.
Here $F$ stands for the spectral function $f(\omega)$.
Using the Bayes's theorem in probability theory\cite{bay},
\begin{equation}
{\mathrm P}[X|YZ] = \frac{{\mathrm P}[Y|XZ]\, {\mathrm P}[X|Z]}{{\mathrm P}[Y|Z]},
\label{eq:01}
\end{equation}
where ${\mathrm P}[X]$ is the probability of an event $X$,
one rewrites the conditional probability ${\mathrm P}[F|DH]$  as,
\begin{equation}
{\mathrm P}[F|DH] \propto {\mathrm P}[D|FH]\, {\mathrm P}[F|H].
\label{eq:H2}
\end{equation}
Here ${\mathrm P}[D|FH]$ is the probability of data for a 
given spectral function, called the likelihood function, and
${\mathrm P}[F|H]$ is 
the probability of the spectral function for a given prior knowledge,
called the prior probability.

The likelihood function is
equivalent to $\chi^{2}$ in the least square method\cite{Brandt}.
For a large number of Monte Carlo measurements of 
a correlation function, the data is expected to obey a gaussian distribution
according to the central limit theorem, which gives
\begin{eqnarray}
{\mathrm P}[D|FH] & = & \frac{1}{Z_{L}}\, e^{-L},
\label{eq:L2}\\
L & = & \frac{1}{2}\sum_{i,j}^{N_{D}}\left(D(\tau_{i})-D_{\!f}(\tau_{i})\right)
C_{ij}^{-1}\left(D(\tau_{j})-D_{\!f}(\tau_{j})\right),
\label{eq:integ}
\end{eqnarray}
with  the normalization constant, $Z_{L} = (2\pi)^{N_{D}}\sqrt{\det C}$,
and the number of temporal points $N_{D}$.
The lattice propagator data averaged over gauge configurations, $D(\tau)$, and
the covariance matrix, $C$, are defined by 
\begin{eqnarray}
D(\tau_{i}) & = & \frac{1}{N_{\it conf}}\sum_{n=1}^{N_{\it conf}}D^{n}(\tau_{i}),\\
C_{ij} & = & \frac{1}{N_{\it conf}(N_{\it conf}-1)}\sum_{n=1}^{N_{\it conf}}
\left(D(\tau_{i})-D^{n}(\tau_{i})\right)
\left(D(\tau_{j})-D^{n}(\tau_{j})\right),
\end{eqnarray}
where $N_{\it conf}$ is the total number of gauge configurations and 
$D^{n}(\tau)$ is the data for the $n$-th gauge configuration.
Finally, $D_{\!f}(\tau)$ is the propagator constructed
from the spectral function $f(\omega)$ and
the kernel $K(\omega , \tau)$ as
\begin{equation}
D_{\!f}(\tau)=\int d\omega\, f(\omega) K(\omega , \tau).
\label{eq:ime}
\end{equation}

The prior probability is written in terms of the entropy 
$S(f)$\cite{cla,deve,axiom,monkey} for a given 
model $m(\omega)$ represented by a real and positive function,
and a real and positive parameter $\alpha$. 
The entropy $S(f)$ becomes zero at its maximum
point where $f(\omega)$ is equal to $m(\omega)$.
Explicitly we have
\begin{eqnarray}
{\mathrm P}[F|Hm\alpha] &=& \frac{1}{Z_{S}(\alpha)}\, e^{\alpha S},
\label{eq:ma2}\\
S(f) &=& \int d\omega \, \left[f(\omega)-m(\omega)
-f(\omega)\log\left(\frac{f(\omega)}{m(\omega)}\right)
\right] \\
&&\longrightarrow \sum_{l=1}^{N_{\omega}}\left[f_{l}-m_{l}
-f_{l}\log\left(\frac{f_{l}}{m_{l}}\right) 
\right], \label{eq:009}
\end{eqnarray}
with the normalization constant, $Z_{S}(\alpha)=(2\pi/\alpha)^{N_{\omega}/2}$,
calculated in Appendix \ref{sec:Zs}.
In (\ref{eq:009})
the continuous spectral function $f(\omega)$ is approximately 
represented by a discrete set of points $f(\omega_l)=f_{l}$ 
with $l=1,\cdots, N_\omega$. 
Hereafter we replace the prior knowledge $H$ in (\ref{eq:H2})
by $H m \alpha$, writing $m$ and $\alpha$ explicitly.
It is worth mentioning that 
this form of the entropy leads to
a positive spectral function in MEM.

Combining (\ref{eq:L2}) and (\ref{eq:ma2}), one obtains
\begin{equation}
{\mathrm P}[F|DHm\alpha] \propto \frac{e^{Q_{\alpha}(f)}}
{Z_{L}Z_{S}(\alpha)}\, ,\ \ \ Q_{\alpha}(f)=\alpha S(f) -L.
\end{equation}
Therefore the condition satisfied by the most probable spectral function
$f_{\alpha}$ for a given $\alpha$ (and model $m(\omega)$)
is given by
\begin{eqnarray}
\left. \frac{\delta Q_{\alpha}(f)}{\delta f}\right|_{f=f_{\alpha}}
=0.
\label{eq:equation}
\end{eqnarray}

The parameter $\alpha$ dictates the relative weight of 
the entropy $S(f)$ and $L$.
One can deal with $\alpha$ dependence of $f_\alpha$ as follows.
One first defines ${\mathrm P}[\alpha|DHm]$\cite{jarr,cla,deve},
the probability of $\alpha$ for given data and all prior knowledge,
which can be transformed as
\begin{eqnarray}
{\mathrm P}[\alpha|DHm] &\propto& 
{\mathrm P}[\alpha|Hm]\int\! {\mathcal D}F\, \frac{e^{Q_{\alpha}(f)}}
{Z_{L}Z_{S}(\alpha)}
\label{eq:9}.
\end{eqnarray}
See Appendix \ref{sec:pag} for details.
In the final result $\widehat{f}(\omega)$,
$\alpha$ is averaged with this weight factor ${\mathrm P}[\alpha|DHm]$,
\begin{equation}
\widehat{f}(\omega) = \int\! d\alpha \,{\mathrm P}[\alpha|DHm]f_{\alpha}(\omega)
/ \int\! d\alpha \,{\mathrm P}[\alpha|DHm].
\end{equation}
This procedure is called Bryan's method\cite{Bry}
and used in this article.
We restrict the range of $\alpha$ in the actual average
as $\alpha_{min} \leq \alpha \leq \alpha_{max}$,
where $\alpha_{min}$ and $\alpha_{max}$ are chosen to satisfy 
$
{\mathrm P}[\widehat{\alpha}|DHm]\geq 10\,{\mathrm P}[\alpha_{min,max}|DHm]
$
with $\widehat{\alpha}$ being the maximum value of ${\mathrm P}[\alpha|DHm]$.
The standard choice of ${\mathrm P}[\alpha|Hm]$ in
(\ref{eq:9}) is either a constant or $1/\alpha$\cite{jarr,deve,Bry}.
In the next section 
we will show that the final result is insensitive to the choice
as long as ${\mathrm P}[\alpha|DHm]$ 
is sharply peaked around $\widehat{\alpha}$,
and therefore we adopt
${\mathrm P}[\alpha|Hm]=$ constant in our main analysis.

In MEM it is not possible to assign error bars to each point in the spectral 
function since the errors between different points are strongly correlated.
Instead we estimate the uncertainty of the 
spectral function averaged over $\omega$ in a certain region
by the method explained in Appendix \ref{sec:err}.
The magnitude of this uncertainty gives an estimate for
the goodness of the given model $m(\omega)$\cite{jarr,hats:rev}.

\subsection{Test}
\newcounter{loman}
\setcounter{loman}{1}
\label{sec:test}

Several tests of MEM have already been carried out in Ref.\cite{hats:rev},
where the dependence of results on
the number of time slices $N_{D}$, the size of errors of data and
the model $m(\omega)$ have been examined using mock-up data created from
test spectral functions. 
The following conclusions have been drawn from the tests:
\begin{enumerate}
\item Decreasing the error of data $D(\tau)$ is more important than
increasing $N_{D}$ for obtaining better estimates of $f(\omega)$
which reproduce the original spectral function more closely.
\item It is better to include information of $f(\omega)$, such
as the asymptotic value, if it is known, into the model $m(\omega)$.
\item If the obtained $f(\omega)$ depends strongly on the model,
a better model in the sense of leading to a $f(\omega)$ 
which is closer to the original spectral function 
gives smaller errors for the averaged $f(\omega)$.
\item The error of the averaged $f(\omega)$ in a certain region 
can be used to measure
the significance of $f(\omega)$ in the region. For example,
if the error of the averaged $f(\omega)$ around a peak is
much smaller than the averaged value, the peak is likely to be true, and
vice versa.
\end{enumerate}

Before applying MEM to actual data, we perform further tests
on (1)the dependence on $N_{D}$ and the temporal separation of data 
$\Delta\tau$, and (2)the dependence on the choice of  P$[\alpha|Hm]$.
For these tests
we use a realistic spectral function in the vector channel
of the $e^{+}e^{-}$ annihilation\cite{hats:rev,real}, which is given by
$f_{in}(\omega)=\rho_{in}(\omega)\omega^{2}$,
where the factor $\omega^{2}$ is expected 
from the dimension of meson spectral function, with
\begin{equation}
\rho_{in}(\omega)=\frac{2}{\pi}\left[F^{2}_{\rho}\frac{\Gamma_{\rho}(\omega)m_{\rho}}
{(\omega^{2}-m_{\rho}^{2})^{2}+\Gamma^{2}_{\rho}(\omega)
m_{\rho}^{2}}+\frac{1}{8\pi}
\left(1+\frac{\alpha_{s}}{\pi}\right)\frac{1}
{1+e^{(\omega_{0}-\omega)/\delta}}\right].
\end{equation}
Here $F_{\rho}$ is the residue of $\rho$ meson resonance defined by
\begin{equation}
\langle0|\bar{d}\gamma_{\mu}u|\rho\rangle=
\sqrt{2}F_{\rho}m_{\rho}\epsilon_{\mu}
=\sqrt{2}f_{\rho}m^{2}_{\rho}\epsilon_{\mu},
\end{equation}
with the polarization vector $\epsilon_{\mu}$, and
$\Gamma_{\rho}(\omega)$ includes the $\theta$-function which represents
the threshold of $\rho \rightarrow \pi\,\pi$ decay as
\begin{equation}
\Gamma_{\rho}(\omega)=\frac{1}{48\pi}\frac{m_{\rho}^{3}}{F_{\rho}^{2}}
\left(1-\frac{4m_{\pi}^{2}}{\omega^{2}}\right)^{\frac{3}{2}}
\theta(\omega-2m_{\pi}).
\end{equation}

\begin{figure}[!t]
\begin{center}
\leavevmode
\epsfxsize=6cm\epsfbox{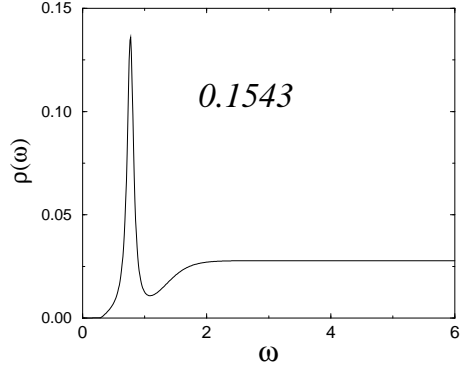} 
\caption{
The input spectral function
$\rho_{in}(\omega)$.
The value in the figure is the area under the curve for $0\le \omega \le 6$.
\label{fig:rhoin}}
\end{center}
\end{figure}

We make dimensionful quantities dimensionless using 
the lattice spacing $a$, 
$\omega \rightarrow \omega a,\ \tau \rightarrow \tau /a$
where $a$ is set to 1 GeV$^{-1}$.
The values of parameters are 
\begin{equation}
m_{\rho}=0.77,m_{\pi}=0.14,F_{\rho}=0.142,
\omega_{0}=1.3,\delta=0.2,\alpha_{s}=0.3,
\end{equation}
where $\alpha_{s}$ is independent of $\omega$ for simplicity.
The shape of $\rho_{in}(\omega)$ for this choice of parameters
is shown in Fig.~\ref{fig:rhoin}.
The value in the figure represents the area of $\rho_{in}(\omega)$
for $0 \le \omega \le 6$.

We make mock-up data $D(\tau)$ from $f_{in}(\omega)$ as follows.
(\roman{loman})
The central value of $D(\tau)$ is given by
integrating the spectral function $f_{in}(\omega)$ and 
a kernel $K(\omega ,\tau)=e^{-\omega\tau}$
over $\omega$ in the same way as $D_{\!f}(\tau)$ in (\ref{eq:integ}).
\setcounter{loman}{2}(\roman{loman}) 
Errors of $D(\tau)$ are generated by gaussian random numbers
with the variance $\sigma(\tau_{i})=b\cdot e^{a\tau_{i}}D(\tau_{i})$,
$a=0.1$, $b = 10^{-10}$,
in order to incorporate the fact that
the error of lattice correlation functions increases as $\tau$ increases.

In this test, no correlation between different $\tau$ is taken into account,
thus the covariance matrix $C$ is set to be diagonal.
The model function is given by $m(\omega) = m_{0}\omega^{2}$ 
with $m_{0} = 0.0277$,
which is motivated by the value of $\rho_{in}(\omega \rightarrow \infty)$.
We set the maximum value of $\omega$, $\omega_{max}=6$, 
and the $\omega$-space is discretized with 
an equal separation $\Delta\omega=0.01$, and $N_{\omega}=600$.
We also 
calculate the area of the MEM result $\rho_{out}(\omega)$ for 
$0\le \omega \le \omega_{max}$ and define 
$r=\sum_{l=1}^{N_{\omega}}(\rho_{in}(\omega_{l})-\rho_{out}(\omega_{l}))^{2}$,
to measure the difference between 
$\rho_{in}$ and $\rho_{out}$.

\begin{figure}[p]
\begin{center}
\begin{tabular}{cccc}
&$N_{D}=16$&$N_{D}=31$&$N_{D}=46$\\
\raisebox{2.cm}{$\Delta\tau=0.5$}&
\leavevmode
\epsfxsize=4.2cm\epsfbox{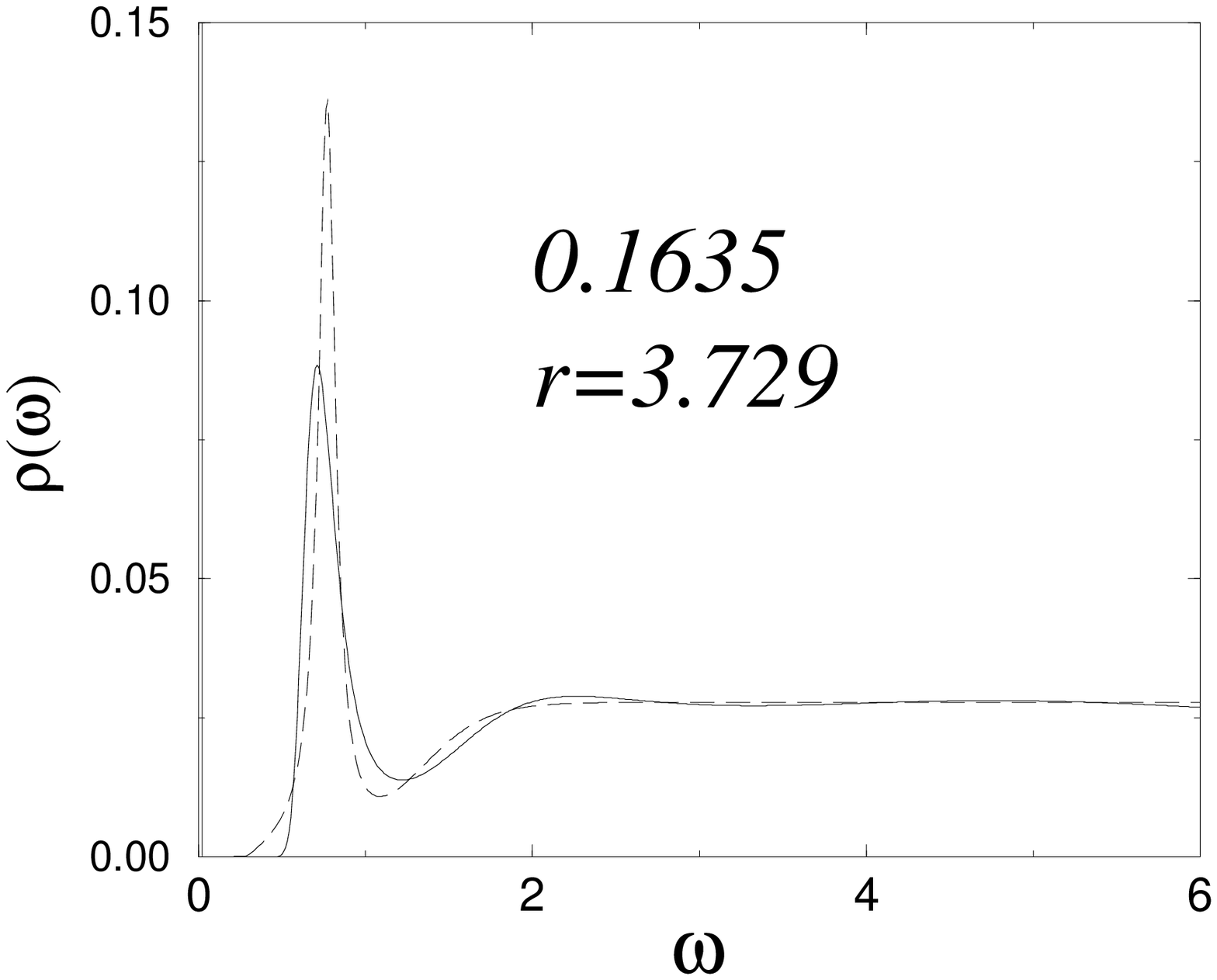} 
&
\leavevmode
\epsfxsize=4.2cm\epsfbox{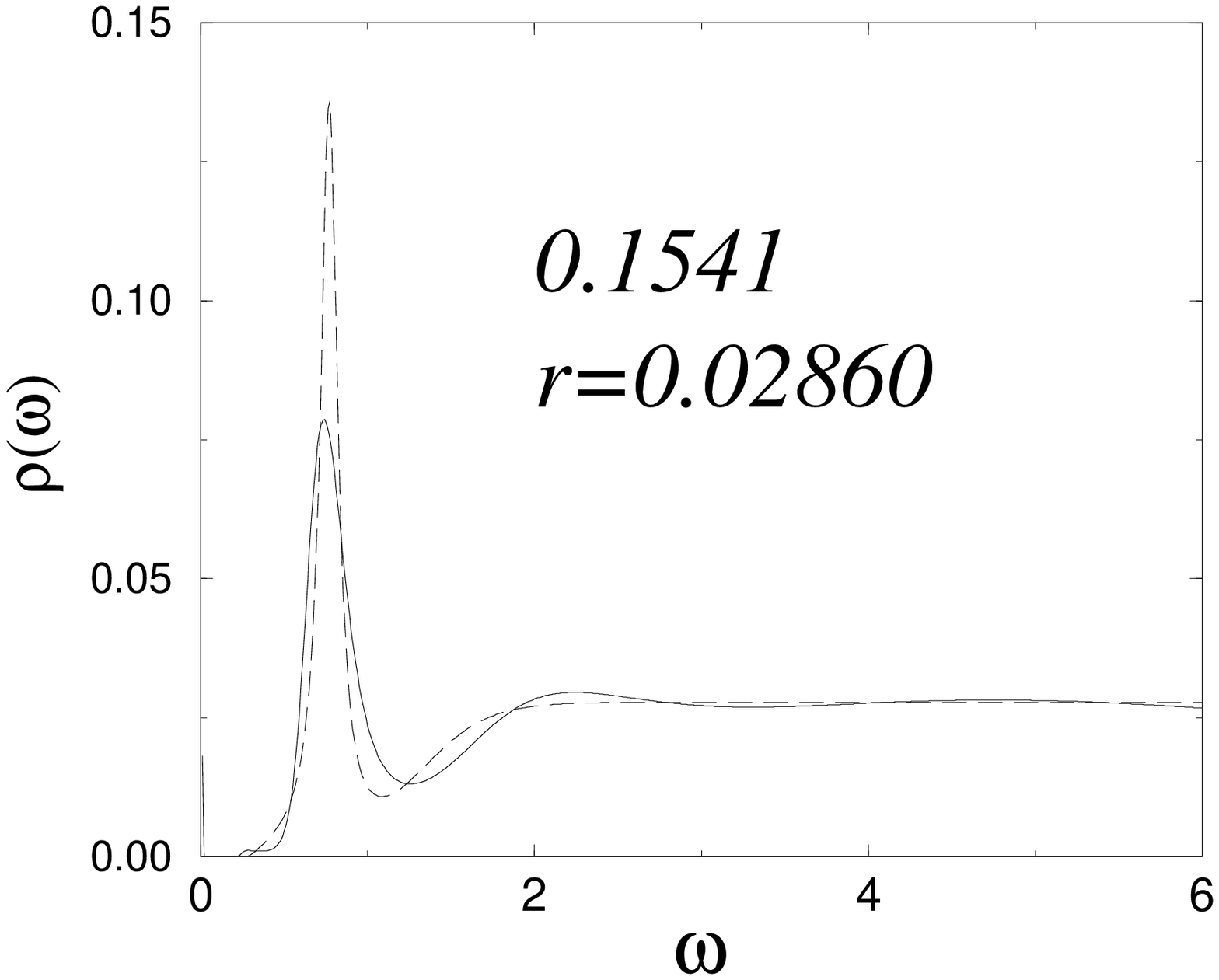} 
&
\leavevmode
\epsfxsize=4.2cm\epsfbox{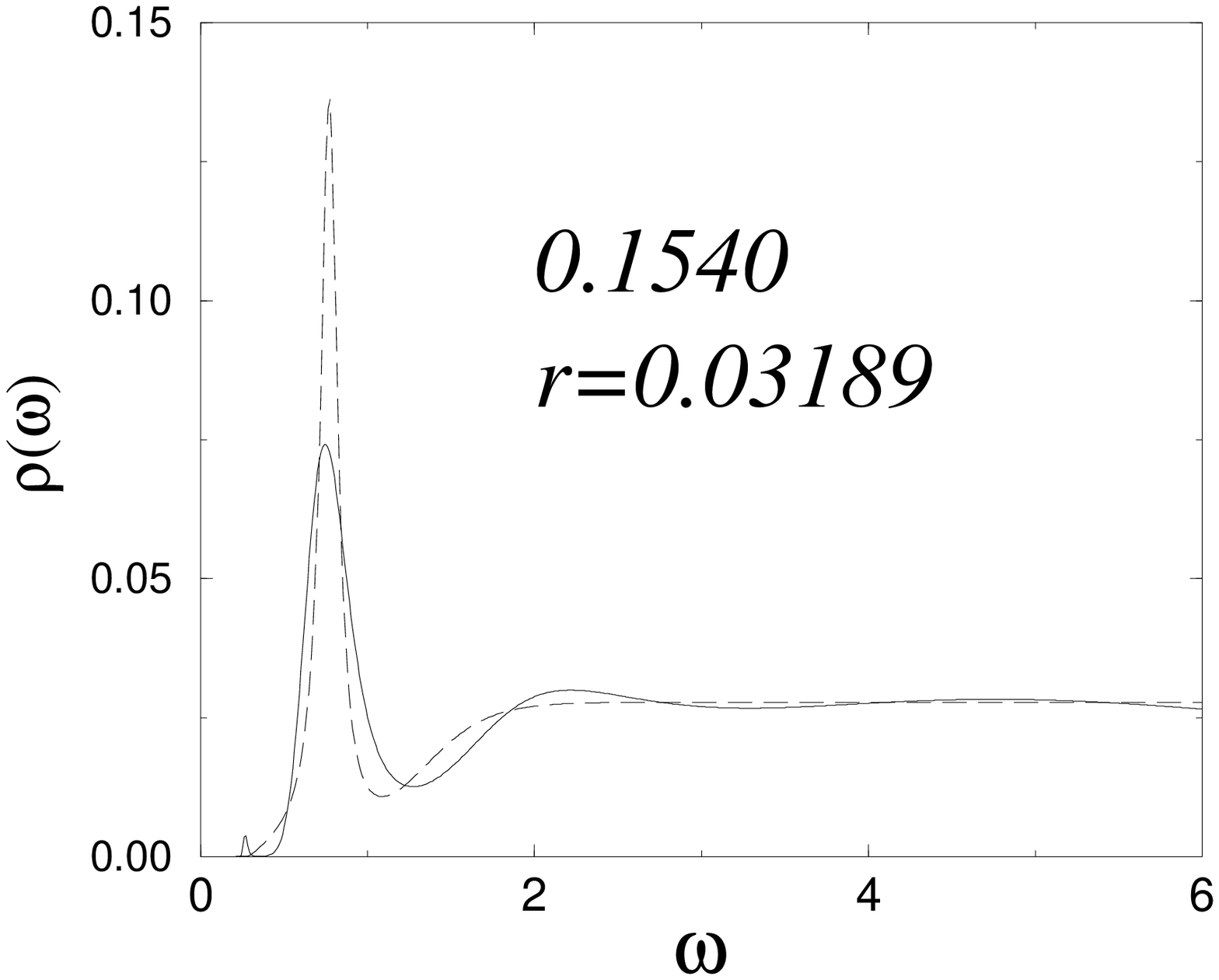} 
\\
\raisebox{2.cm}{$\Delta\tau=0.33$}&
\leavevmode
\epsfxsize=4.2cm\epsfbox{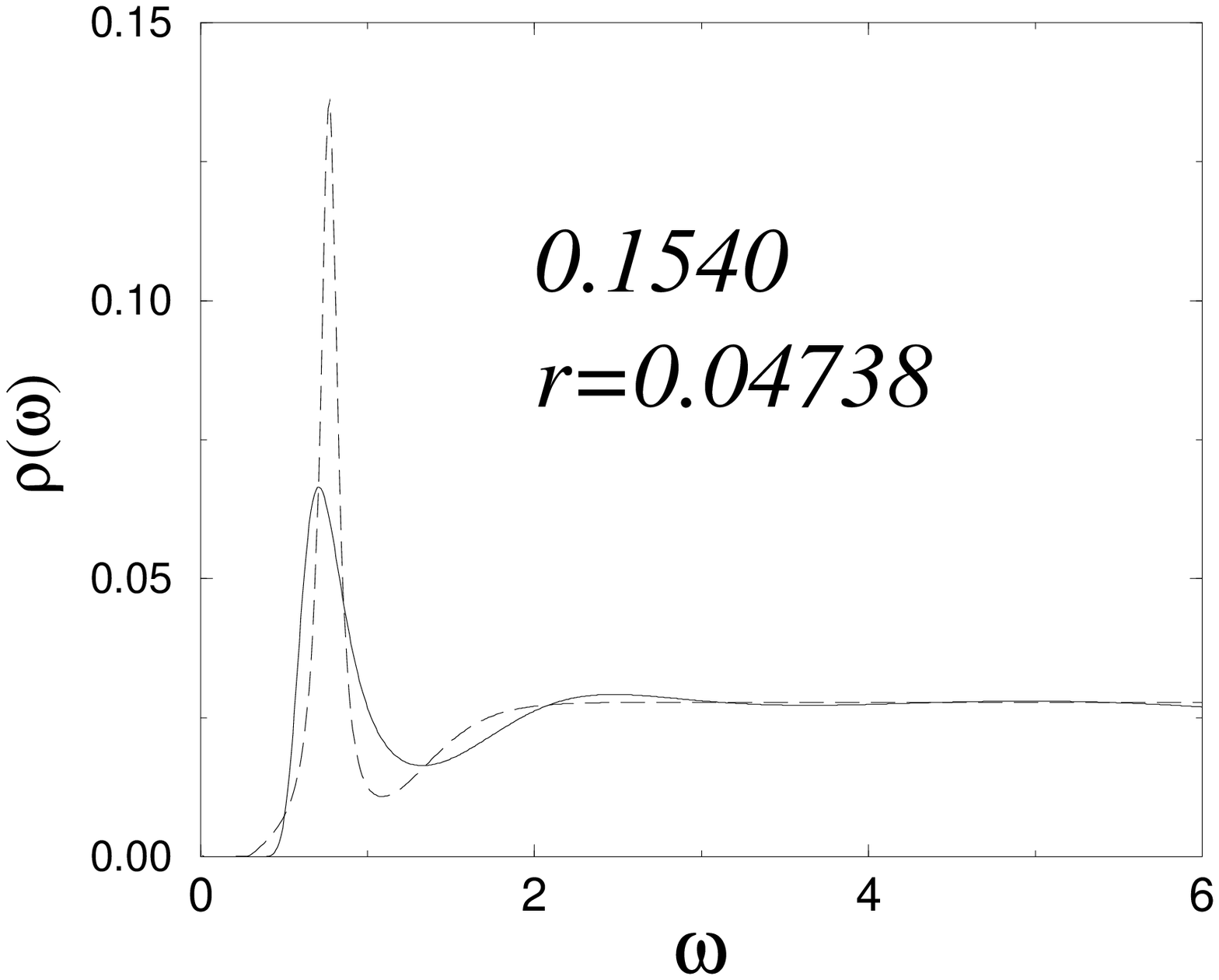} 
&
\leavevmode
\epsfxsize=4.2cm\epsfbox{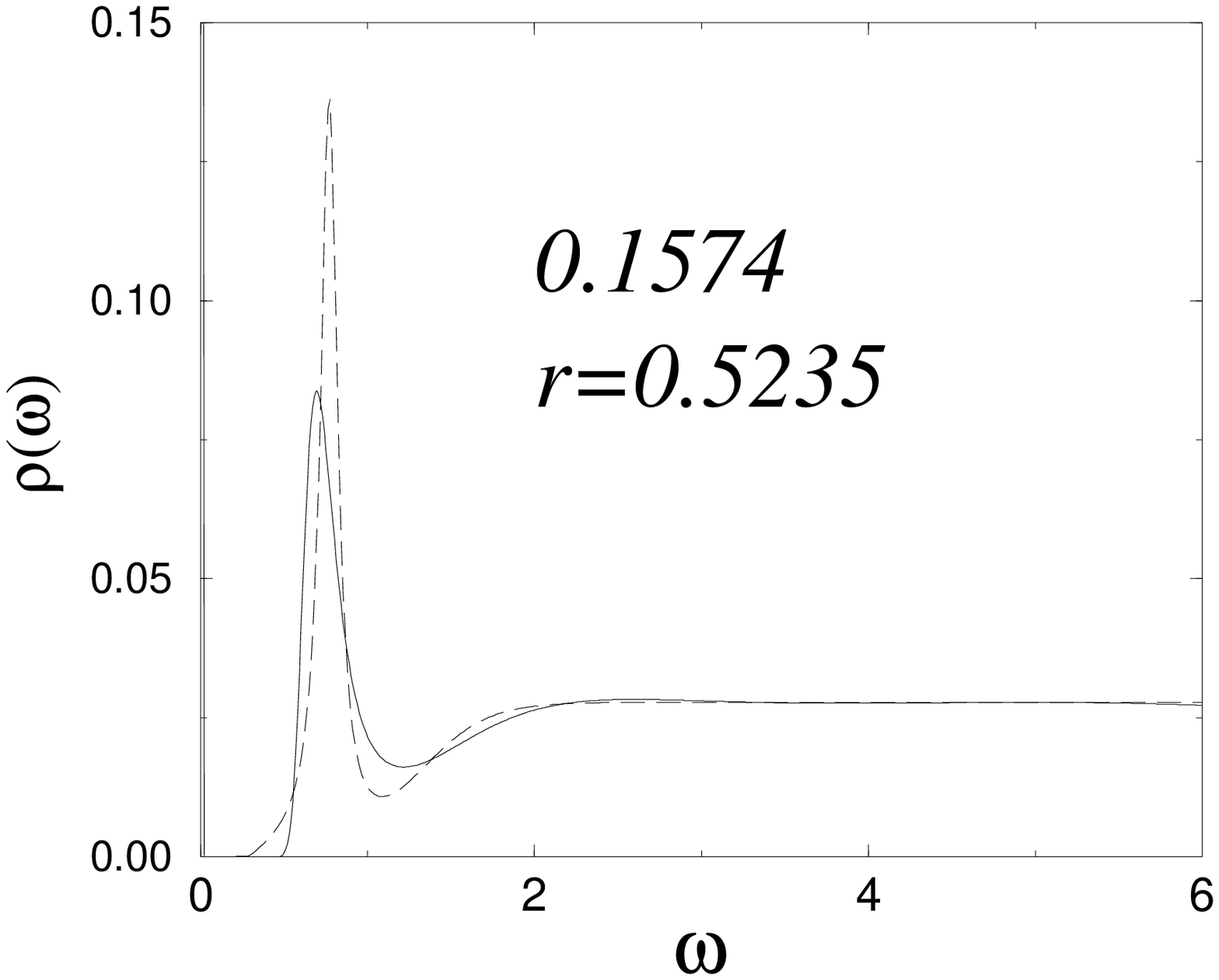} 
&
\leavevmode
\epsfxsize=4.2cm\epsfbox{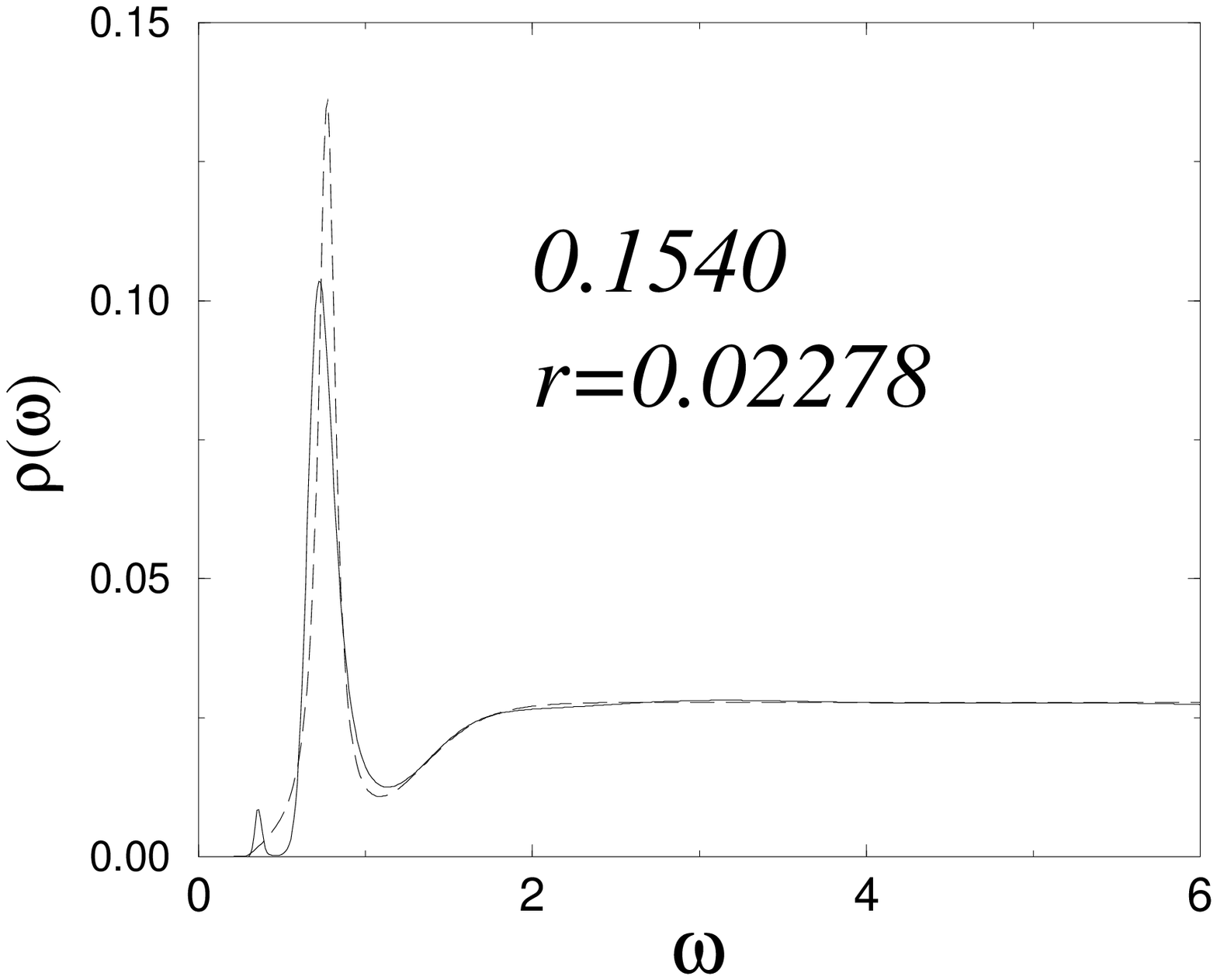}
\end{tabular}
\end{center}
\caption{The output spectral function $\rho_{out}(\omega)$
obtained by MEM for 
different $\Delta\tau$ and $N_{D}$ is shown by solid lines.
The input $\rho_{in}(\omega)$ is shown by long dashed lines.
The values in each figure are the area of $\rho_{out}(\omega)$ and 
$r=\sum_{l=1}^{N_{\omega}}(\rho_{in}(\omega_{l})-\rho_{out}(\omega_{l}))^{2}$.
\label{fig:4}}
\end{figure}

\begin{figure}[!p]
\begin{center}
\leavevmode
\epsfxsize=6cm\epsfbox{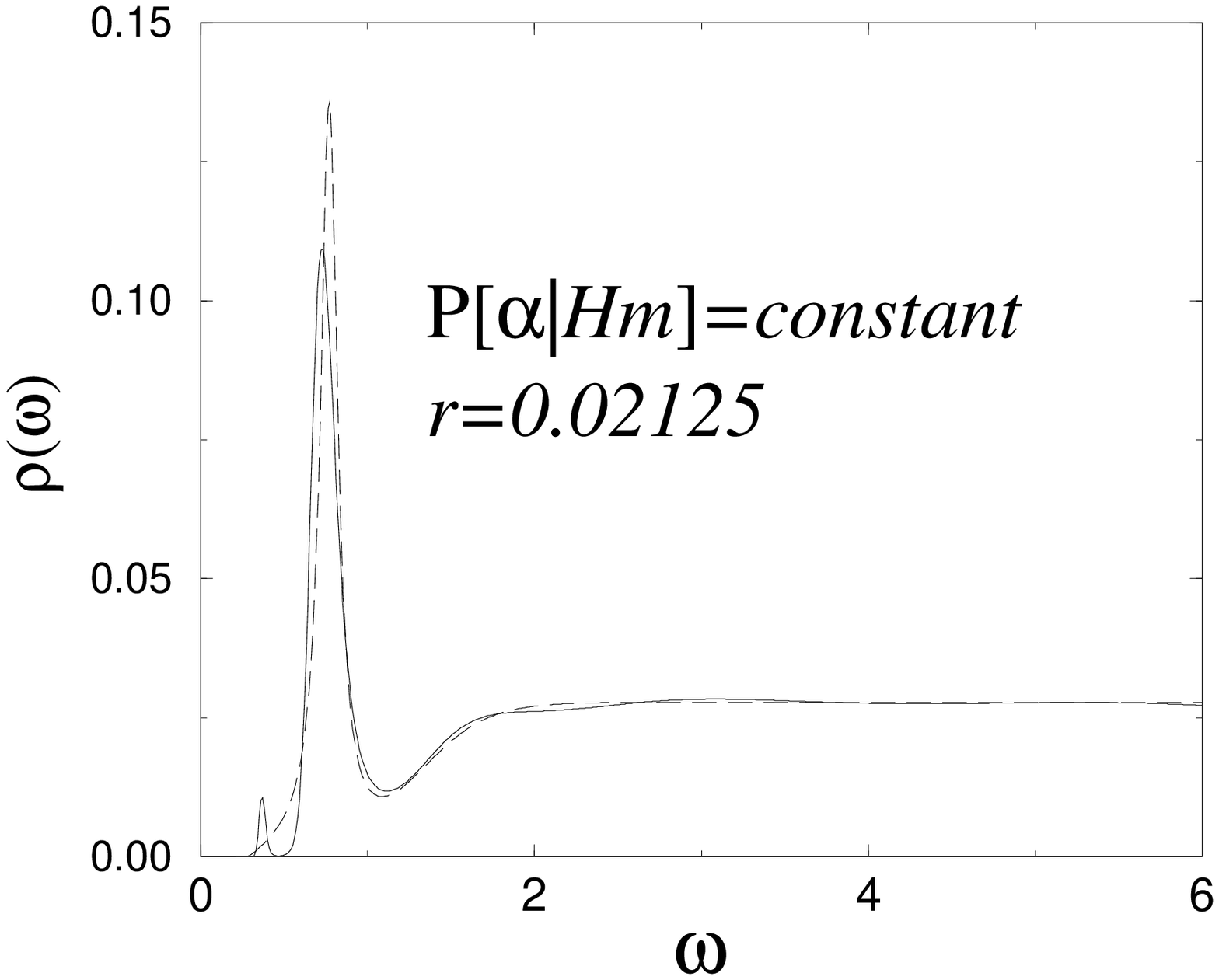} 
\leavevmode
\epsfxsize=6cm\epsfbox{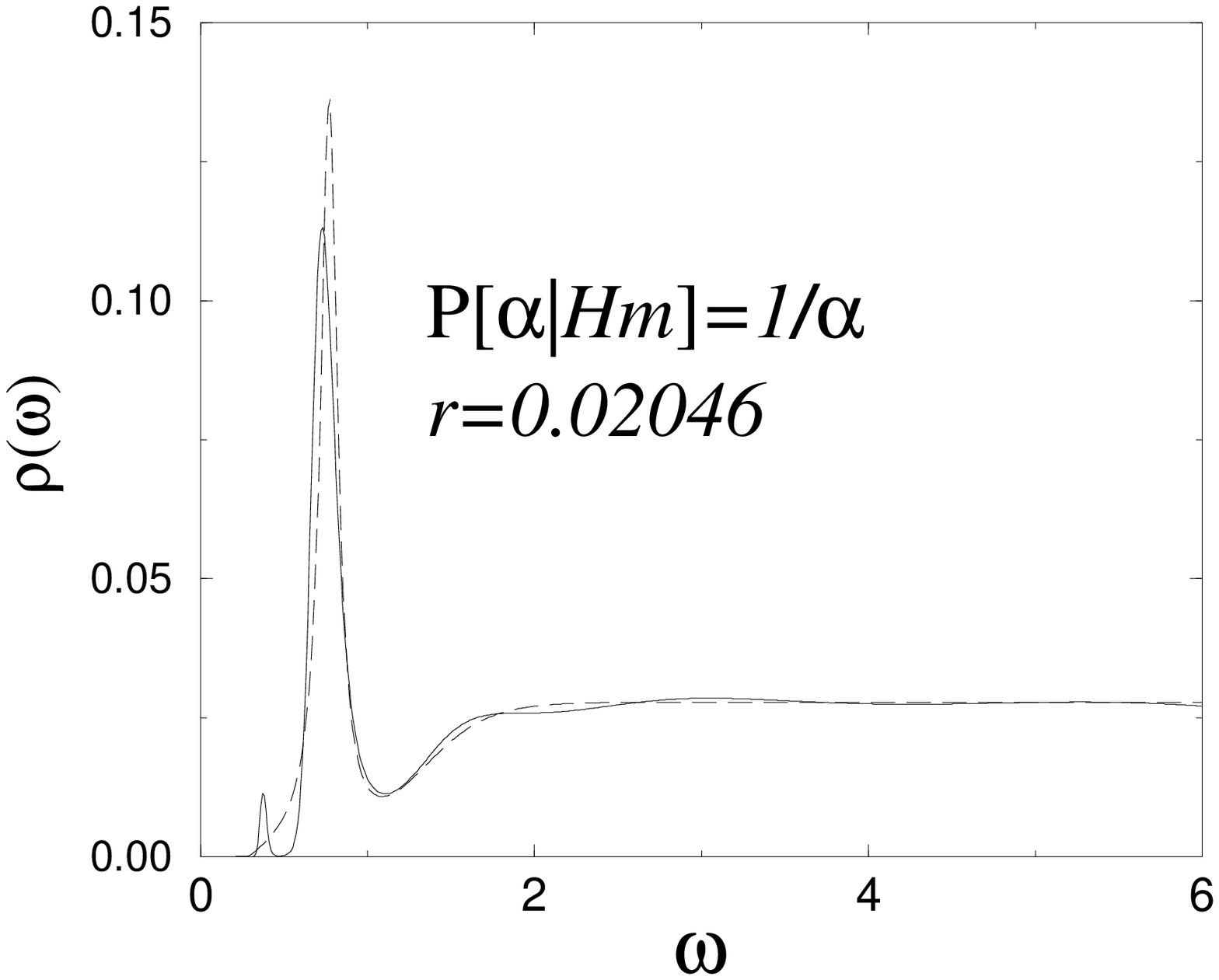} 
\\
\leavevmode
\epsfxsize=6cm\epsfbox{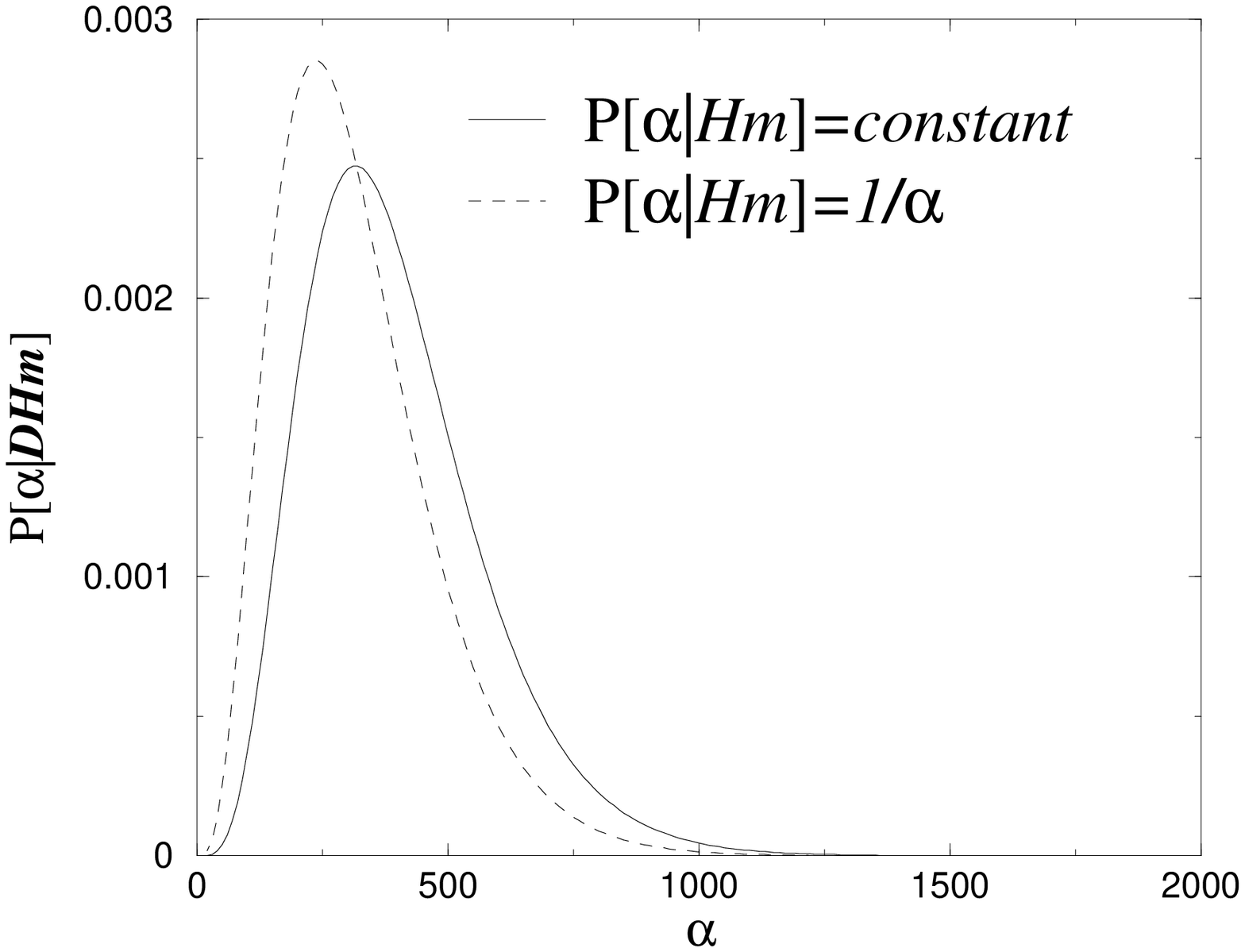} 
\end{center}
\caption{Influence of the choice of ${\mathrm P}[\alpha|Hm]$.
The left figure is for ${\mathrm P}[\alpha|Hm]=$ constant, and 
the right for ${\mathrm P}[\alpha|Hm]=1/\alpha$.
The figure below shows the corresponding ${\mathrm P}[\alpha|DHm]$
normalized to unity for which
data with $\Delta\tau=0.33$ and $N_{D}=46$ is used. 
The input $\rho_{in}(\omega)$ is shown by the long dashed lines,
and 
$r=\sum_{l=1}^{N_{\omega}}(\rho_{in}(\omega_{l})-\rho_{out}(\omega_{l}))^{2}$
represents the difference from $\rho_{in}(\omega)$.
\label{fig:3}}
\end{figure}

We summarize the result of $\rho_{out}(\omega)$ in various cases
as follows.\\
(1) To investigate the dependence of $\rho_{out}(\omega)$
on $\Delta\tau$ and $N_D$,
we extract $\rho_{out}(\omega)$ by MEM,
from data with $\Delta\tau =0.5,0.33$ and $N_{D}=16,31,46$,
as shown in Fig.~\ref{fig:4}.
Data at large $\tau$ are necessary to reconstruct
$\rho_{out}(\omega)$ at small $\omega$ correctly,
as seen from the fact that
a false peak sometimes appears around $\omega =0$
from data with $\Delta\tau=0.5$ and $N_D=16$
($\tau_{max}=\Delta\tau(N_D -1)=7.5$) or
with $\Delta\tau=0.33$ and $N_D=31$ ($\tau_{max}=10$).
Once $\tau_{max}$ becomes large enough (larger than 15 in this case),
a smaller $\Delta\tau$ is better for the result, as seen from the 
comparison between results from
data with $\Delta\tau =0.5$ and $\Delta\tau =0.33$
at $N_D=46$.\\
(2) We also check the dependence of $\rho_{out}(\omega)$
on two forms of ${\mathrm P}[\alpha|Hm]$, either
${\mathrm P}[\alpha|Hm] =$ constant or $1/\alpha$.
As shown in Fig.~\ref{fig:3},
the two choices give almost identical shapes of $\rho_{out}(\omega)$,
though the weight factor ${\mathrm P}[\alpha|DHm]$ is rather different
between the two cases.

Our investigations add further information on the parameter dependence
of the result in MEM, which we summarize as the following three points:
\begin{enumerate}
\item[5.] $\tau_{max} = \Delta\tau (N_D -1)$ must be sufficiently large
for a reliable result of $f(\omega)$.
\item[6.] Once $\tau_{max}$ is taken large enough, 
smaller $\Delta\tau$ is better.
\item[7.] The result $\rho_{out}(\omega)$ is insensitive to
the choice of P$[\alpha|Hm]$.
\end{enumerate}

\section{lattice QCD data and parameters in MEM analysis}
\label{sec:para}

We now apply MEM to the lattice correlation functions previously obtained 
in quenched QCD\cite{CPdata,pre} with the plaquette action for gluons and
the Wilson action for quarks.
The simulation was performed at four values of $\beta$, 
corresponding to $a^{-1}=2$--4 GeV for the continuum extrapolation,
on $32^3\times56$ to $64^3\times112$ lattices with the spatial size 
about 3~fm.
Simulation parameters are compiled in Table~\ref{tab:1}.
At each $\beta$, five values of the hopping parameter $\kappa$,
which correspond to $m_{\pi}/m_{\rho} \approx$ 0.75, 0.7, 0.6, 0.5
and 0.4, were employed for the chiral extrapolation.
The values of hopping parameters are numbered from heavy to light 
in Table~\ref{tab:1}.
For example,
we call $\kappa$ corresponding to the lightest and the heaviest quark masses 
as $K51$.
Except for an additive renormalization factor,
the average quark mass is equal to
the average inverse hopping parameter $K^{-1}$ 
given by
\begin{equation}
K^{-1} = \frac{1}{2}\left(\kappa_{1}^{-1} + \kappa_{2}^{-1}\right),
\end{equation}
where $\kappa_{1}$ and $\kappa_{2}$ are the hopping parameters 
of quark and anti-quark in the meson.

In our MEM analysis, we employ pseudoscalar and vector  
meson correlation functions, defined by
\begin{equation}
\sum_{{\mathbf x}}\langle \bar{d}\,\Gamma u(\tau,{\mathbf x})
\,(\bar{d}\,\Gamma u)^{\dagger}(0,{\mathbf 0}) \rangle=
\int d\omega \, f(\omega) K(\omega , \tau) ,
\end{equation}
where $\Gamma$ is $\gamma_{5}(\gamma_{\mu})$ for pseudoscalar(vector) meson,
$f(\omega)$ is a spectral function and $K(\omega , \tau)$ is a kernel. 
We use only point source data to
satisfy the condition that $f(\omega) \ge 0$.
Since the spectral function of the meson propagator has dimension 2,
we define a dimensionless function $\rho (\omega)$ as
\begin{equation}
f(\omega) = \rho(\omega)\omega^{2}.
\end{equation}

\begin{figure}[t]
\begin{tabular}{cccc}
\raisebox{2.5cm}{PS}&
\leavevmode
\epsfxsize=4.5cm\epsfbox{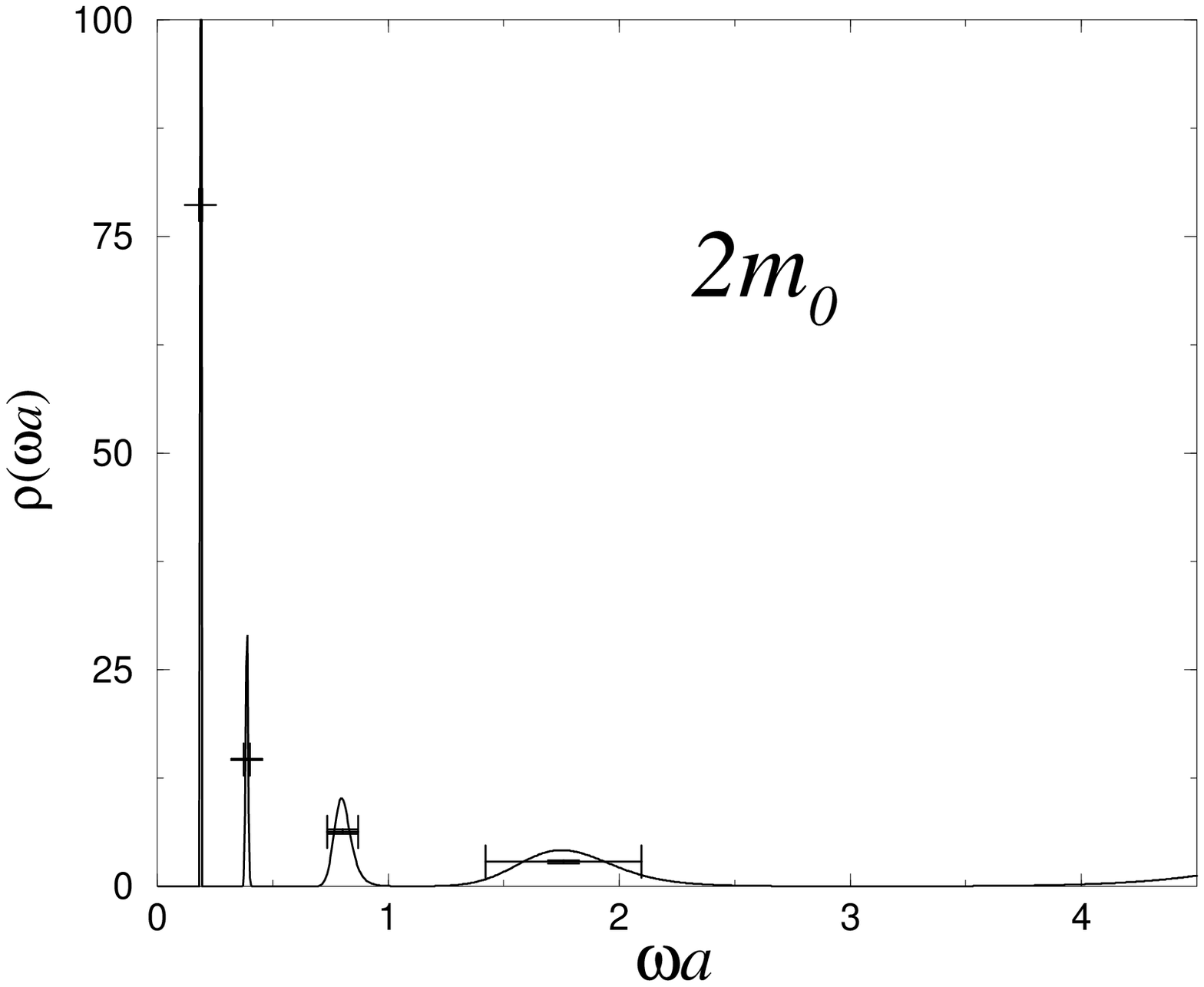} 
&
\leavevmode
\epsfxsize=4.5cm\epsfbox{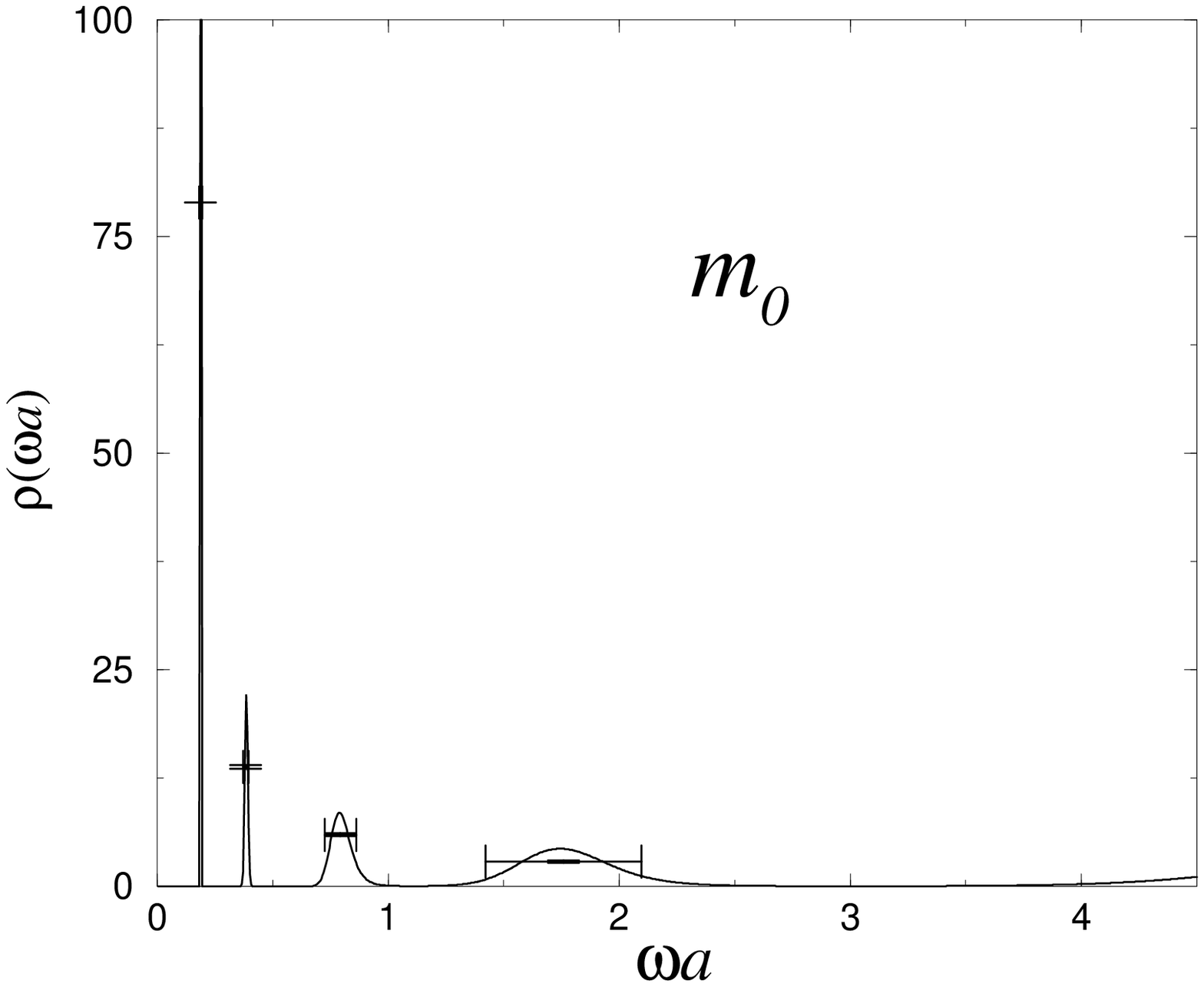} 
&
\leavevmode
\epsfxsize=4.5cm\epsfbox{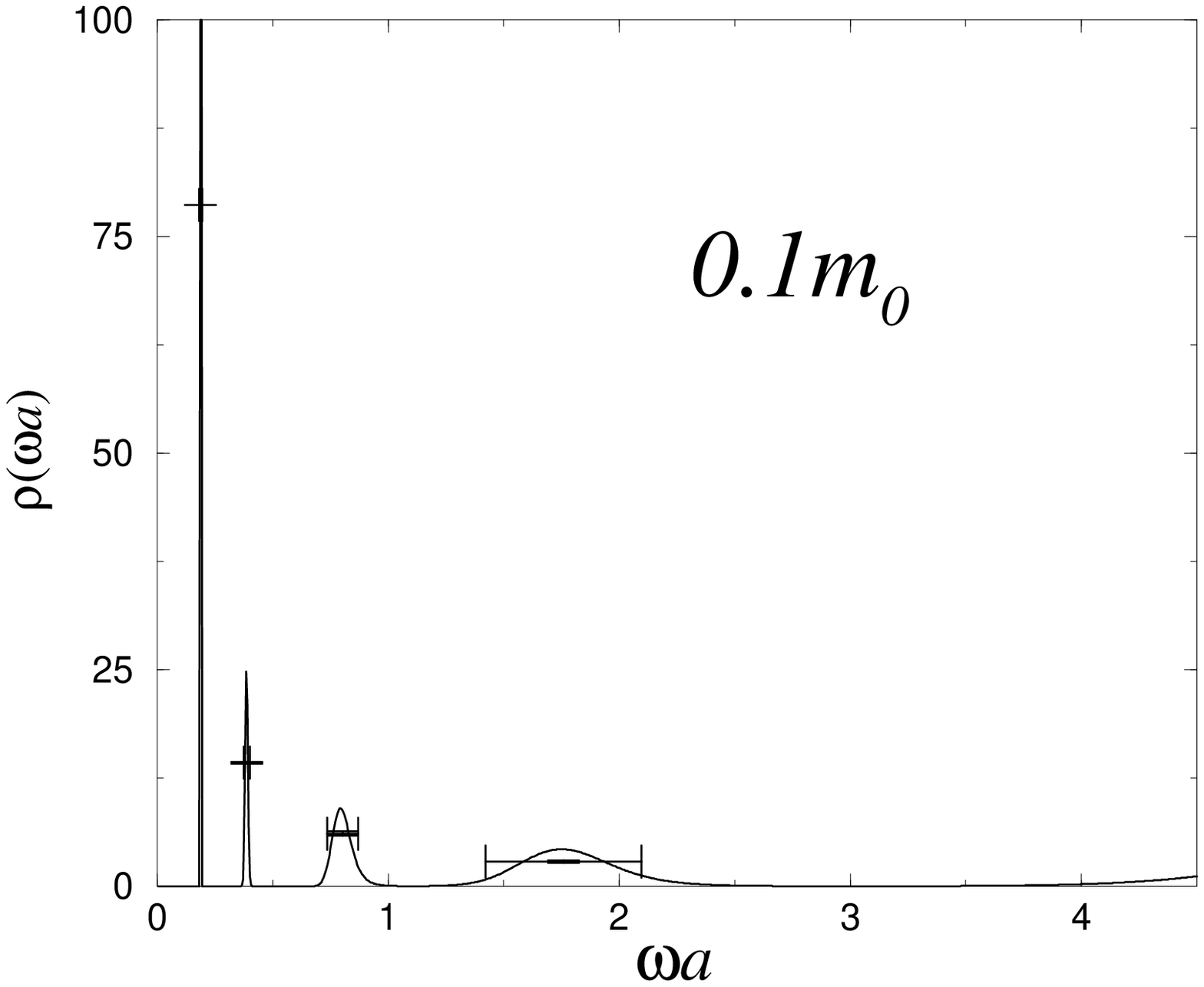} 
\\
\raisebox{2.5cm}{V}&
\leavevmode
\epsfxsize=4.5cm\epsfbox{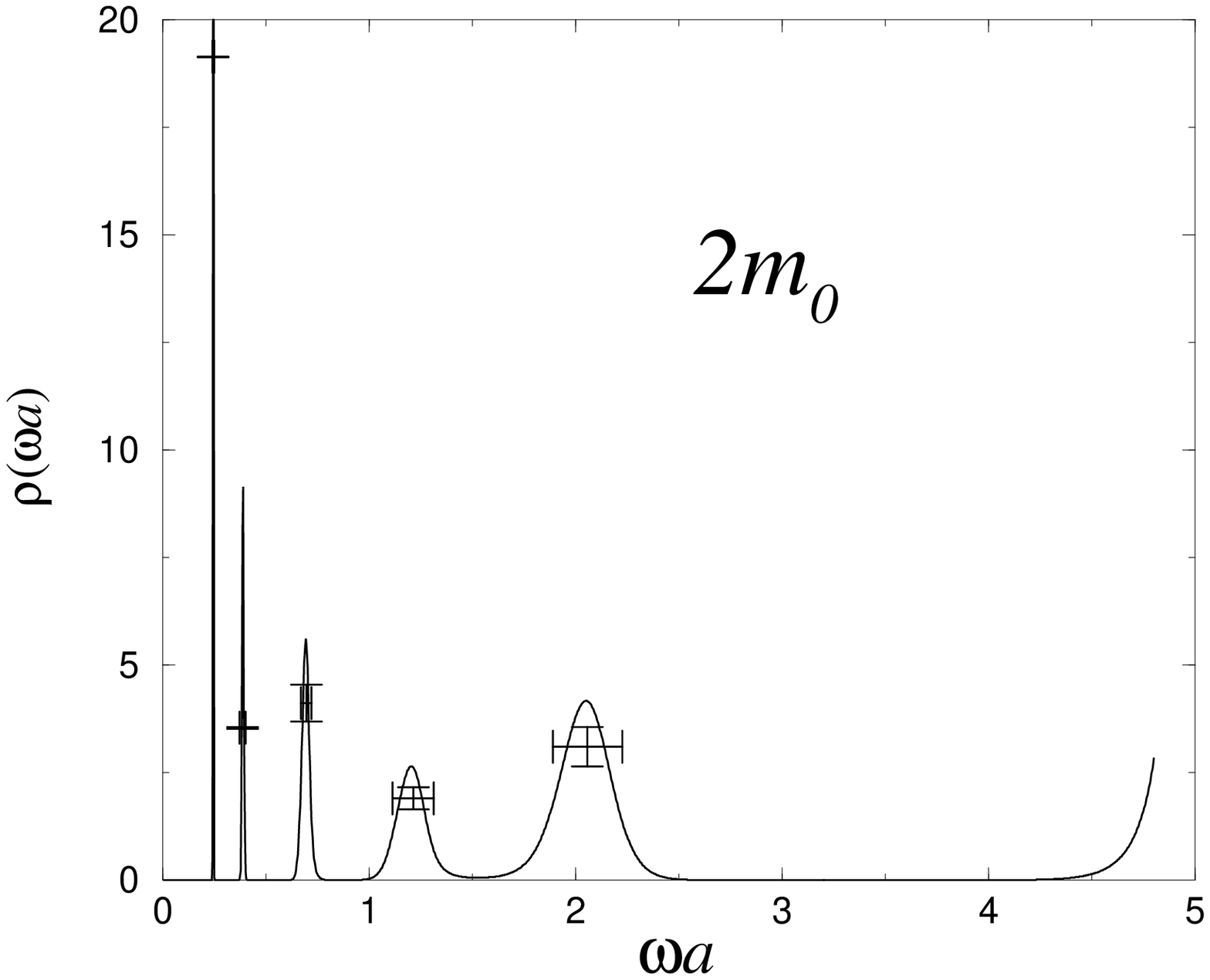} 
&
\leavevmode
\epsfxsize=4.5cm\epsfbox{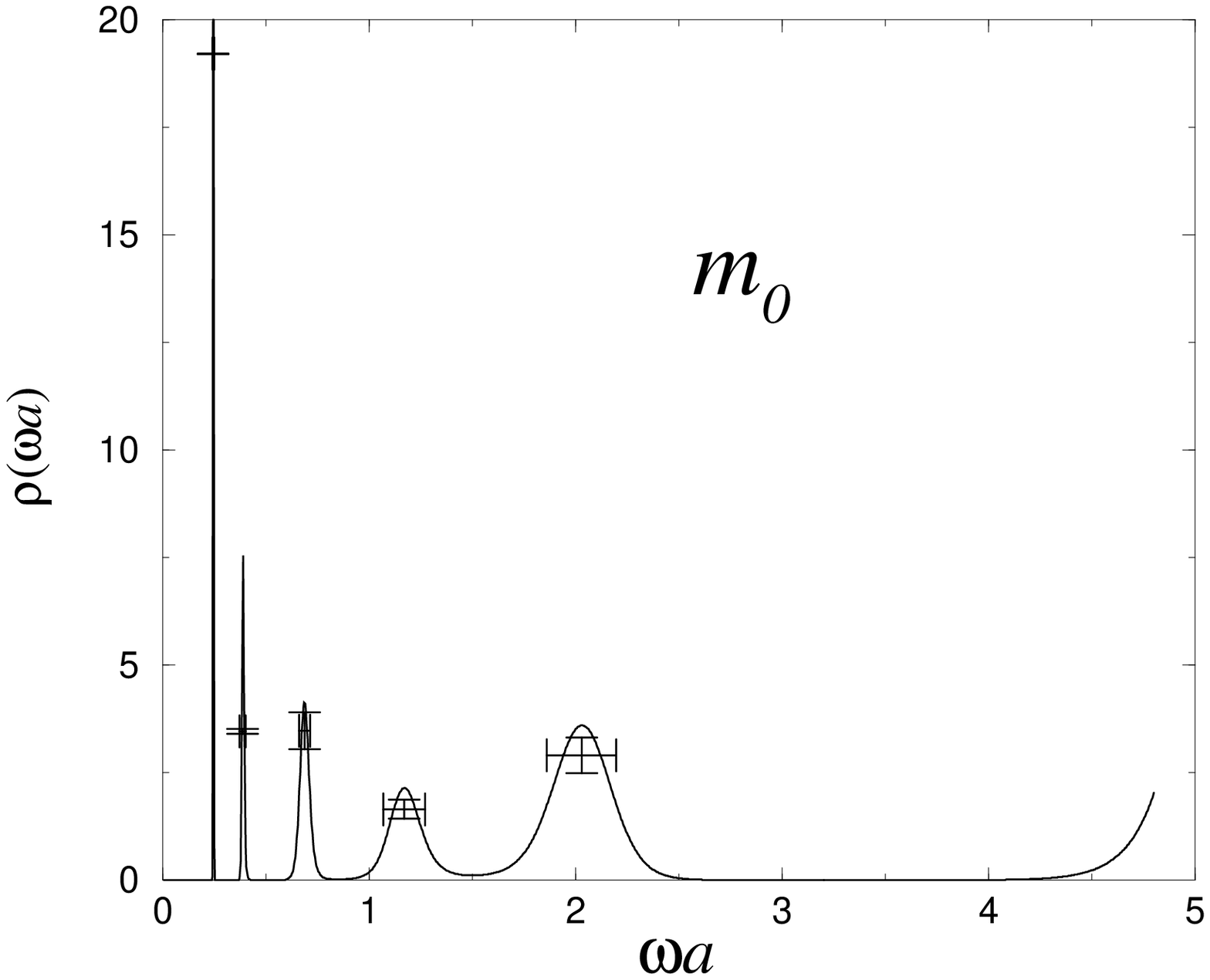} 
&
\leavevmode
\epsfxsize=4.5cm\epsfbox{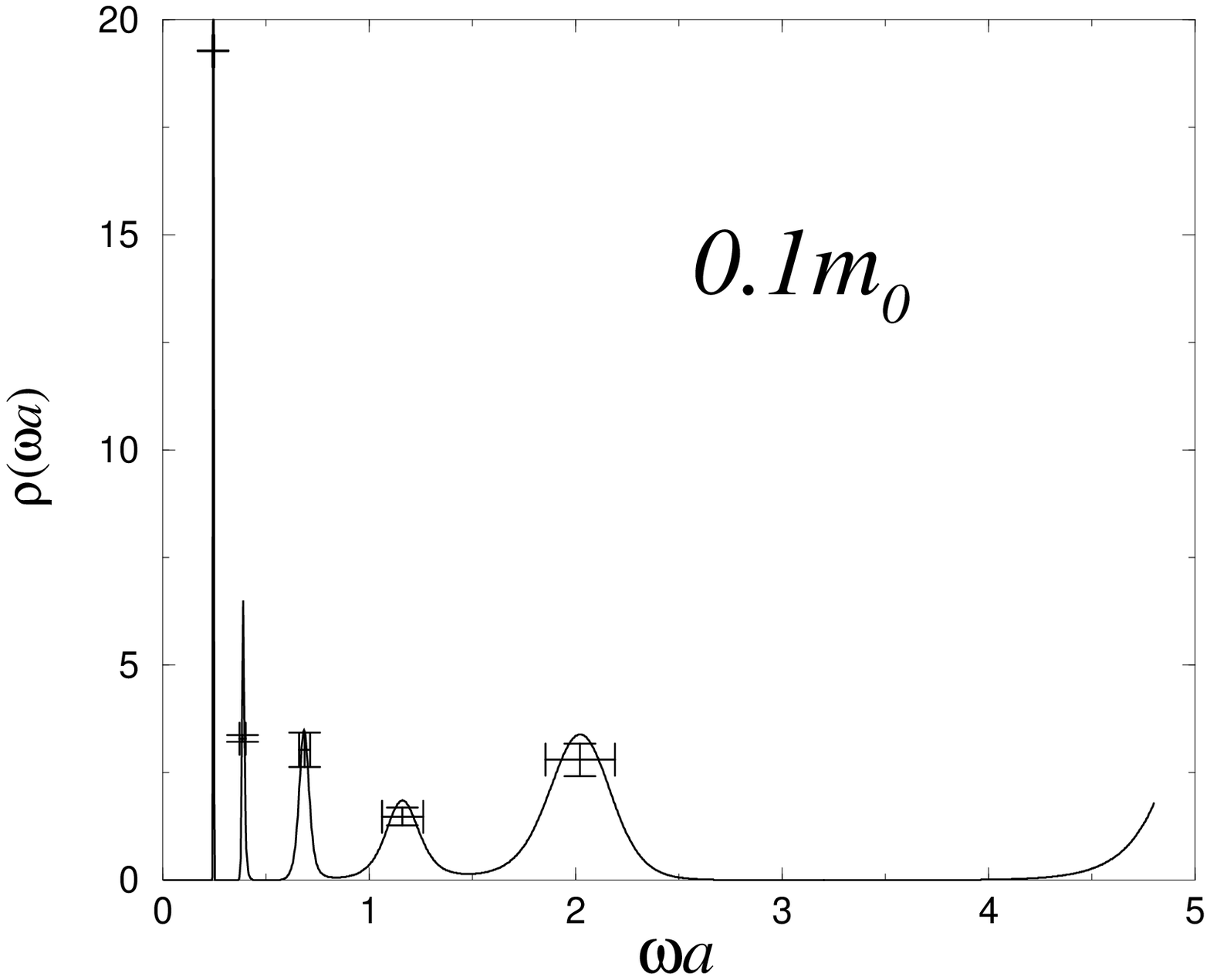} 
\end{tabular}
\caption{
Model($m_{0}$) dependence for 
pseudoscalar (PS) and vector (V) channels at $\beta = 6.47$ and $K11$.
\label{fig:07}
}
\end{figure}

The model is chosen to be $m(\omega) = m_0\omega^2$ and
the value of $m_0$ is taken equal to
the asymptotic value of $\rho(\omega)$ in
perturbation theory\cite{hats:rev} given by
\begin{equation}
m_{0} = \frac{C_{1}}{4\pi^{2}}\left(1+C_{2}\frac{\alpha_{s}}{\pi}
\right)
\left(\frac{1}{Z^{2}}
\prod_{i=1}^{2}\frac{1}{2\kappa_{i}}\right)
\label{eq:mod},
\end{equation}
where $\alpha_{s}$ is the strong coupling constant,
the coefficients $C_i$'s are perturbatively calculated
in continuum QCD\cite{pert},
and $Z$ is the renormalization constant for 
pseudoscalar (PS) or vector (V) operator.
The spectral function from our data is 
insensitive to the value of $m_0$,
as shown in Fig.~\ref{fig:07},
where
$f(\omega)$ obtained with three different models
are plotted
for pseudoscalar and vector mesons at $\beta =6.47$ and $K11$.
In the figure
the horizontal bars indicate the region over which
the result is averaged, 
while the vertical bars indicate the uncertainty in the 
averaged value of the result.
Both averaged spectral functions and their uncertainty 
are almost identical for different models.
Because of this property,
we simply take $\alpha_s =0.21$ and employ
the non-perturbative $Z_{{\mathrm V}}$ and the perturbative $Z_{{\mathrm PS}}$
calculated at $\beta=5.90$ in (\ref{eq:mod})
for all $\beta$.
The normalization factor $1/2\kappa$ is used also 
for the pseudoscalar meson with tadpole-improved $Z_{{\mathrm PS}}$.
Values of $Z$'s as well as $C_i$'s are given in 
Table~\ref{tab:3}.

Other parameters in the MEM analysis such as $N_D$ and $(\omega a)_{max}$,
are determined as follows.
We take $N_D$ as large as possible unless the error of data becomes too
large for a reliable result,
and
we choose $(\omega a)_{max}\gg\pi$ and increase it until the result
becomes stable.
Both parameters are also given in Table~\ref{tab:3}.
For $\Delta\omega$, which should be smaller than 1/$T$,
we take $\Delta\omega =10^{-4}$ around the peak of the ground state
to determine the ground state mass accurately,
while $\Delta\omega =2.5\times 10^{-3}$ away from the peak.

\section{Results}
\label{results}

In this section, we present our results of spectral function
for pseudoscalar and vector meson propagators, from which we extract
physical quantities such as masses and decay constants.

\subsection{Spectral functions}
\begin{figure}[p]
\begin{center}
\begin{tabular}{cc}
\vspace*{-0.3cm}
\leavevmode
\epsfxsize=6.2cm\epsfbox{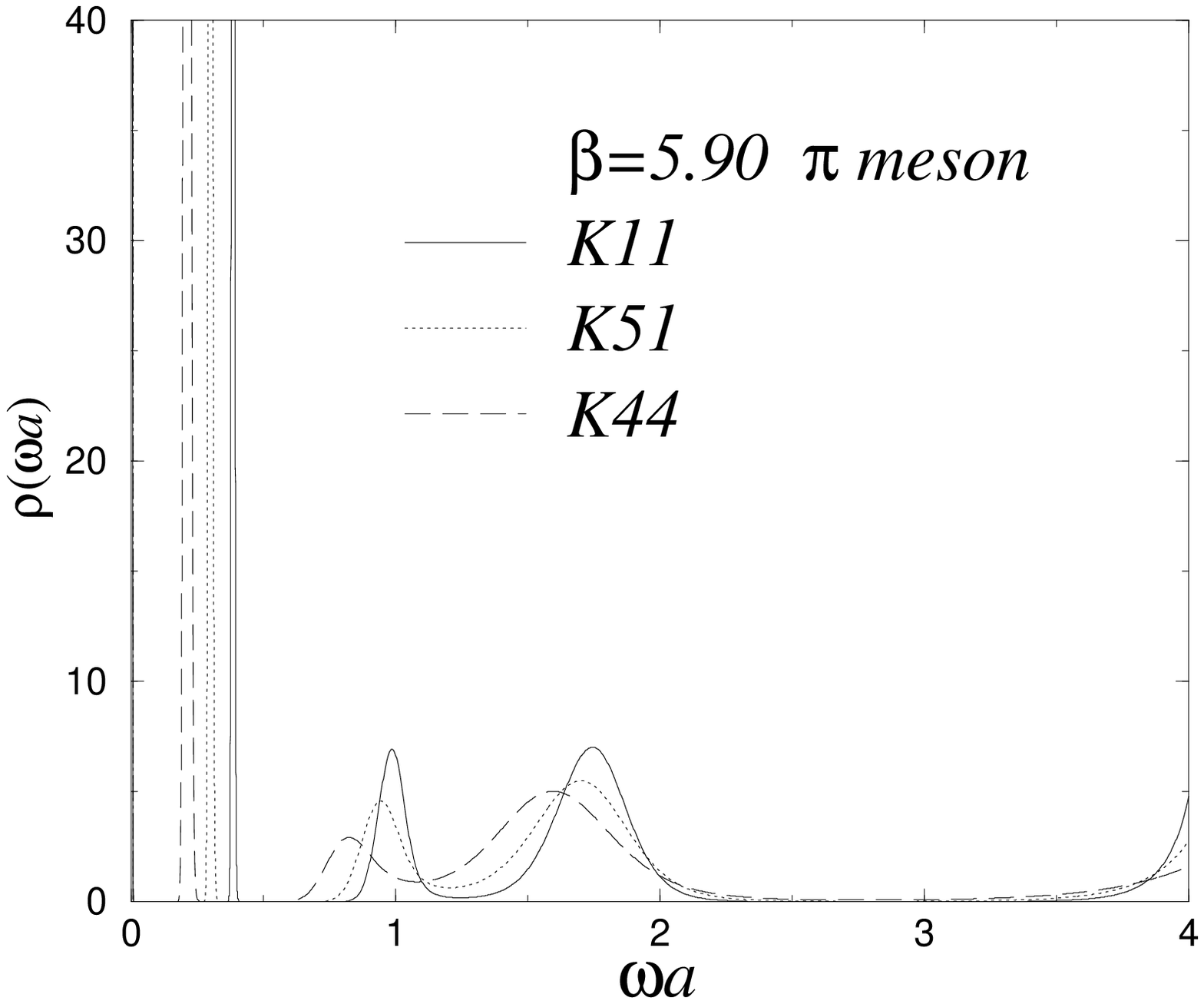} 
&
\hspace{0.5cm}
\leavevmode
\epsfxsize=6.2cm\epsfbox{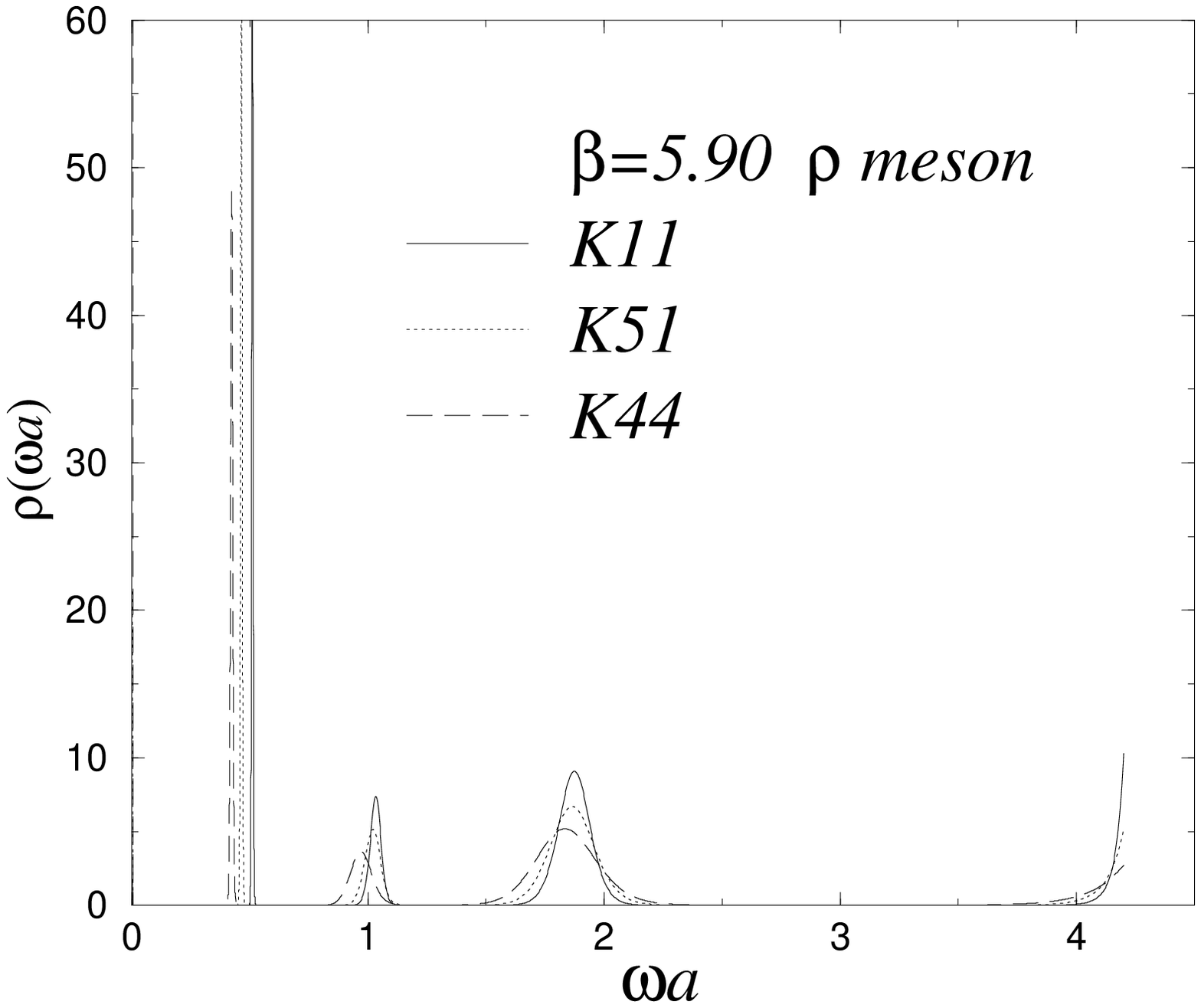} 
\\
\vspace{-0.3cm}
\leavevmode
\epsfxsize=6.2cm\epsfbox{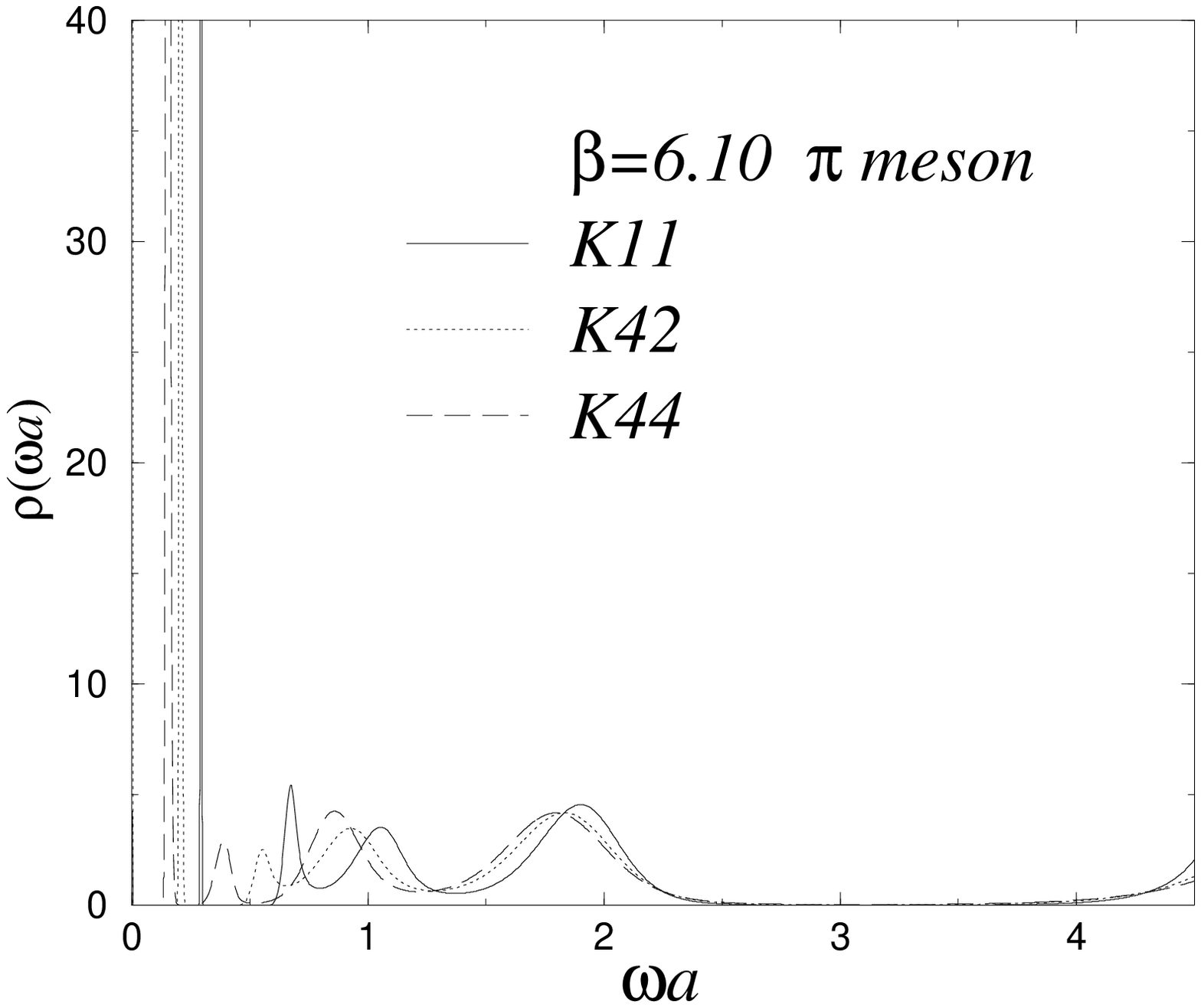} 
&
\hspace{0.5cm}
\leavevmode
\epsfxsize=6.2cm\epsfbox{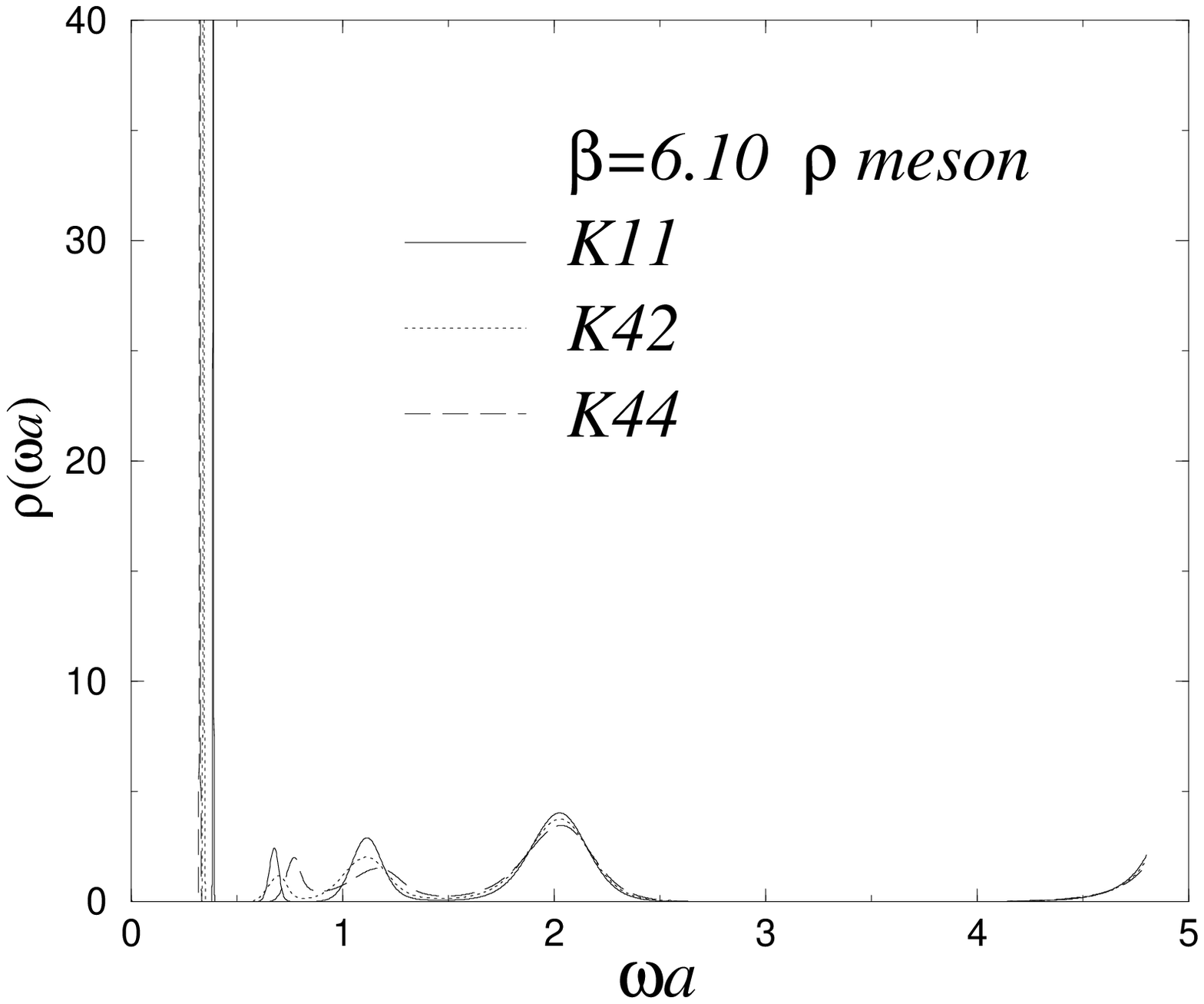} 
\\
\vspace{-0.3cm}
\leavevmode
\epsfxsize=6.2cm\epsfbox{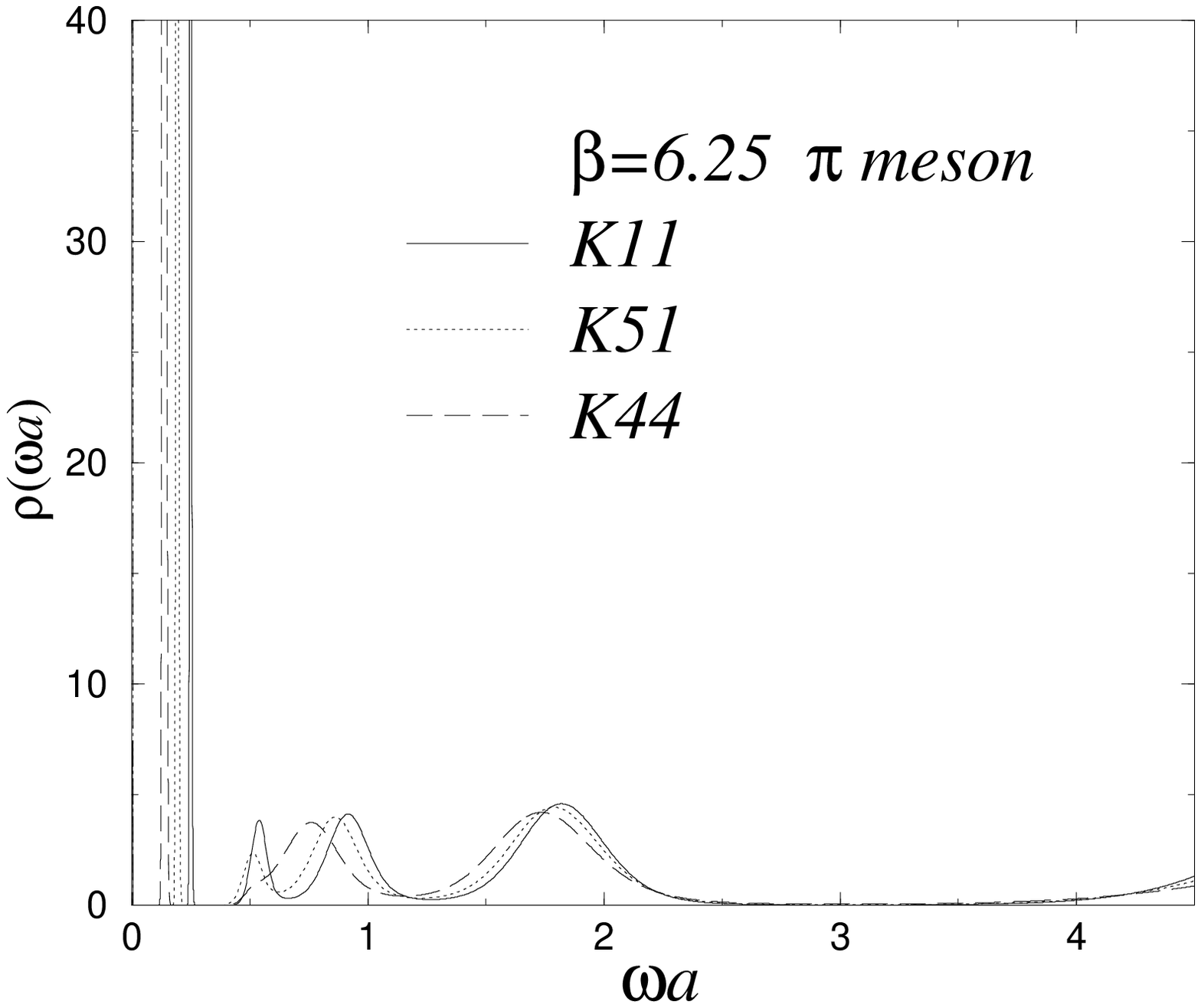} 
&
\hspace{0.5cm}
\leavevmode
\epsfxsize=6.2cm\epsfbox{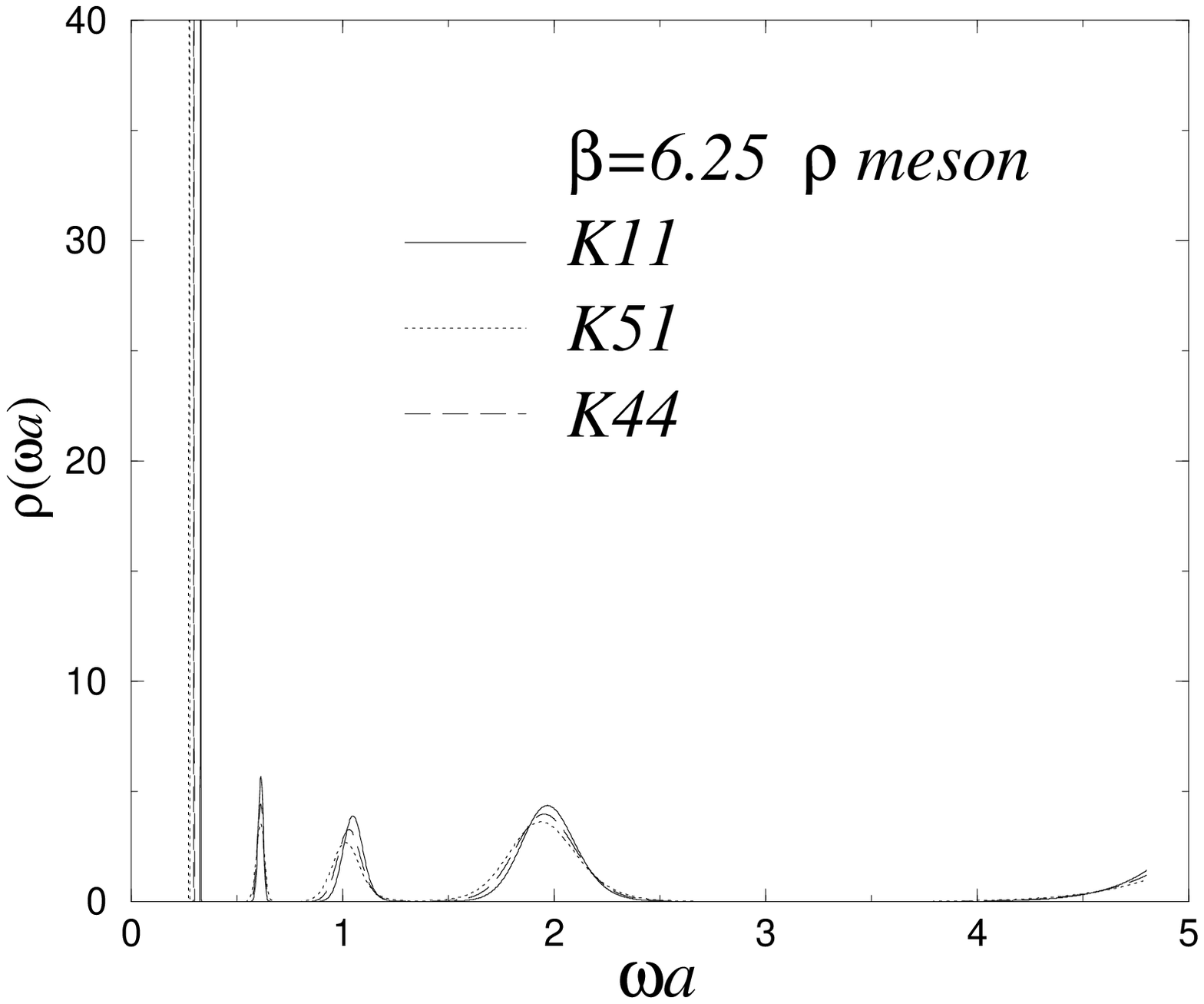} 
\\
\leavevmode
\epsfxsize=6.2cm\epsfbox{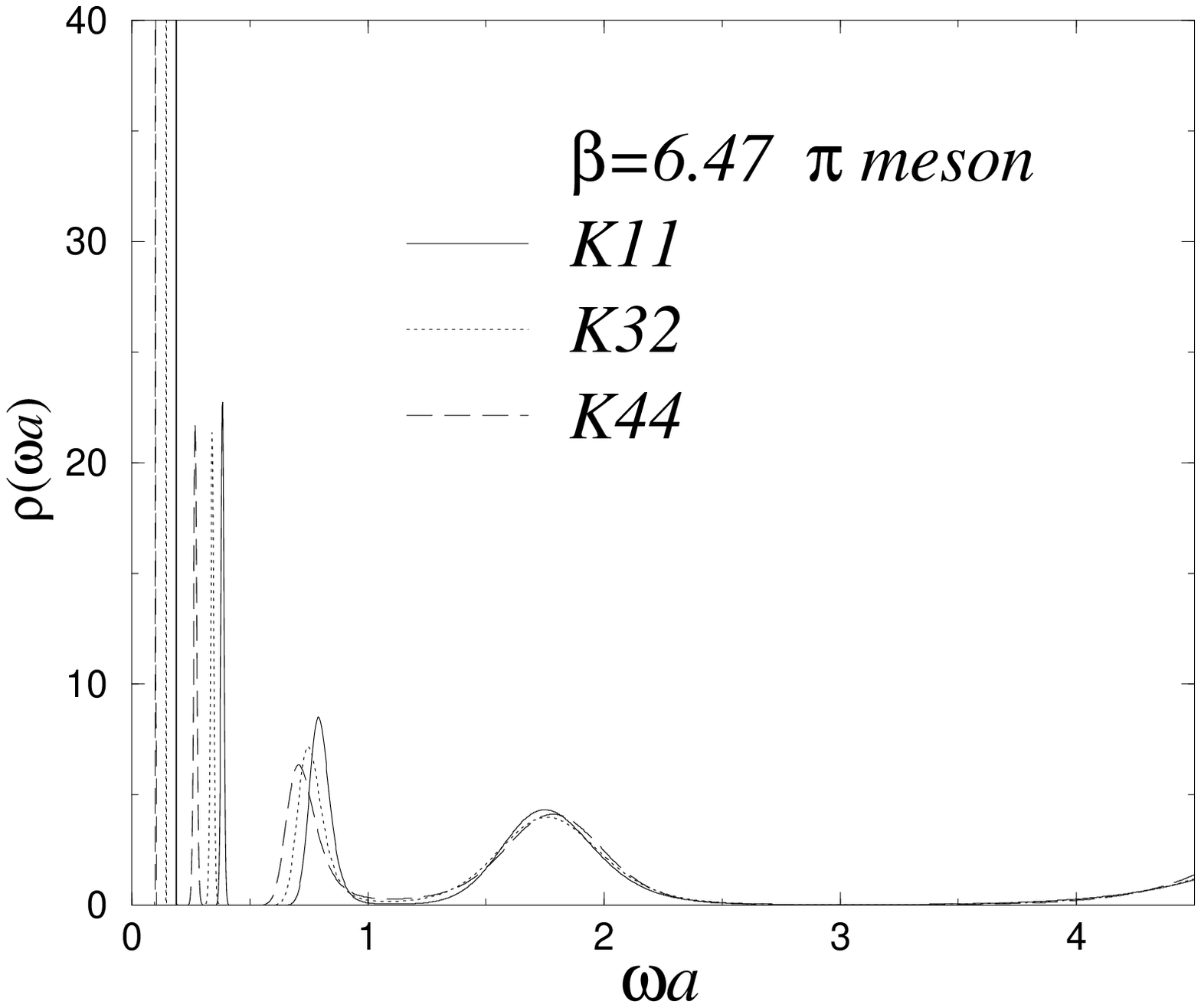} 
&
\hspace{0.5cm}
\leavevmode
\epsfxsize=6.2cm\epsfbox{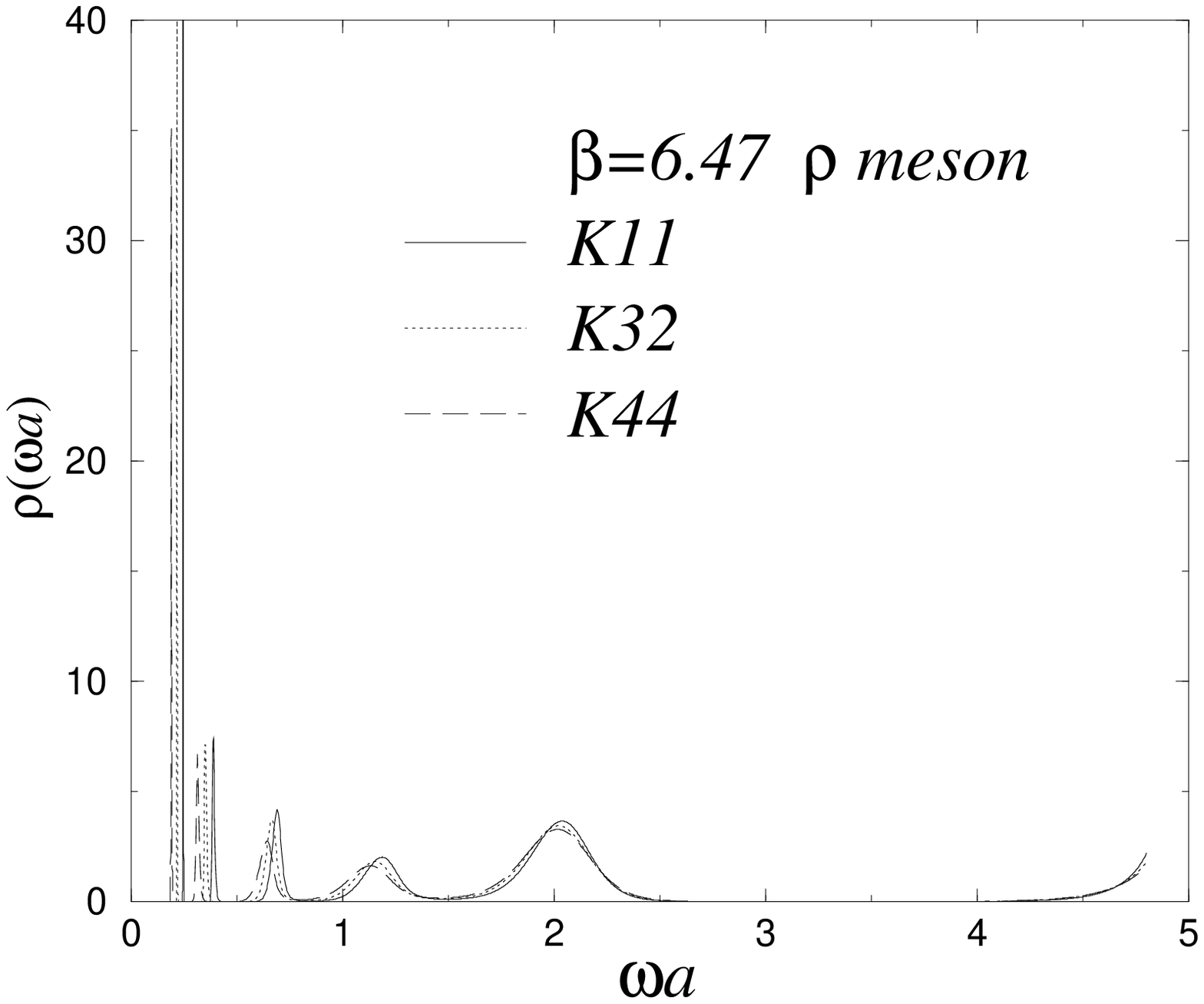} 
\\
\end{tabular}
\end{center}
\vspace{-.3cm}
\caption{Spectral functions at all $\beta$ obtained by MEM for 
different values of $K^{-1}$.
On the left hand side the $\pi$ meson spectral function and
on the right hand side the $\rho$ meson spectral function are shown.
The state at $\omega a\approx 2$ is considered as unphysical
since its position does not move with $\beta$.
\label{fig:5}}
\end{figure}

Our results of $\rho (\omega)$ obtained from meson propagators
by MEM for three different $K^{-1}$ at all $\beta$ are compiled 
in Fig.~\ref{fig:5}.
The lowest peak corresponds to the ground state, the next peak corresponds to 
the first excited state and so on.
At fixed $\beta$, positions for these peaks move toward smaller $\omega$
as the quark mass decreases. This shows that meson masses decrease 
with decreasing quark mass, as expected.
The number of peaks increases
from $\beta =5.90$ to $\beta =6.47$ for both pseudoscalar and vector channels,
since 
more states with higher energy appear in spectral functions
for larger lattice cut-off (smaller lattice spacing).
All peak positions move to smaller values as $\beta$ increases,
except the peaks at $\omega a \approx 1.7$ for the pseudoscalar channel and 
at $\omega a \approx 2$ for the vector channel.
Thus masses in the physical limits stay finite,
except those of the latter peaks which become infinite.
We discuss these unphysical states in more detail in the next section.

\subsection{Meson masses}
From peak positions of the spectral function,
we determine masses of excited states as well as the ground state.
Errors of these masses are estimated by 
the single elimination jackknife method.

\begin{figure}[p]
\begin{center}
\leavevmode
\epsfxsize=12cm\epsfbox{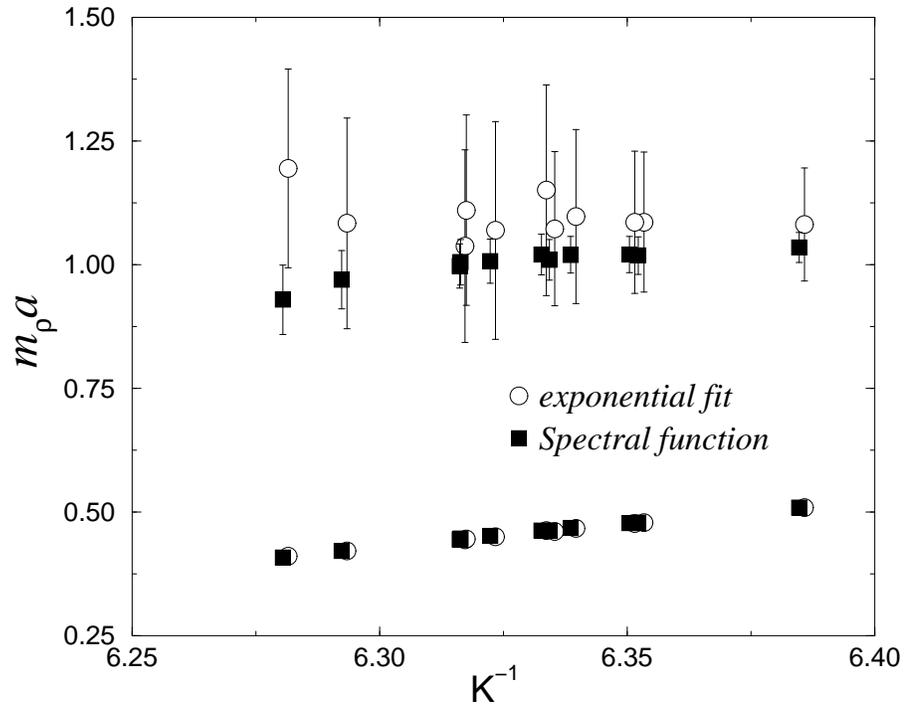} 
\end{center}
\caption{
Comparison of the $\rho$ meson mass 
for the ground and the first excited state
from the spectral function and the one from 
the double exponential fit.
Circles are slightly shifted to larger $K^{-1}$.
\label{fig:tab4}
}
\end{figure}

In order to check whether the peaks in the spectral function
really correspond to particle states in correlation functions,
we also extract masses of the ground and the first excited states
by fitting correlation functions with a double exponential form.
In order to obtain the mass of the first excited state reliably,
correlation functions from both point and smeared sources 
for the $\rho$ meson are simultaneously fitted.
Results at $\beta=5.90$ are given in Table~\ref{tab:6} and Fig.~\ref{fig:tab4},
where errors are again evaluated with the single elimination jackknife method,
together with those obtained by MEM.
We find that the ground state masses from the two methods agree very well,
and the first excited state masses are consistent with each other within the 
statistical error. It is noted, however, that
the error for the first excited state
by MEM is much smaller than the one by the usual fit.

\begin{figure}[!h]
\begin{center}
\begin{tabular}{lccc}
\raisebox{3.0cm}{$\beta=5.90$}&&
\leavevmode
\epsfxsize=6.3cm\epsfbox{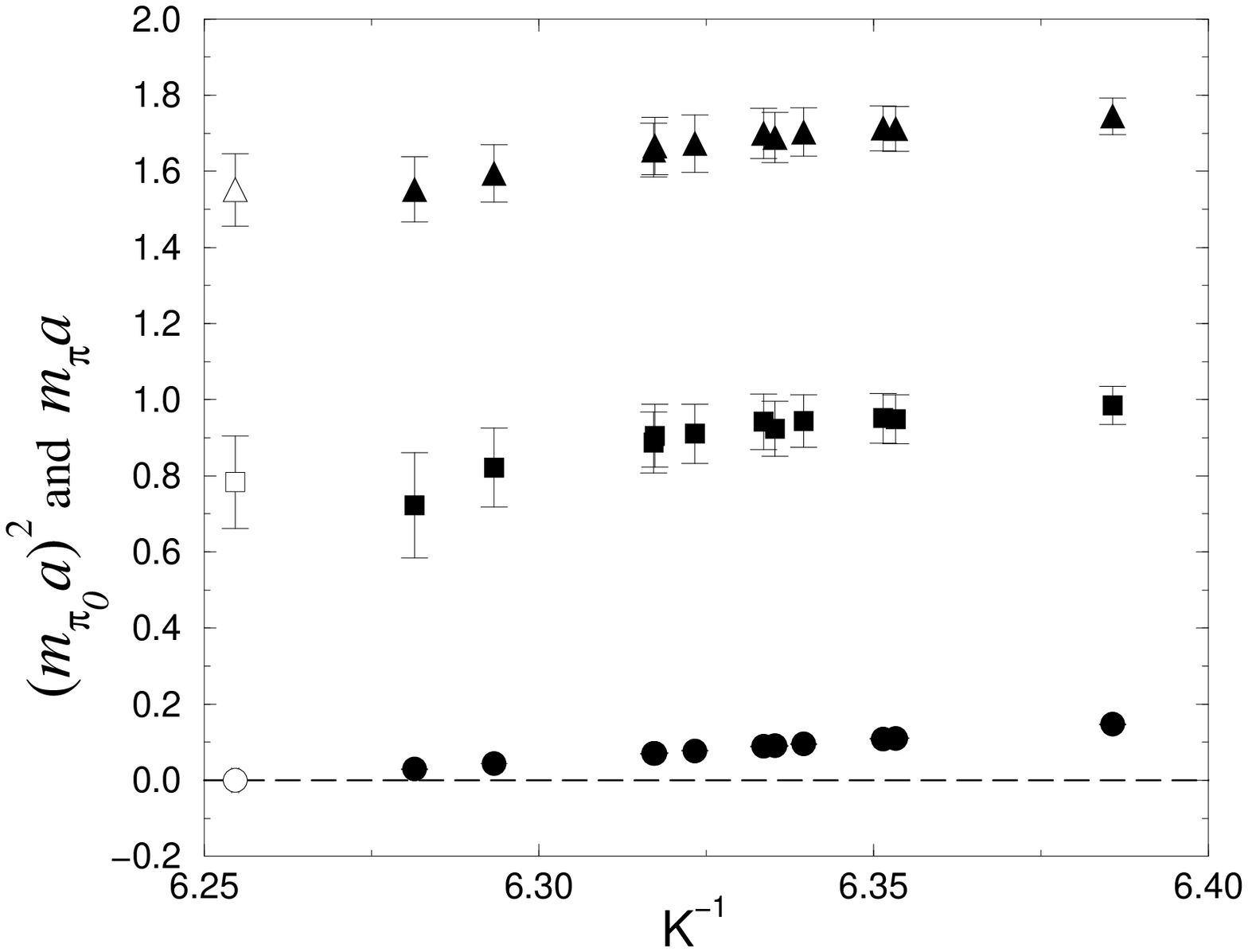}
&
\leavevmode
\epsfxsize=6.1cm\epsfbox{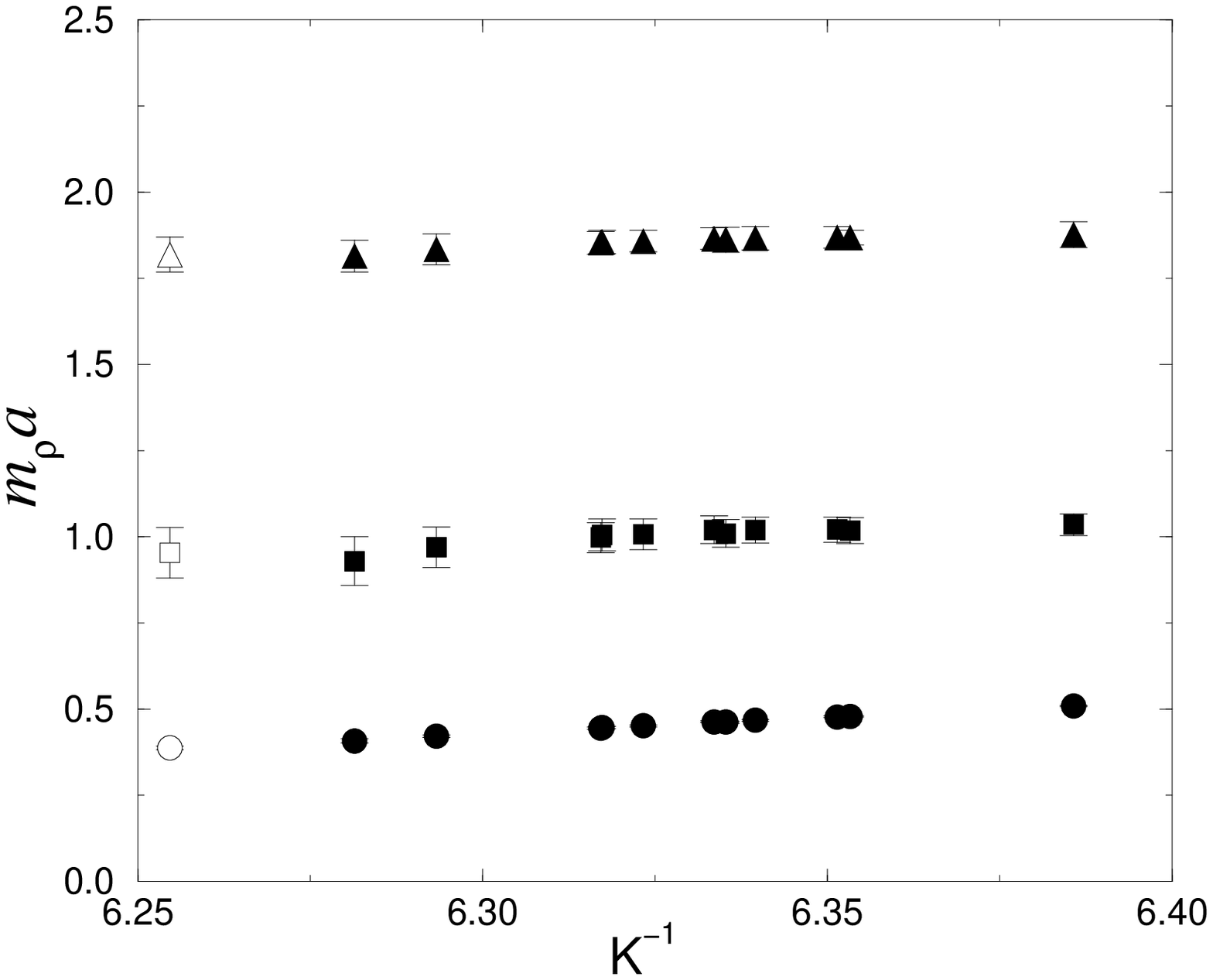}
\\
\raisebox{3.0cm}{$\beta=6.10$}&&
\leavevmode
\epsfxsize=6.3cm\epsfbox{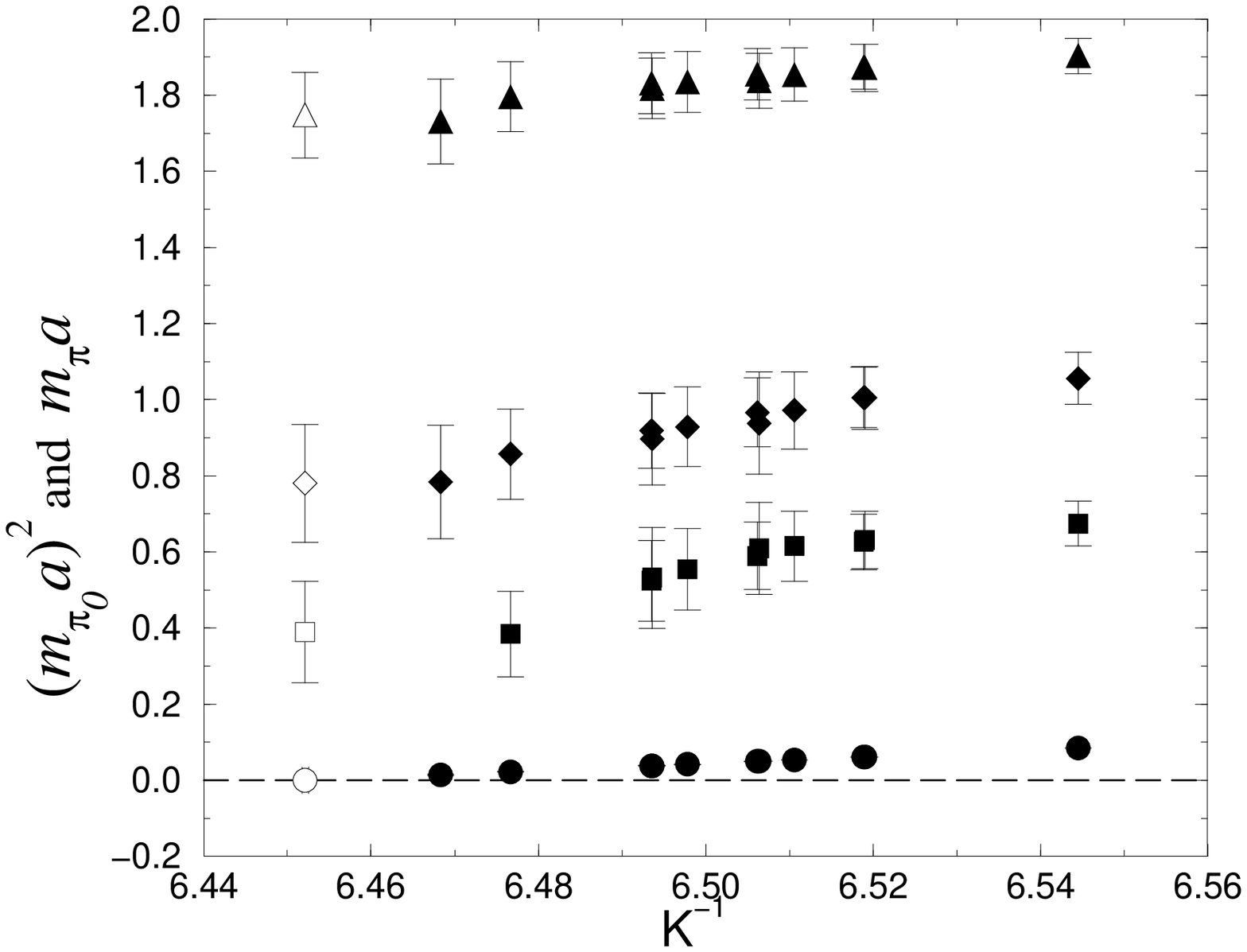}
&
\leavevmode
\epsfxsize=6.1cm\epsfbox{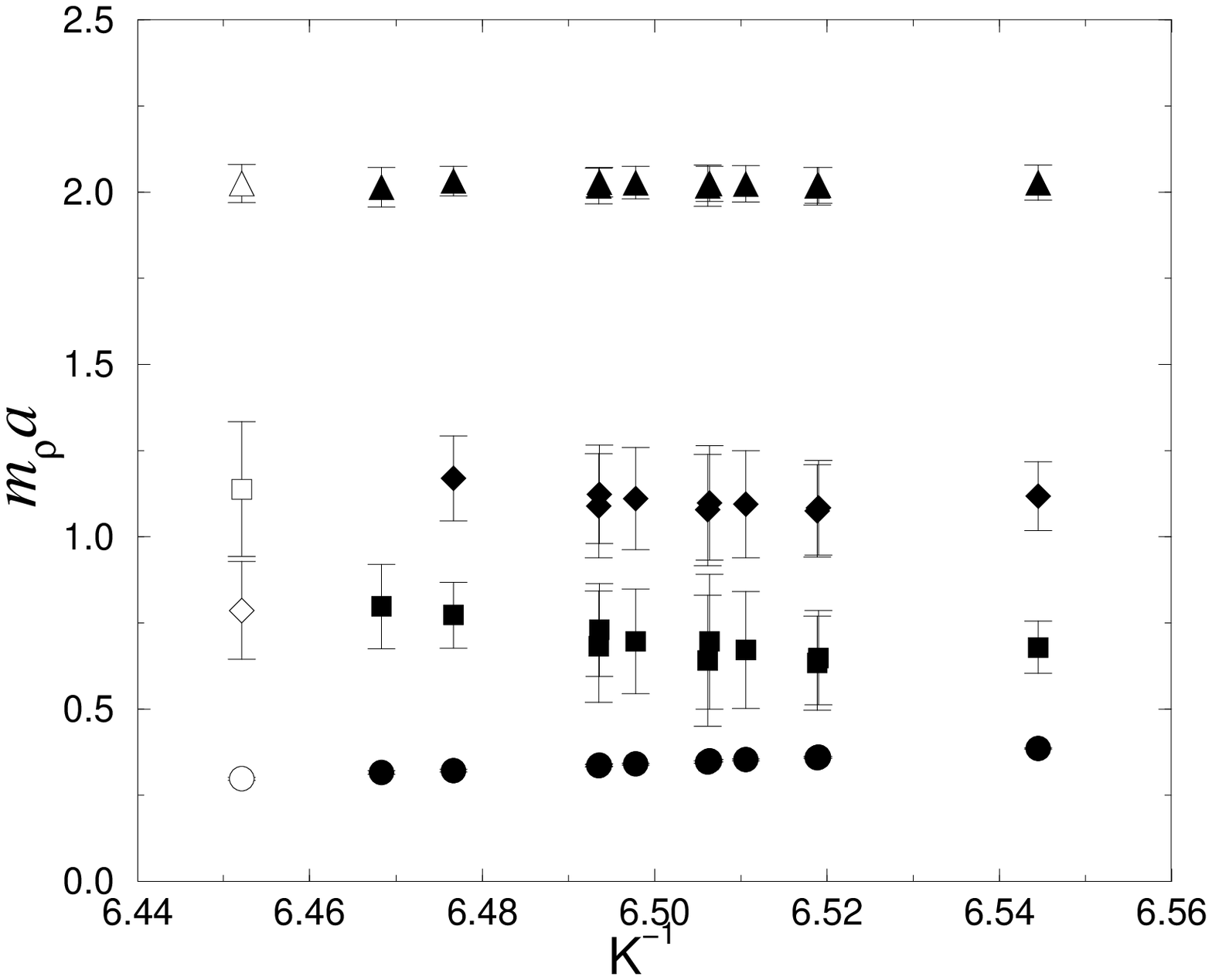}
\\
\raisebox{3.0cm}{$\beta=6.25$}&&
\leavevmode
\epsfxsize=6.3cm\epsfbox{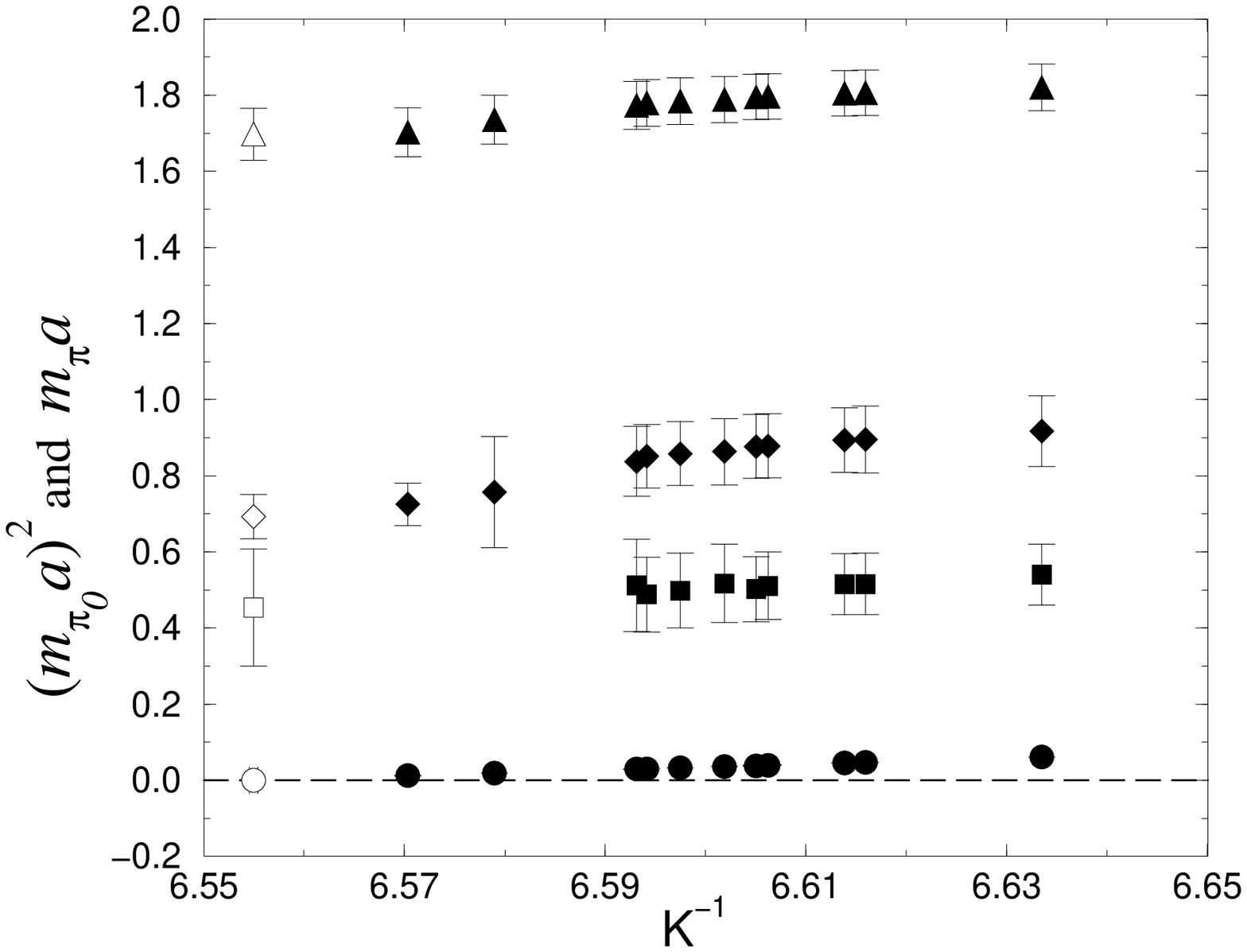}
&
\leavevmode
\epsfxsize=6.1cm\epsfbox{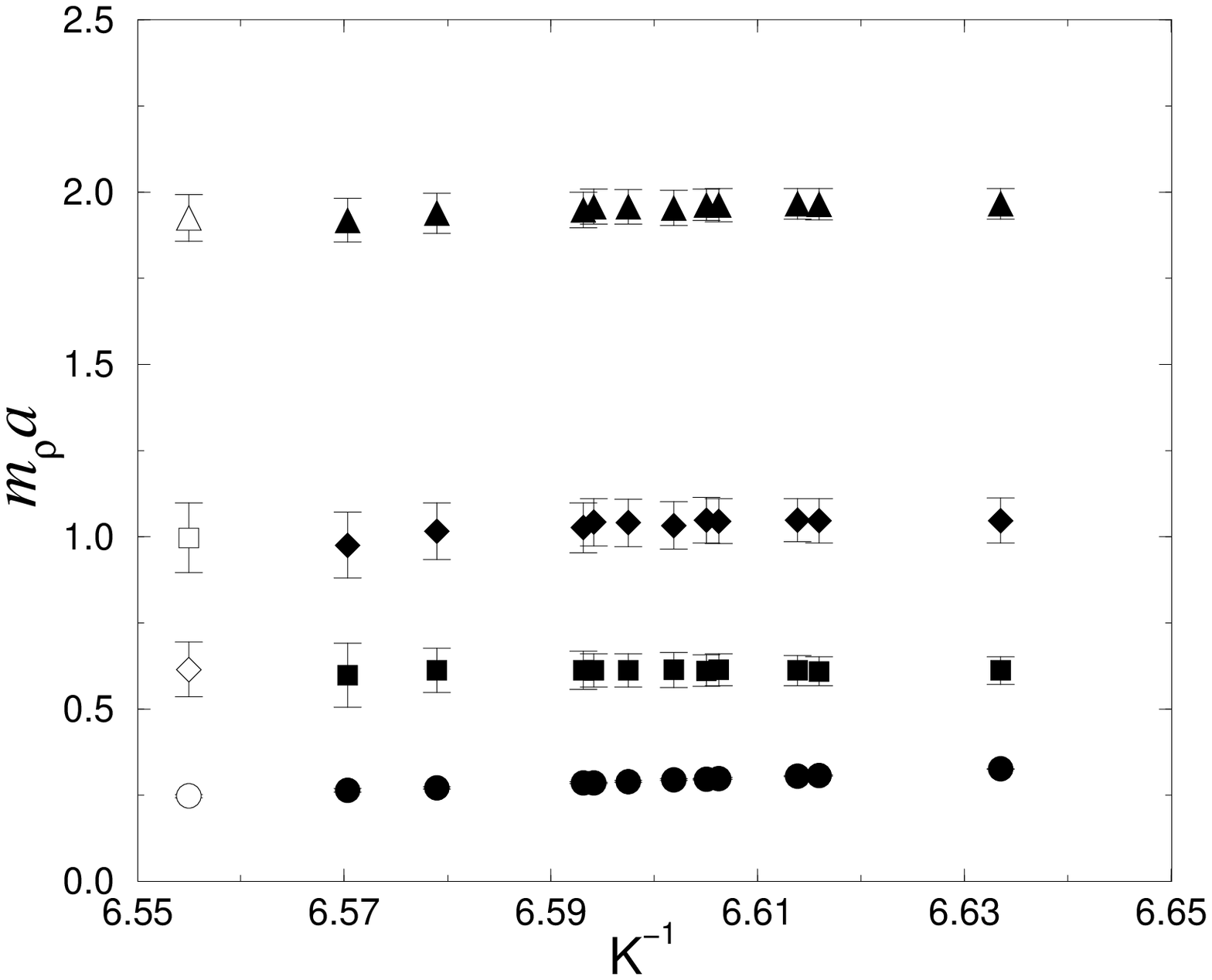}
\\
\raisebox{3.0cm}{$\beta=6.47$}&&
\leavevmode
\epsfxsize=6.3cm\epsfbox{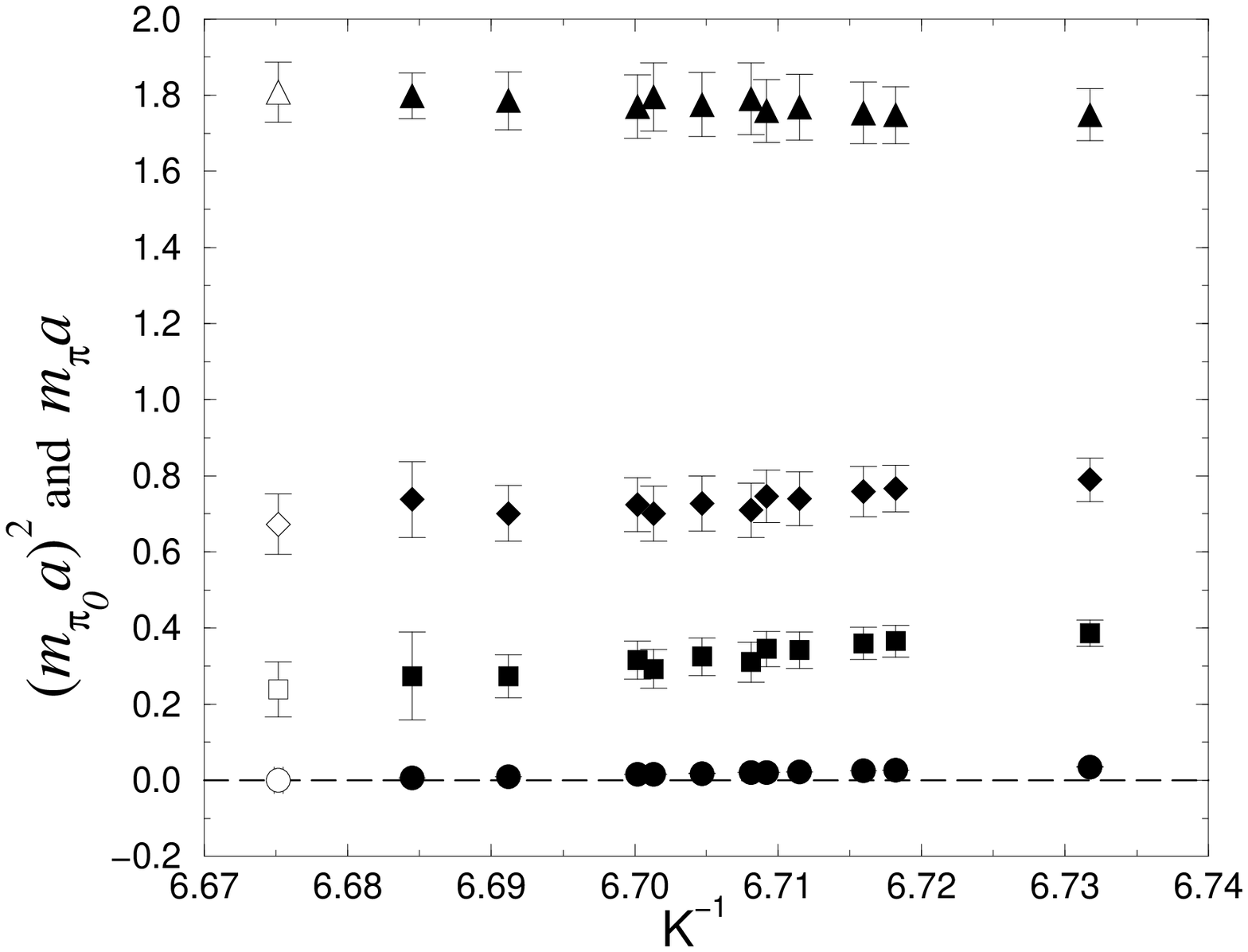}
&
\leavevmode
\epsfxsize=6.1cm\epsfbox{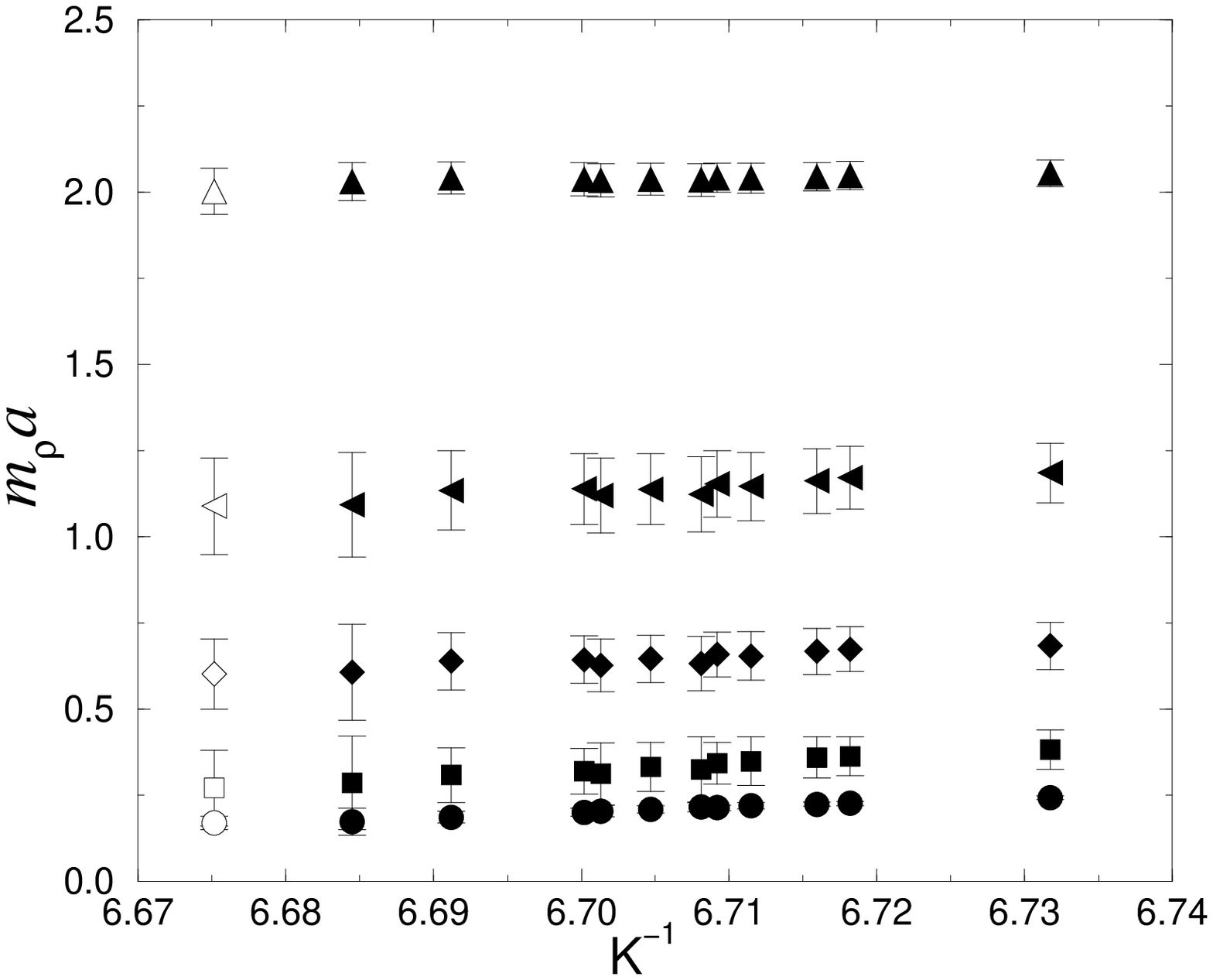}
\\
\end{tabular}
\end{center}
\caption{
Masses and their chiral extrapolations at all $\beta$.
On the left hand side the $\pi$ meson mass and
on the right hand side the $\rho$ meson mass are shown.
Circles, squares, diamonds and left triangles
represent 
the ground, the first excited, the second excited 
and the third excited state mass, respectively.
The state shown by up triangles is considered unphysical 
as discussed in the text.
Open symbols stand for the values in the chiral limit.
\label{fig:mass6.47}
}
\end{figure}

We determine the chiral limit and the critical hopping parameter $\kappa_{c}$
where the ground state of the $\pi$ meson mass vanishes
by extrapolating $(m_{\pi}a)^{2}$ linearly in $K^{-1}$.
For other states including the excited states of $\pi$ mesons,
the masses $ma$ themselves obtained from the spectral function 
are extrapolated linearly in $K^{-1}$ to the chiral limit.
The chiral extrapolation at each $\beta$ is shown
in Fig.~\ref{fig:mass6.47}.
Some excited state peaks 
do not appear in the spectral functions 
obtained from some jackknife samples.
These masses are excluded from the chiral extrapolation and
are not plotted in the figures.
The lattice spacing $a$ is fixed by setting
the ground state mass for the $\rho$ meson in the chiral limit to the
experimental value, $m_\rho =$ 770 MeV.
All dimensionful quantities are normalized by the $\rho$ meson mass
in the chiral limit.

\begin{figure}[b]
\begin{center}
\leavevmode
\epsfxsize=9cm\epsfbox{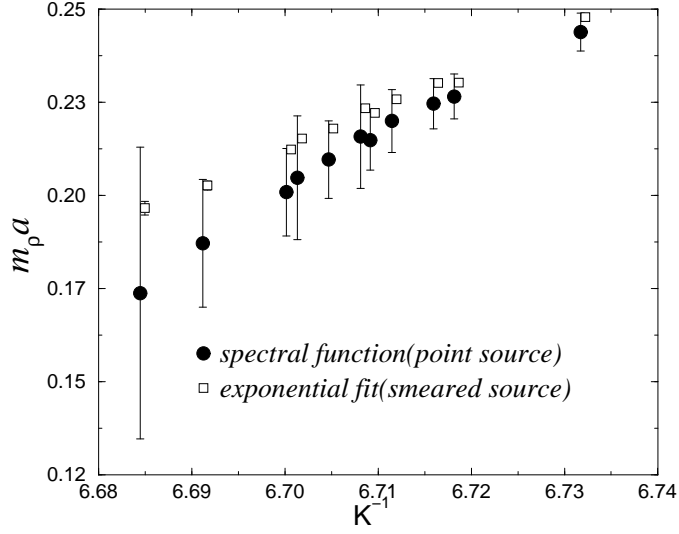} 
\end{center}
\caption{
Ground state masses of the $\rho$ meson obtained by different analyses 
at $\beta =$ 6.47.
Squares are slightly shifted to larger $K^{-1}$.
\label{fig:b=6.47}
}
\end{figure}

Masses in the chiral limit
are compiled in Table~\ref{tab:7},
together with the result of the standard analysis~\cite{CPdata,pre} 
for the lattice spacing,
which agrees with the values from the present MEM analysis.
At $\beta = 6.47$, our lattice spacing has a larger error.
This is caused by 
large errors of point source data at this $\beta$.
As shown in Fig.~\ref{fig:b=6.47},
the ground state masses for each $K^{-1}$ agree with the previous results
from exponential fit of smeared source data\cite{pre}.

\begin{figure}[h!]
\begin{center}
\begin{tabular}{cc}
\leavevmode
\epsfxsize=7cm\epsfbox{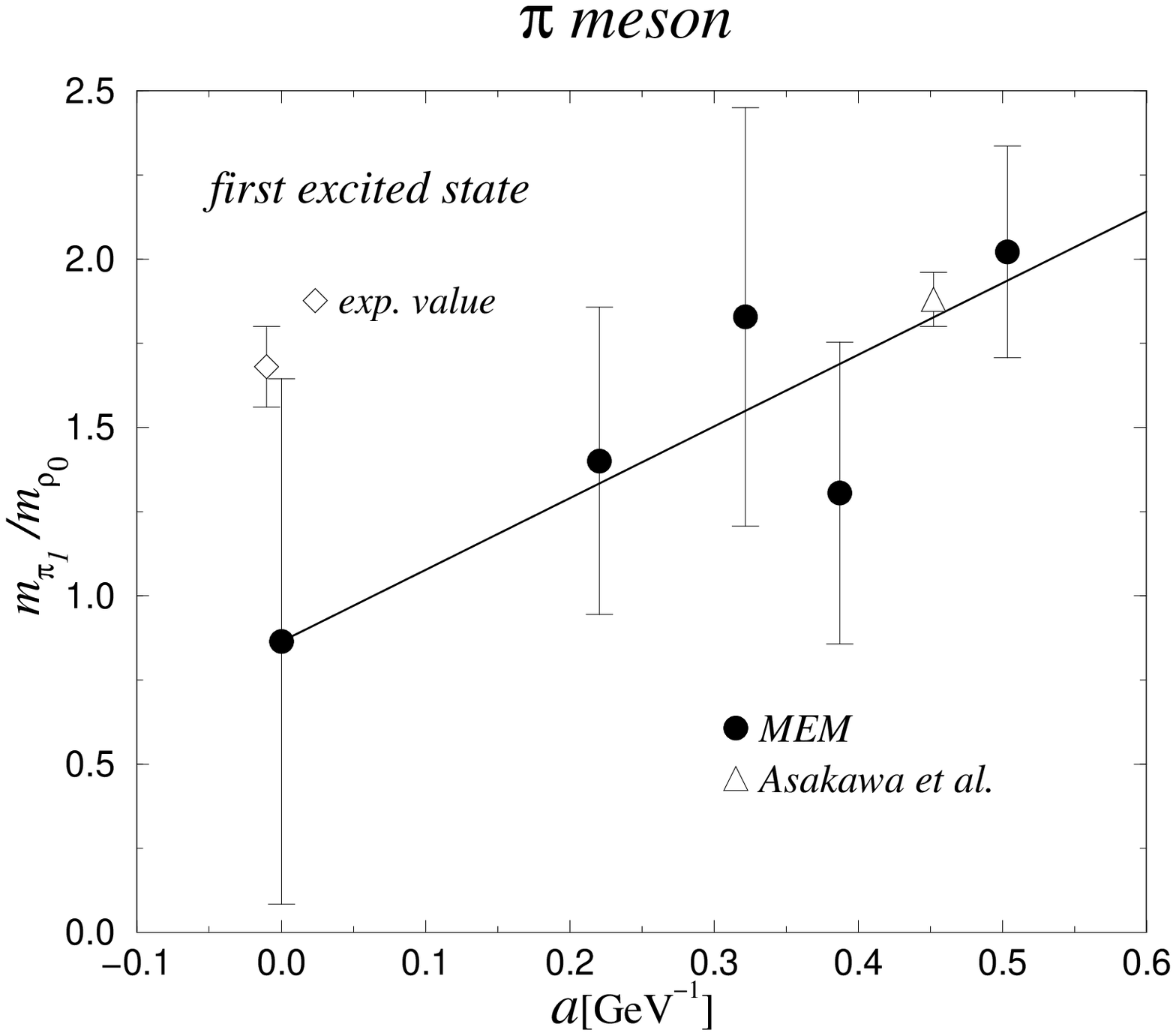} 
&
\leavevmode
\epsfxsize=7cm\epsfbox{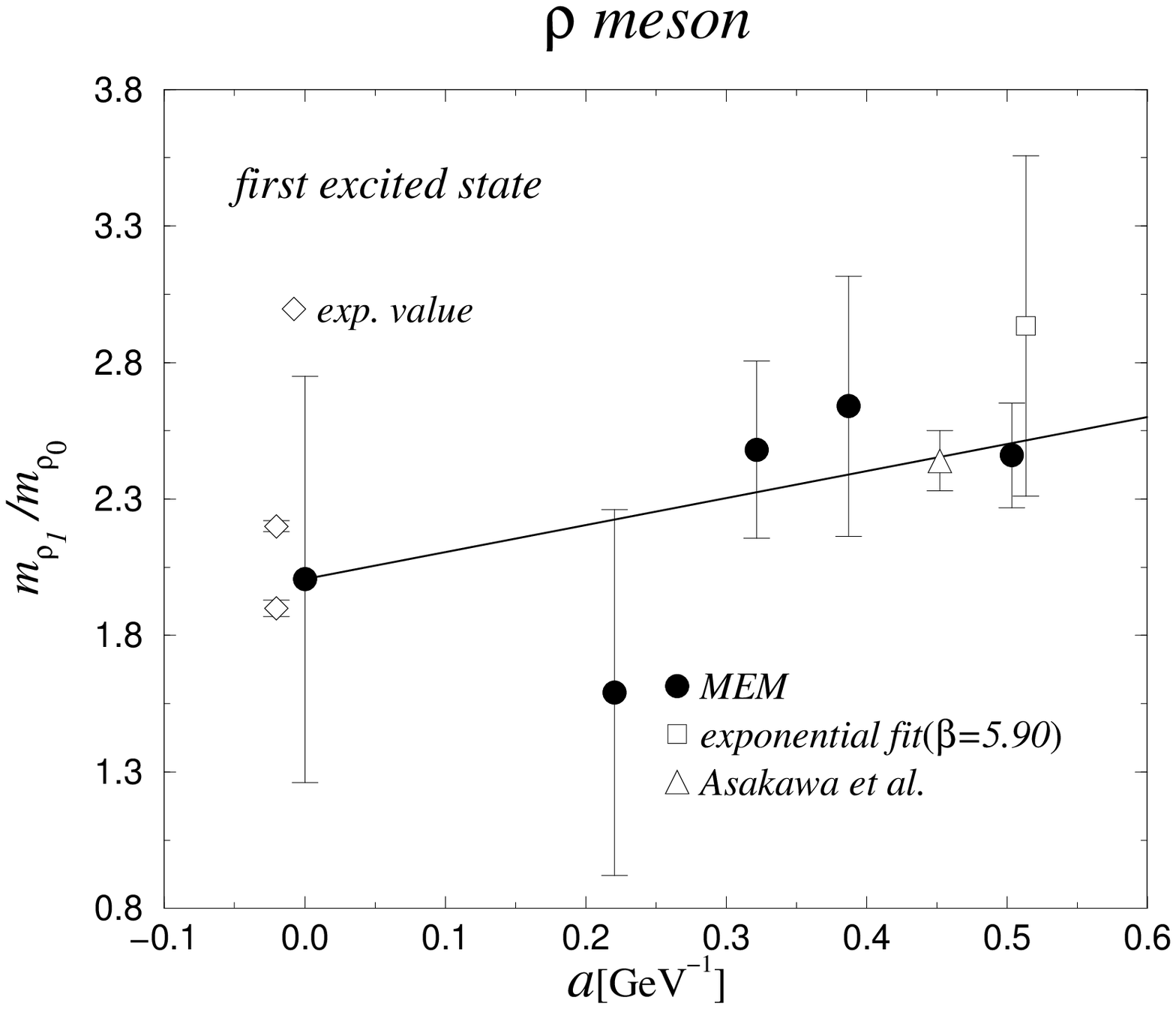} 
\\
\leavevmode
\epsfxsize=7cm\epsfbox{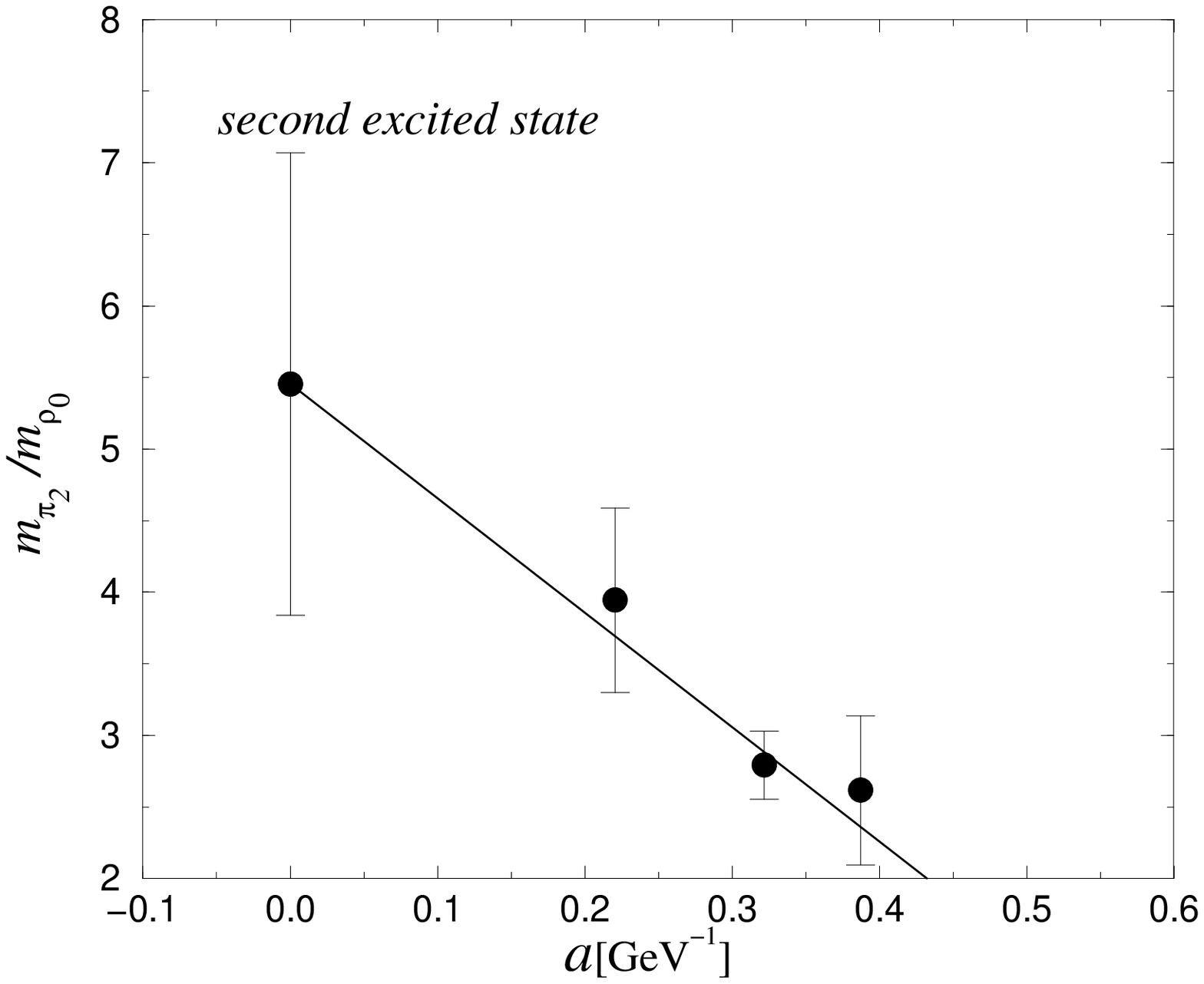} 
&
\leavevmode
\epsfxsize=7cm\epsfbox{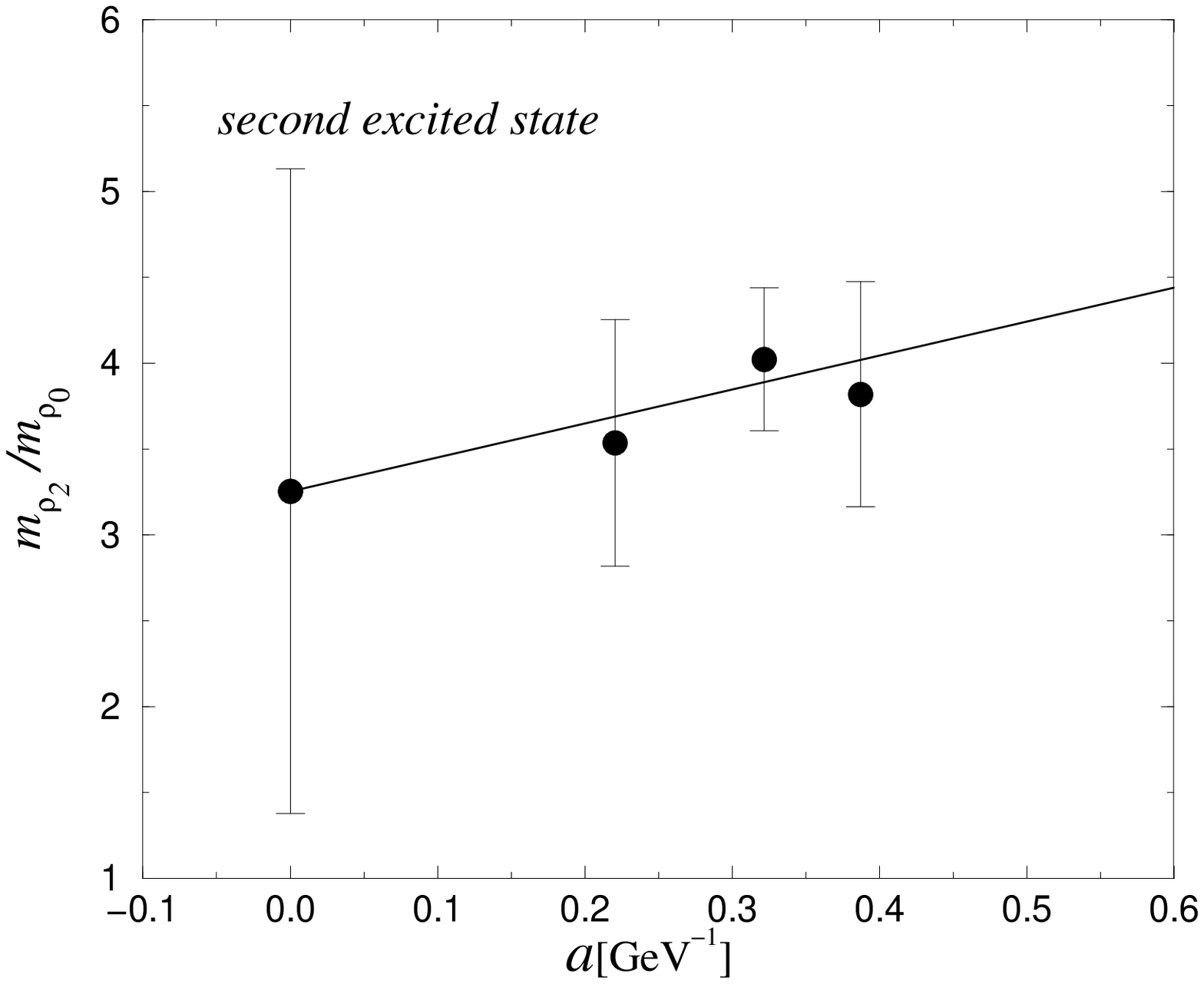} 
\end{tabular}
\end{center}
\caption{
Continuum extrapolation of masses of physical excited states.
For the first excited state,
open diamonds and triangles
represent the experimental value, and that 
obtained by Asakawa {\it et al.}~[5].
For the $\rho$ meson
the open square shows the result of the double exponential
fit at $\beta =$ 5.90.
\label{fig:7}
}
\end{figure}

Masses of excited states in the chiral limit
are extrapolated to the continuum limit, except unphysical states mentioned 
before, as shown in Fig.~\ref{fig:7}.
We see that the mass of the first excited state is consistent with 
the one reported in Ref.~\cite{hats:rev} for both $\pi$ and $\rho$ mesons.
Note that the error for the first excited state of $\rho$ meson
from the double exponential fit at 
$\beta =5.90$ (square) 
is too large for a reasonable continuum extrapolation.
Mass ratios in the continuum limit are given in Table~\ref{tab:fig7}.
The mass ratio of the first excited state to the ground state for
the $\pi$ meson in the continuum limit is 0.86(77), which should be compared
with the experimental value 1.68(12),
while the ratio for the $\rho$ meson is 2.00(74) 
in comparison to the experimental value of 
1.90(3) or 2.20(2)
(there are two candidates for the first excited state of the $\rho$ meson
in experiment).
The first excited state masses for both mesons are  
consistent with experimental values albeit the errors are quite large.
For the $\rho$ meson
we are not able to decide whether 
the first excited state is
$\rho(1450)$ or $\rho(1700)$ due to the large error of our result.

\subsection{Decay constants}

From the spectral function
we can also extract the decay constant
for the ground state of $\pi$ and $\rho$ mesons,
$f_{\pi}$ and $f_{\rho}$, defined by
\begin{eqnarray}
\langle0|(\bar{d}\gamma_{5}u)_{lat}|\pi_{0},{\mathbf p}=0\rangle&=&
\frac{\sqrt{2}f_{\pi}m^{2}_{\pi}}
{(m_{u}+m_{d})_{lat}^{AWI}}
\frac{1}{Z_{{\mathrm A}}}
\prod_{i=1}^{2}
\sqrt{\frac{1}
{1-3\kappa_{i}/4\kappa_{c}}}
,\\
\langle0|(\bar{d}\gamma_{\mu}u)_{lat}|\rho_{0},{\mathbf p}=0\rangle&=&
\sqrt{2}f_{\rho}m^{2}_{\rho}\epsilon_{\mu}\frac{1}{Z_{{\mathrm V}}}
\prod_{i=1}^{2}
\sqrt{\frac{1}{2\kappa_{i}}}.
\end{eqnarray}
We employ the one-loop result with tadpole-improvement 
for the renormalization factor $Z_{A}$~\cite{tad}
given by 
$Z_{{\mathrm A}}=1-0.316\alpha_{V}(1/a)$,
and the bare quark masses $(m_{u}+m_{d})_{lat}^{AWI}$
derived from the axial ward identity\cite{pre}.
For the vector meson decay constant, 
we use a non-perturbative value for $Z_{{\mathrm V}}$\cite{pre}.

Decay constants can be extracted from the correlation function
as follows.
For the pseudoscalar meson we have,
\begin{eqnarray}
\sum_{{\mathbf x}}
\langle0|\bar{d}\gamma_{5} u(\tau,{\mathbf x})
(\bar{d}\gamma_{5} u)^{\dagger}(0,{\mathbf 0})|0\rangle&=&
\sum_{n}\langle0|\bar{d}\gamma_{5} u|\pi_{n}\rangle
\langle\pi_{n}|(\bar{d}\gamma_{5} u)^{\dagger}|0\rangle
\frac{e^{-E_{n}\tau}}{2E_{n}}\\
&\longrightarrow&
|\langle\pi_{0}|\bar{d}\gamma_{5} u|0\rangle|^{2}\,
\frac{e^{-m_{\pi_{0}}\tau}}{2m_{\pi_{0}}},
\ \ \ \tau \rightarrow \infty,
\end{eqnarray}
where $E_{n}$ is $n$-th excited state energy,
and a similar expression for the vector meson.
Under the assumption that the ground state peak of spectral function is sharp,
these correlation functions are related to the area of 
the spectral function around the ground state peak according to
\begin{eqnarray}
\int_{ground} d\omega\,\rho_{{\scriptscriptstyle{\mathrm PS}}}
(\omega)\omega^{2} 
&=& \frac{f_{\pi}^{2}m^{3}_{\pi}}
{\left ((m_{u}+m_{d})_{lat}^{AWI}\right)^{2}}
\frac{1}{Z^{2}_{{\mathrm A}}}
\prod_{i=1}^{2}
\left(\frac{1}
{1-3\kappa_{i}/4\kappa_{c}}
\right),
\label{eq:fpi} \\
\int_{ground} d\omega\,\rho_{{\scriptscriptstyle{\mathrm V}}}
(\omega)\omega^{2} 
&=& f_{\rho}^{2}m^{3}_{\rho}\frac{1}{Z^{2}_{{\mathrm V}}}
\prod_{i=1}^{2}
\left(\frac{1}{2\kappa_{i}}\right).
\label{eq:frho}
\end{eqnarray}
For the first excited state, 
we also extract decay constants from the area of the spectral function
around the first excited state under the same assumption as 
for the ground state.

\begin{figure}[p]
\begin{center}
\begin{tabular}{cc}
\leavevmode
\epsfxsize=7.5cm\epsfbox{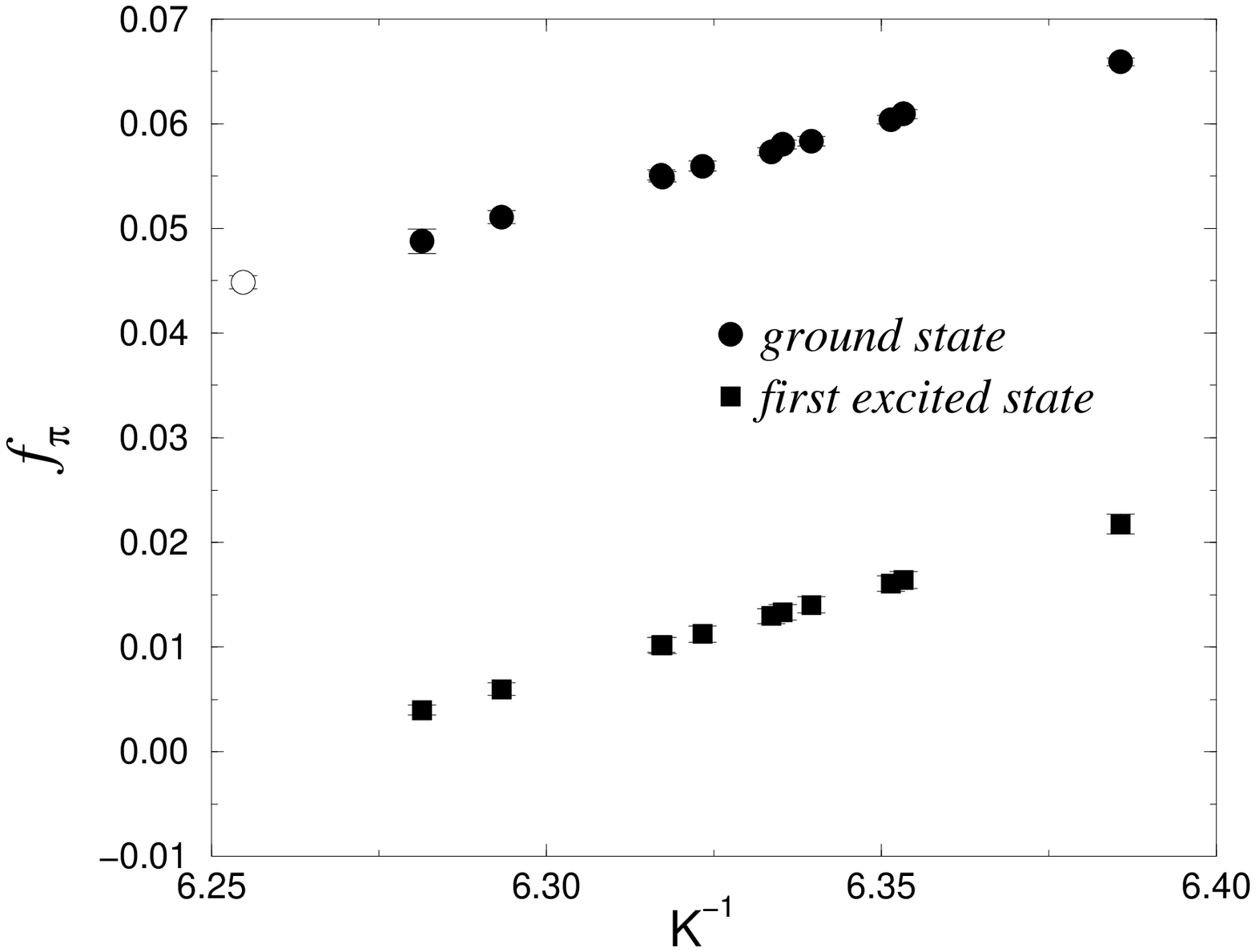} 
&
\leavevmode
\epsfxsize=7.3cm\epsfbox{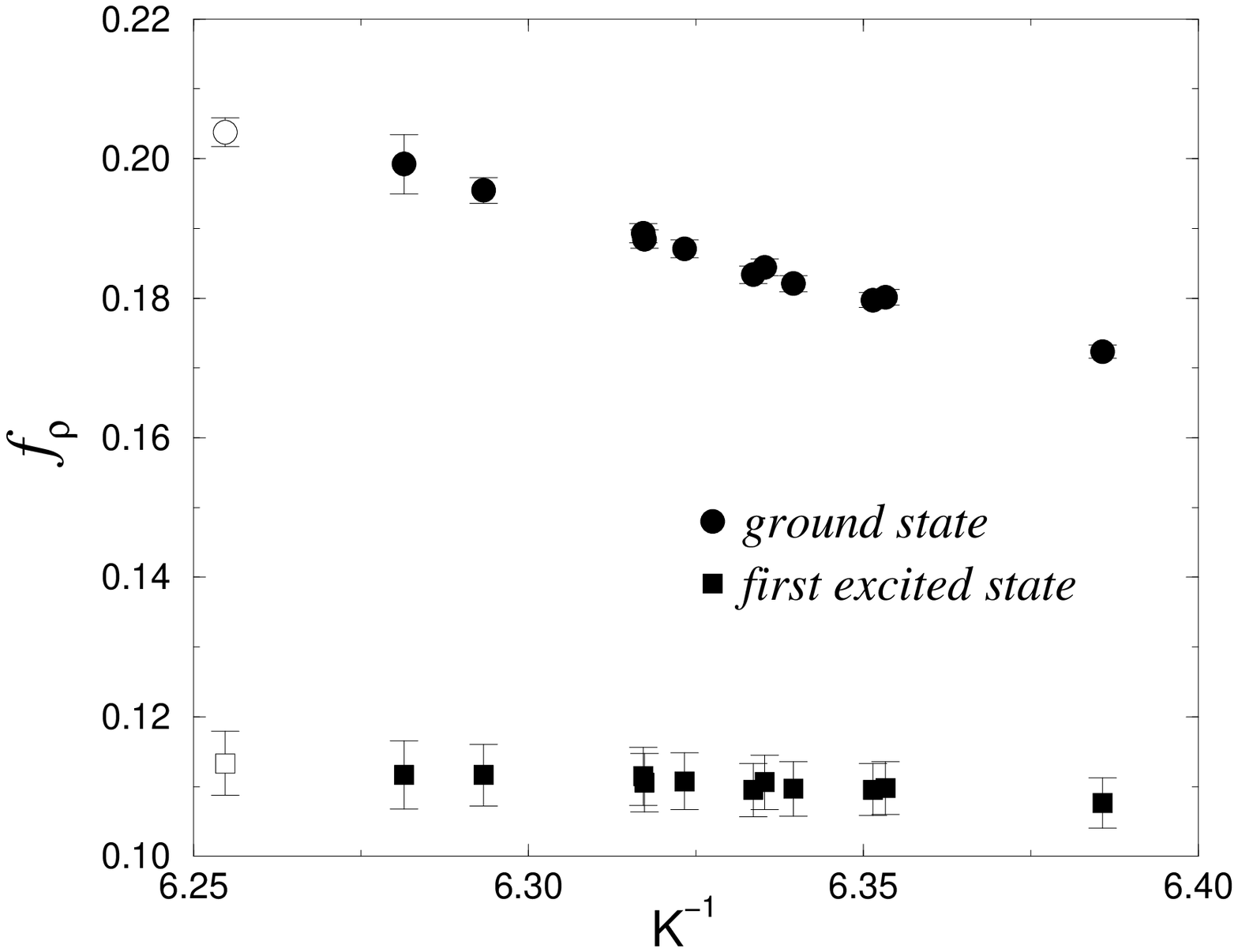} 
\end{tabular}
\end{center}
\caption{
Chiral extrapolations of pseudoscalar and vector meson
decay constants at $\beta =5.90$.
\label{fig:dec.chi}
}
\end{figure}

Decay constants obtained from the above relations
are extrapolated linearly in $K^{-1}$ to the chiral limit, 
as shown for $\beta = 5.90$ in Fig.~\ref{fig:dec.chi}, and
results are also given in Table~\ref{tab:7}.
The decay constant for the first excited state of 
the $\pi$ meson should vanish in the chiral limit
according to (\ref{eq:fpi}),
because $m_{\pi}$ remains finite in the chiral limit for excited states.
This property can be seen from the figure. 

\begin{figure}[!h]
\begin{center}
\begin{tabular}{cc}
\leavevmode
\epsfxsize=7.5cm\epsfbox{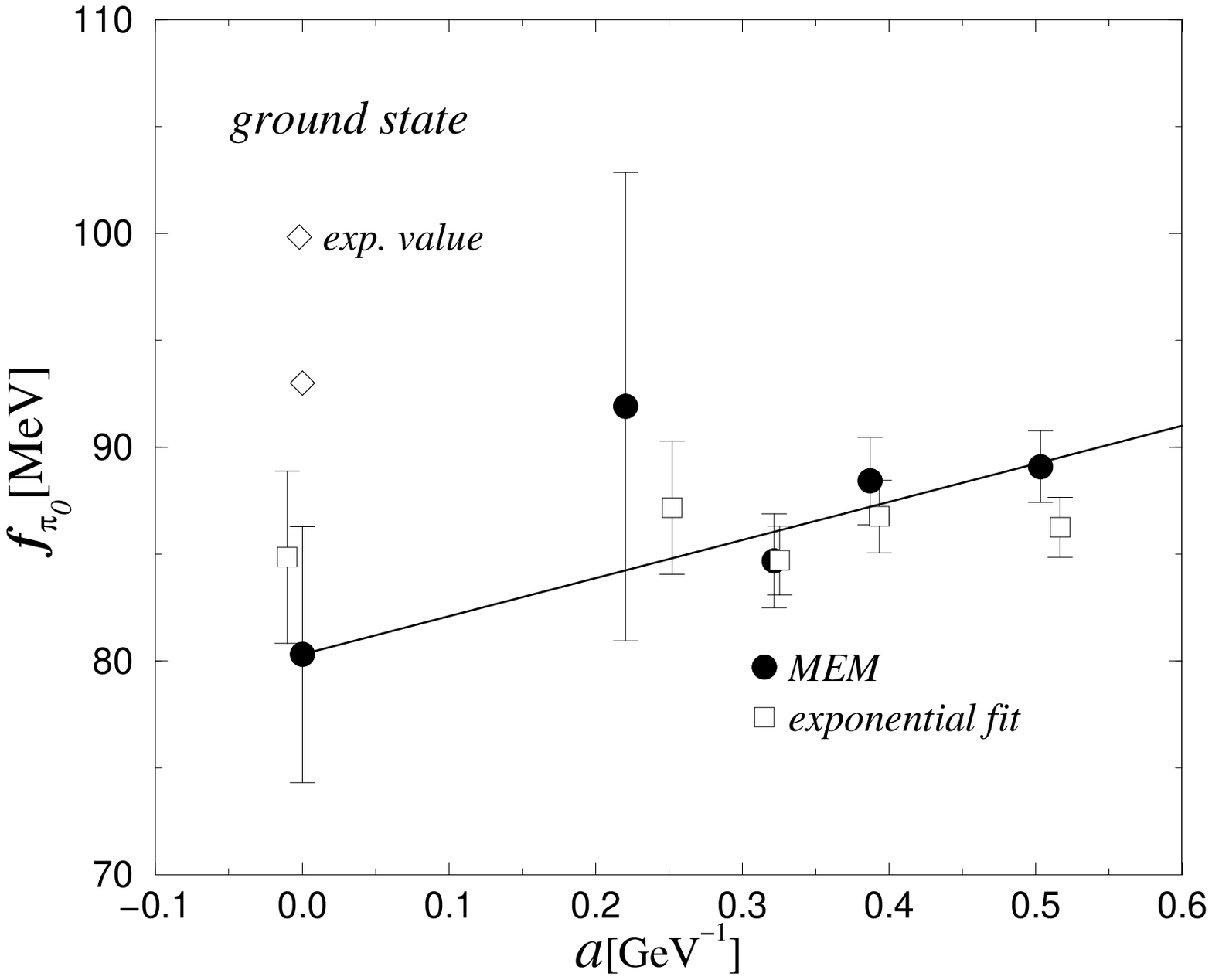} 
&
\leavevmode
\epsfxsize=7.5cm\epsfbox{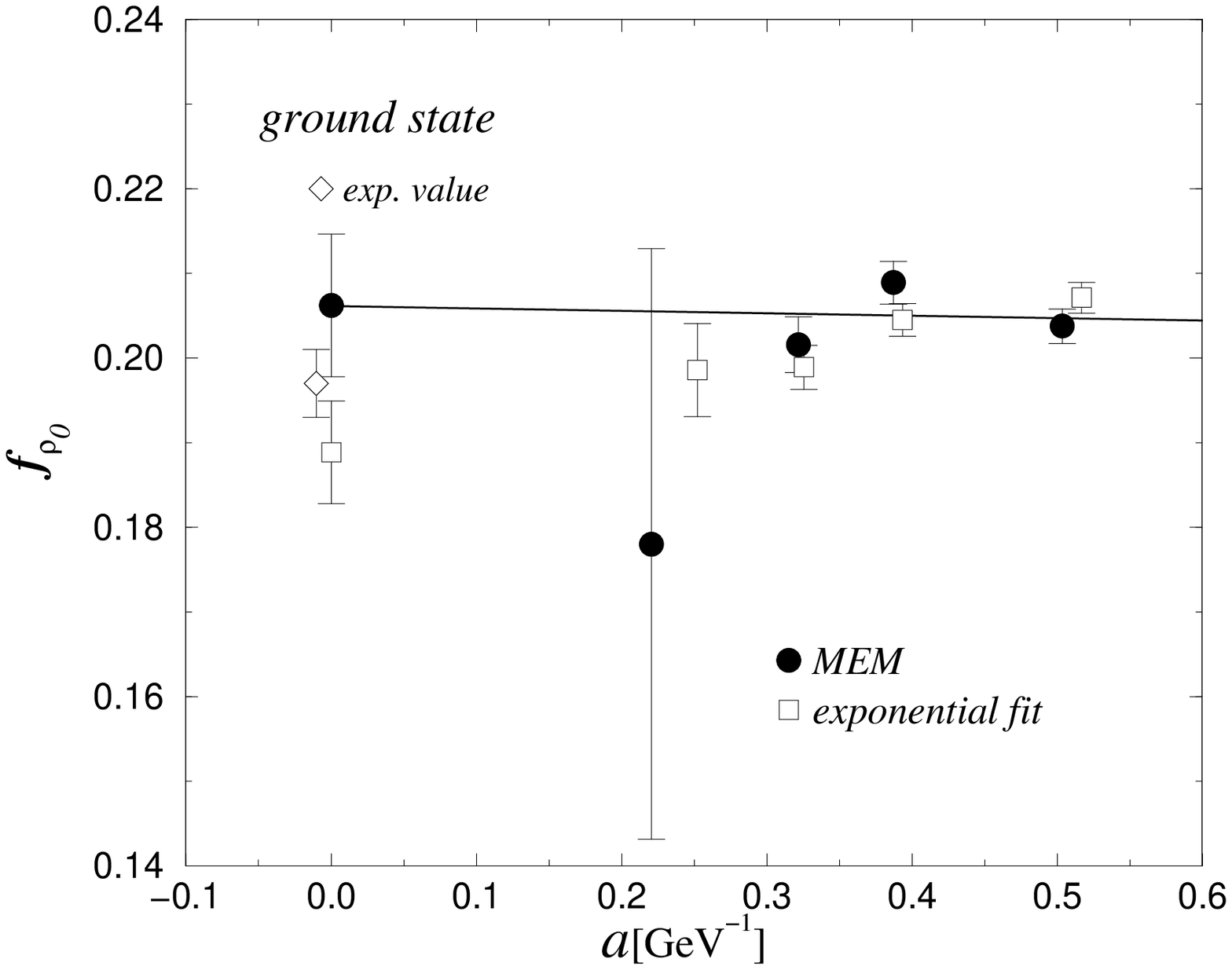} 
\\
\multicolumn{2}{c}{
\leavevmode
\epsfxsize=7.5cm\epsfbox{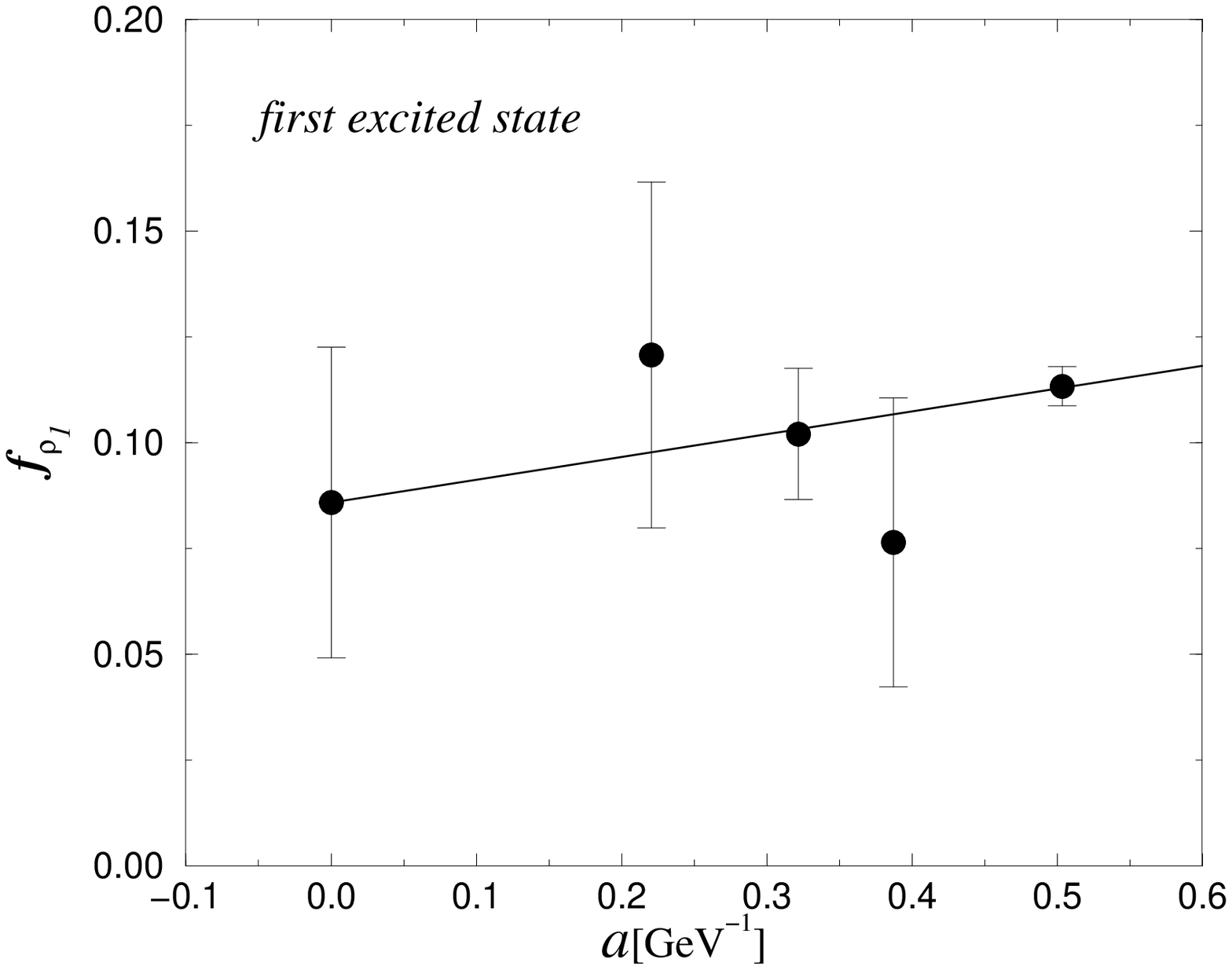}
}\\
\end{tabular}
\end{center}
\caption{Continuum extrapolations of pseudoscalar and vector meson
decay constants.
Open diamonds show
experimental values for
the ground state.
Open squares represent the previous results from standard analysis\protect\cite{pre}.
\label{fig:decay}}
\end{figure}

The continuum extrapolation is shown
in Fig.~\ref{fig:decay}, and the results in the continuum limit 
are compiled in Table~\ref{tab:figdecay}.
For the ground state, the decay constant for $\pi$ and $\rho$ mesons
are consistent with previous results (squares)\cite{pre}.
In the continuum limit we find 
$f_{\pi_{0}} = 80.3(5.9)$ MeV, which is smaller than the experimental value 
93 MeV,
and $f_{\rho_{0}} = 0.2062(84)$,
which is slightly larger than the experimental value 0.198(4), and 
the first excited state decay constant for $\rho$ meson
$f_{\rho_{1}} = 0.085(36)$.

\subsection{Remark on spectral widths}

\begin{figure}[p]
\begin{center}
\leavevmode
\epsfxsize=7.5cm\epsfbox{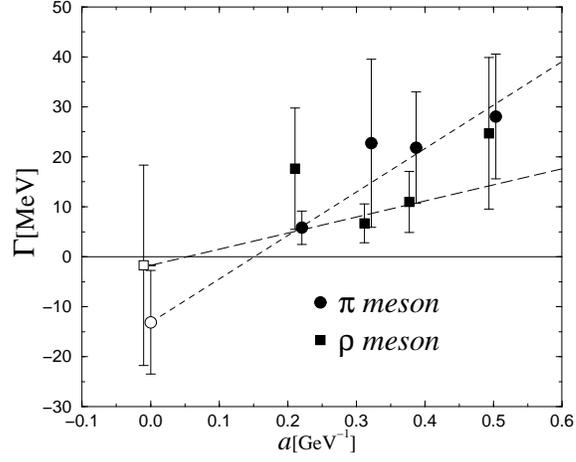}
\caption{Widths for the ground state
peak of $\pi$ and $\rho$ mesons and their continuum extrapolation
.
\label{fig:wid}}
\end{center}
\end{figure}

The width for the ground state peak should be zero for the $\pi$ meson,
and should be very small for the $\rho$ meson in the quenched approximation.
Therefore the width for the ground state
in spectral functions, if non-zero, is likely to be an artifact of MEM.
The width $\Gamma$ of the ground 
state peak for $\pi$ and $\rho$ mesons
are extrapolated to the chiral limit,
and are compiled in Table~\ref{tab:7}.
As shown in Fig.~\ref{fig:wid}, these widths are very small 
and almost consistent with zero within errors, as expected.

On the other hand, other states have larger widths.
At this moment it is difficult to conclude whether
these widths are physical or artifacts of MEM.
In order to decide the nature of these widths,
further researches are needed.

\section{Unphysical states and fermion doublers}
\label{sec:doubler}

\begin{figure}[p]
\begin{center}
\begin{tabular}{cc}
\leavevmode
\epsfxsize=7.5cm\epsfbox{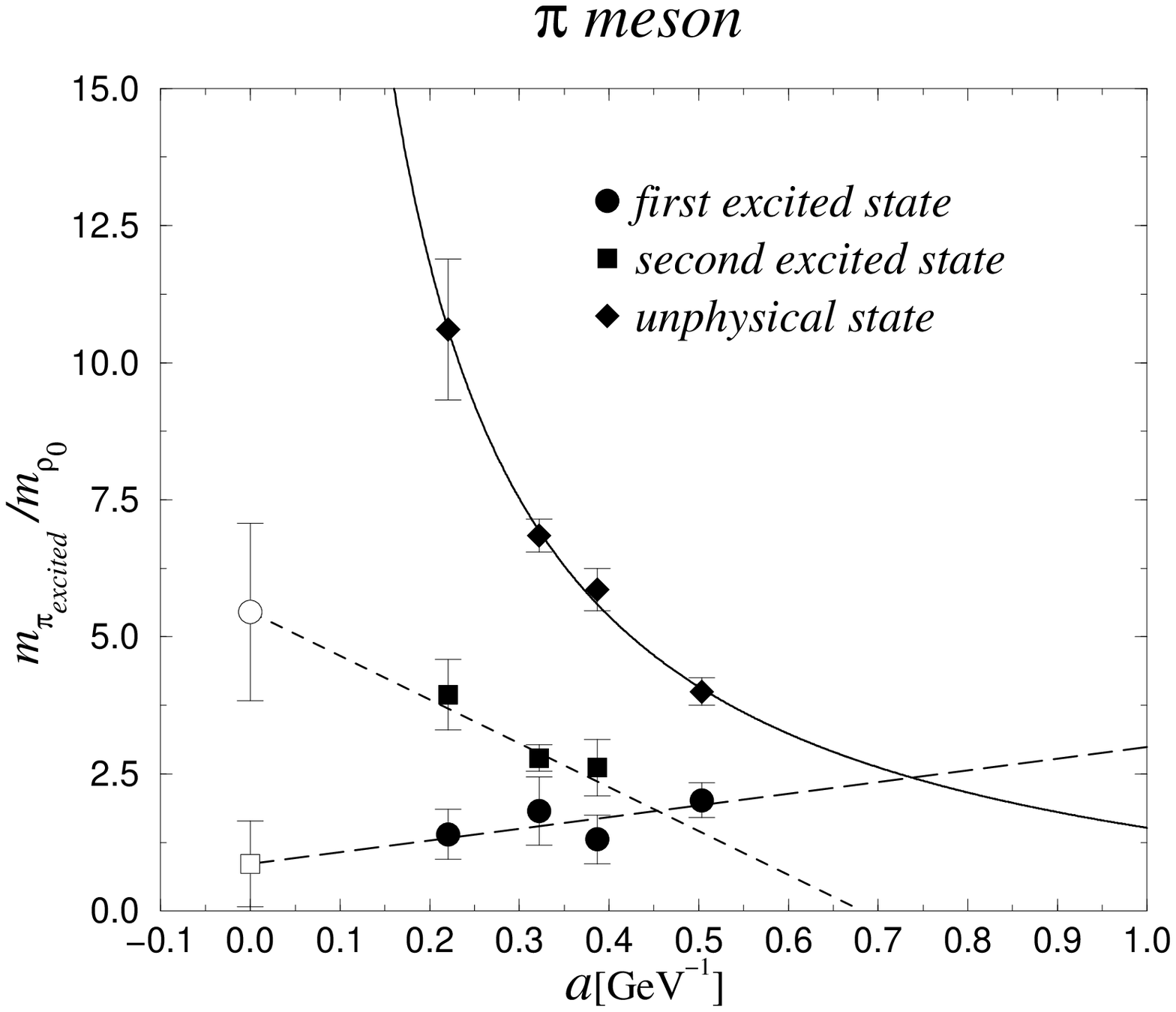} 
&
\leavevmode
\epsfxsize=7.5cm\epsfbox{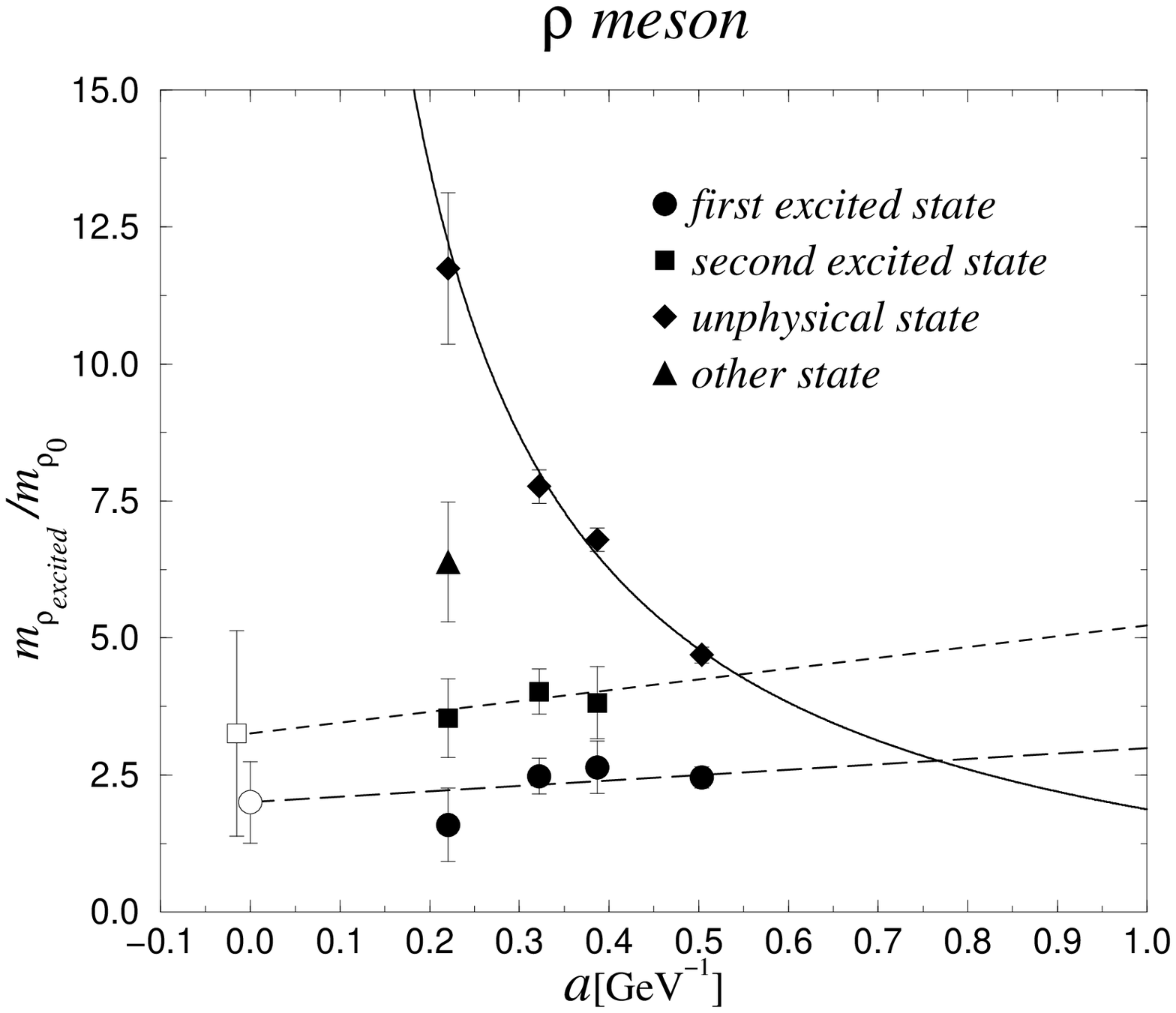} 
\end{tabular}
\end{center}
\caption{Combination for the excited mass fit and the unphysical state fit
of $\pi$ and $\rho$ mesons.
\label{fig:V}}
\end{figure}

As mentioned in the previous section,
the state in the pseudoscalar channel at $\omega a \approx 1.7$ and
the one in the vector channel at $\omega a \approx 2$ 
appear with a large width in the spectral functions at all $\beta$.
A similar state has been also observed in the Wilson quark action
at $\beta =6.0$ ($a^{-1}=$ 2.2 GeV) of the
plaquette gauge action\cite{hats:rev} and
at $\beta =4.1$ ($a^{-1}=$ 1.1 GeV)
of a tree-level Symanzik improved gauge action\cite{diquark}.
We consider this state to be unphysical since its mass diverges toward the 
continuum limit.
In fact the mass of this state can be fitted by $C_1/a + C_2$ 
in Fig.~\ref{fig:V} (see Table~\ref{tab:fig8} for numerical details),
together with a linear continuum extrapolation for physical excited state.
We also see from this figure
that no physical excited states appear in the spectral function
if its mass is larger than that of the unphysical state.
At first sight, the state at $\omega a\approx 1$ seems to be a candidate
of another unphysical state.
We think, however, that this state is physical,
since the position of the peak moves as $\beta$ varies
and moreover such a state has not been observed 
at a different lattice spacing\cite{diquark}.

We argue that the unphysical state is a bound state of two
fermion doublers of the Wilson quark action as follows.
The pole mass of a free quark with Wilson parameter $r=1$ is given by
\begin{equation}
M(n) = \frac{1}{a}\log (1+ma+2n)\ \ \ \ \ n=0,1,2,3\ ,
\label{eq:rew}
\end{equation}
where $n=0$ corresponds to the physical quark,
and $n\ne 0$ represent doublers with $n$ of the 3 spatial momenta components
equal to $\pi /a$.
At $r=1$ the time doubler does not propagate 
due to its infinite mass.
In the chiral limit
the mass for the  $n=1$ doubler is given by
$M(1)a\approx 1.1$, therefore,
in this free case, the mass of two $n=1$ doublers is 
$2\times M(1)\,a\approx 2.2$.
Note that, for the meson correlation functions with zero spatial momentum,
states consisting of, e.g., a physical quark and a doubler cannot
contribute.

In the interacting case,
the mass for the bound state made of two doublers
is expected to decrease from 2.2 in the free theory due to binding energy,
which would depend on the quantum number of the state.
This may explain the difference between the peak position 
at $\omega a \approx 1.7$ for the pseudoscalar channel
and at $\omega a \approx 2$ for
the vector channel.

From the consideration above we conclude that the unphysical state is 
a bound state of two $n=1$ doublers.
We note that bound states of $n\ge 2$ doublers do not appear
in the spectral function (in fact there is no peak at $\omega a =$
$3.2 \approx 2\times M(2)\,a$ and $3.9 \approx 2\times M(3)\,a$).

\section{Conclusion}
\label{sec:conc}

In this study, we have applied the maximum entropy method to
high-precision quenched lattice QCD data
to extract the spectral functions for pseudoscalar and vector mesons.
Masses for excited states as well as the ground state are obtained
from the position of peaks in the spectral function,
and decay constants are determined from the area under them.

Masses of the ground and the first excited state
agree with those obtained by the usual double exponential fit,
showing the reliability of MEM,
while the first excited state mass from the spectral function
has much smaller errors, demonstrating the superiority of MEM.

We have been able to make a continuum extrapolation for the first excited state
for $\pi$ and $\rho$ mesons, obtaining the masses $m_{\pi_{1}}=660(590)$ MeV
and $m_{\rho_{1}}=1540(570)$ MeV.
While the errors are admittedly large, this is the first time that such
an extrapolation has been attempted.
For the ground state decay constant for $\pi$ and $\rho$ mesons 
we have found that the result of MEM analysis is consistent with 
standard analysis.

We have found a state in the meson spectral function
at $\omega a\approx 2$ for all $\beta$,
and have argued that it is 
an unphysical bound state of two fermion doublers.
If this interpretation is correct,
this will be the first time that the doubler state has been identified 
numerically in lattice QCD simulations.
Further confirmation of this interpretation can be made
by changing the Wilson parameter $r$ from unity,
by analyzing the KS fermion data with MEM,
or by considering meson correlation functions with a momentum of $\pi /a$.

A future extension of MEM analysis is an application to
unquenched data to see dynamical quark effects in 
the spectral function;
decays and scatterings of intermediate states may be
detected from possible widths in the spectral function.
It will also be interesting to see the change of 
the spectral function before and after the phase transition
at finite temperatures.

\section*{Acknowledgements}
This work is supported in part by Grants-in-Aid of the Ministry of Education 
(Nos.~10640246, 
10640248, 
11640250, 
11640294, 
12014202, 
12304011, 
12640253, 
12740133, 
13640260  
). 
The numerical calculations for the present work have been carried out 
at the Center for Computational Physics, University of Tsukuba.

\appendix

\section*{Technical details of MEM}
\label{sec:App}
We collect technical details of
probability theory and MEM in this Appendix.
\setcounter{section}{0}
\renewcommand{\thesubsection}{\Alph{subsection}}
\renewcommand{\theequation}{\Alph{subsection}\arabic{equation}}

\subsection{The Bayes's theorem}

In this section we list some results of
probability theory and the Bayes's theorem used 
in MEM.
The Bayes's theorem in probability theory\cite{bay} is given by,
\begin{equation}
{\mathrm P}[X|Y] = \frac{{\mathrm P}[Y|X]\,{\mathrm P}[X]}{{\mathrm P}[Y]},
\label{eq:1}
\end{equation}
where ${\mathrm P}[X]$ is the probability of an event $X$, and
${\mathrm P}[X|Y]$ is the conditional probability of $X$ given $Y$.
These probabilities satisfy
\begin{equation}
{\mathrm P}[X] = \int dY\,{\mathrm P}[X|Y]\,{\mathrm P}[Y],
\label{eq:2}
\end{equation}
and the condition for normalization,
\begin{equation}
\int dX\,{\mathrm P}[X] = 1,
\end{equation}
\begin{equation}
\int dX{\mathrm P}[X|Y] = 1.
\label{eq:3}
\end{equation}
In this article, we use ${\mathrm P}[X|YZ]$ which is 
the conditional probability of $X$ given $Y$ and $Z$.
For P$[X|YZ]$,
(\ref{eq:1}), (\ref{eq:2}) and (\ref{eq:3}) are rewritten, respectively, as,
\begin{equation}
{\mathrm P}[X|YZ] = \frac{{\mathrm P}[Y|XZ]\,
{\mathrm P}[X|Z]}{{\mathrm P}[Y|Z]},
\label{eq:4}
\end{equation}
\begin{equation}
{\mathrm P}[X|Z] = \int dY\,{\mathrm P}[X|YZ]\,{\mathrm P}[Y|Z],
\label{eq:012}
\end{equation}
\begin{equation}
\int dX\,{\mathrm P}[X|YZ] = 1.
\label{eq:6}
\end{equation}

The most probable spectral function 
is obtained by maximizing the conditional probability ${\mathrm P}[F|DH]$
(in this section prior knowledge $Hm\alpha$ is rewritten as $H$ again
for simplicity), 
and satisfies the condition,
\begin{equation}
\frac{\delta {\mathrm P}[F|DH]}{\delta F} = 0.
\label{eq:condition}
\end{equation}
We rewrite ${\mathrm P}[F|DH]$ by the Bayes's theorem as,
\begin{equation}
{\mathrm P}[F|DH] = \frac{{\mathrm P}[D|FH]\,
{\mathrm P}[F|H]}{{\mathrm P}[D|H]}.
\label{eq:H}
\end{equation}
The probability ${\mathrm P}[D|FH]$ and ${\mathrm P}[F|H]$ is
the likelihood function and the prior probability, respectively.

Integrating (\ref{eq:H}) over $F$
and using (\ref{eq:6}), one finds that
\begin{equation}
{\mathrm P}[D|H] = \int {\mathcal D}F\,{\mathrm P}[D|FH]\,{\mathrm P}[F|H],
\end{equation}
where ${\mathcal D}F$ is the measure of spectral functions.
From this point of view, ${\mathrm P}[D|H]$ is a normalization factor
related to the likelihood function and the prior probability,
and we do not need to take account of it.

\setcounter{equation}{0}
\subsection{Transformation of covariance matrix}
In this section we introduce the method which easily deals with a non-diagonal 
covariance matrix.
If $C$ is not a diagonal matrix, one can transform $C$ into a diagonal 
from through
$
C=R \sigma^{2}R^{-1}
$, where $R$ 
is the transformation matrix and $\sigma^{2}$ is the eigenvalue matrix of $C$.
Kernel $K_{li}=K(\omega_{l} , \tau_{i})$ and data $D_{i}=D(\tau_{i})$ 
are transformed by $R$ as,
\begin{equation}
\widetilde{K}_{li} = \sum_{i^{'}=1}^{N_{D}} K_{li^{'}} R_{i^{'}i}\ ,
\end{equation}
\begin{equation}
\widetilde{D_{i}} = \sum_{i^{'}=1}^{N_{D}} D_{i^{'}} R_{i^{'}i}.
\end{equation}
After this transformation, 
the likelihood function $L$ defined in (\ref{eq:integ})
is written as,
\begin{equation}
L=\frac{1}{2}\sum_{i=1}^{N_{D}}
\left(
\widetilde{D}_{i} - \sum_{l=1}^{N_{\omega}} f_{l} \widetilde{K}_{li}
\right)^{2}/\sigma^{2}_{i}.
\end{equation}
This transformation does not require any changes in other parts of MEM.

\setcounter{equation}{0}
\subsection{The normalization constant of the prior probability}
\label{sec:Zs}

The factor $Z_{S}(\alpha)$
defined in (\ref{eq:ma2})
is the normalization constant of 
the prior probability.
In order to calculate $Z_{S}(\alpha)$,
we introduce a variable $X_{l}$ which
makes the curvature of $S(f)$ flat,
and expand $S(f)$ by transforming $f_{l}$ into $X_{l}$
and applying the gaussian approximation to $X(f)$ around $X(m)$,
\begin{eqnarray}
S(f) &\approx& S(m) + \sum_{l=1}^{N_{\omega}}\delta X_{l}
\left.\frac{\partial S}{\partial X_{l}}\right|_{X(m)}
+\frac{1}{2}\sum_{l,l^{'}=1}^{N_{\omega}}\delta X_{l}\delta X_{l^{'}}
\left.\frac{\partial^{2} S}{\partial X_{l}\partial X_{l^{'}}}\right|_{X(m)}\\
&=&
S(m) + \sum_{ll^{'}}^{N_{\omega}}\delta X_{l}
\frac{\partial f_{l^{'}}}{\partial X_{l}}
\left.\frac{\partial S}{\partial f_{l^{'}}}\right|_{m}
+\frac{1}{2}\sum_{kk^{'}ll^{'}}^{N_{\omega}}\delta X_{l}\delta X_{l^{'}}
\frac{\partial f_{k}}{\partial X_{l}}
\frac{\partial f_{k^{'}}}{\partial X_{l^{'}}}
\left.\frac{\partial^{2} S}{\partial f_{k}\partial f_{k^{'}}}\right|_{m},
\end{eqnarray}
where $\delta X_{l}=X_{l}(f)-X_{l}(m)$.
From the property of $X_{l}$ we choose
\begin{equation}
\frac{df_{l}}{dX_{l^{'}}}=\sqrt{f_{l}}\delta_{ll^{'}}.
\label{eq:transform}
\end{equation}
Since
\begin{equation}
S(m) = 0,\ \ \ 
\left.\frac{\partial S}{\partial f_{l}}\right|_{m} = 0,\ \ \ 
\left.\frac{\partial^{2} S}{\partial f_{l}\partial f_{l^{'}}}\right|_{m}=
-\frac{1}{f_{l}}\delta_{ll^{'}},
\end{equation}
we take the gaussian form for $S(f)$,
\begin{equation}
S(f) \approx 
-\frac{1}{2}\sum_{l=1}^{N_{\omega}}(\delta X_{l})^{2}.
\end{equation}
The measure ${\mathcal D}F$ is derived from the so-called 
`Monkey Argument`\cite{hats:rev,cla,monkey} and 
related to the metric of $S(f)$.
It is written as,
\begin{equation}
{\mathcal D}F = \prod_{l=1}^{N_{\omega}}\frac{df_{l}}{\sqrt{f_{l}}}.
\end{equation}
${\mathcal D}F$ is transformed by (\ref{eq:transform}) such as, 
${\mathcal D}F \rightarrow \prod_{l=1}^{N_{\omega}}dX_{l}$.
We can easily integrate over $f_{l}$ and obtain the normalization constant,
\begin{eqnarray}
Z_{S}(\alpha)&=&\int {\mathcal D}F\, e^{\alpha S(f)}\\
&\approx& \int \prod_{l=1}^{N_{\omega}}dX_{l}\, exp
\left[-\frac{1}{2}\alpha\sum_{l}^{N_{\omega}}(\delta X_{l})^{2}
\right]\\
&=& \left(\sqrt{\frac{2\pi}{\alpha}}\right)^{N_{\omega}}.
\end{eqnarray}

\setcounter{equation}{0}
\subsection{Uniqueness of MEM solution}

In this section we explain that the condition satisfied by the most 
probable spectral function,
(\ref{eq:condition})
has only one solution,
and has no local minimum.
The likelihood function $L$ satisfies
\begin{equation}
\sum_{l,l^{'}=1}^{N_{\omega}} z_{l}\frac{\partial^{2} (-L)}
{\partial f_{l}\partial f_{l^{'}}}z_{l^{'}}
=-\sum_{i=1}^{N_{D}} \frac{\widetilde{z}_{i}^{2}}{\sigma^{2}_{i}}
\le 0,\ \  
{\mathrm with} 
\ \ \widetilde{z}_{i}=\sum_{l=1}^{N_{\omega}} z_{l} {K}_{li},
\end{equation}
where $z_{l}$'s are non-zero real vectors and 
$\widetilde{z}_{i}$'s are real vectors.
The entropy and a real and positive parameter $\alpha$
satisfy
\begin{equation}
\sum_{l,l^{'}=1}^{N_{\omega}} z_{l}\frac{\partial^{2} \alpha S(f)}
{\partial f_{l}\partial f_{l^{'}}}z_{l^{'}}
=
-\alpha\sum_{l=1}^{N_{\omega}}\frac{z^{2}_{l}}{f_{l}}
< 0,
\end{equation}
where we have used $0\le f_{l}<\infty$ and $0<\alpha <\infty$.
The matrix $\partial^{2} Q_{\alpha}(f)/\partial f_{l}\partial f_{l^{'}}$
is negative definite,
\begin{equation}
\sum_{l,l^{'}=1}^{N_{\omega}}z_{l}\frac{\partial^{2} Q_{\alpha}(f)}
{\partial f_{l}\partial f_{l^{'}}}z_{l^{'}} < 0.
\end{equation}
Using Rolle's theorem,
one can verify that (\ref{eq:condition}) has only one solution corresponding
to the global maximum of $Q_{\alpha}(f)$,
if it exists\cite{hats:rev}.
Roughly speaking, since 
the curvature of $Q_{\alpha}(f)$ is always negative,
$Q_{\alpha}(f)$ has only one maximum.

\setcounter{equation}{0}
\subsection{The calculation of ${\mathrm P}[\alpha|DHm]$}
\label{sec:pag}
In order to search for the most probable value of $\alpha$,
we need to evaluate the conditional probability ${\mathrm P}[\alpha|DHm]$.
This conditional probability is used in Bryan's method\cite{Bry}
as the weight factor of averaging over $\alpha$.
In order to calculate ${\mathrm P}[\alpha|DHm]$,
we transform ${\mathrm P}[\alpha|DHm]$
by the Bayes's theorem and (\ref{eq:012}) as,
\begin{eqnarray}
{\mathrm P}[\alpha|DHm] &=& 
{\mathrm P}[D|Hm\alpha]\,{\mathrm P}[\alpha|Hm]/{\mathrm P}[D|Hm]
\label{eq:7}\\
&=& 
{\mathrm P}[\alpha|Hm]\int\!{\mathcal D}F\, 
{\mathrm P}[D|FHm\alpha]\,{\mathrm P}[F|Hm\alpha]/
{\mathrm P}[D|Hm]
\label{eq:8}\\
&\propto&
{\mathrm P}[\alpha|Hm]\int\! {\mathcal D}F\, \frac{e^{Q_{\alpha}(f)}}
{Z_{L}Z_{S}(\alpha)}.
\end{eqnarray}
Under the assumption that 
${\mathrm P}[F|DHm\alpha]$ is sharply peaked around 
the most probable spectral function $f_{\alpha}$,
$Q_{\alpha}(f)$ is expanded in the variable $X_{l}(f)$ 
used in Appendix \ref{sec:Zs}
and the gaussian approximation around $X_{l}(f)=X_{l}(f_{\alpha})$,
\begin{eqnarray}
Q_{\alpha}(f) 
&\approx&
Q_{\alpha}(f_{\alpha}) + \sum_{l=1}^{N_{\omega}}\delta X_{l}
\left. \frac{\partial Q_{\alpha}}{\partial X_{l}}\right|_{X(f_{\alpha})}
+\frac{1}{2}
\sum_{l,l^{'}=1}^{N_{\omega}}\delta X_{l}\delta X_{l^{'}}
\left. \frac{\partial^{2} Q_{\alpha}}{\partial X_{l}\partial X_{l^{'}}}
\right|_{X(f_{\alpha})}\\
&=&
Q_{\alpha}(f_{\alpha}) + \sum_{ll^{'}}^{N_{\omega}}\delta X_{l}
\frac{\partial f_{l^{'}}}{\partial X_{l}}
\left. \frac{\partial Q_{\alpha}}{\partial f_{l^{'}}}\right|_{f_{\alpha}}
+\frac{1}{2}
\sum_{kk^{'}ll^{'}}^{N_{\omega}}\delta X_{l}\delta X_{l^{'}}
\frac{\partial f_{k}}{\partial X_{l}}
\frac{\partial f_{k^{'}}}{\partial X_{l^{'}}}
\left. \frac{\partial^{2} Q_{\alpha}}{\partial f_{k}\partial f_{k^{'}}}
\right|_{f_{\alpha}},
\end{eqnarray}
where $\delta X_{l} = X_{l}(f)-X_{l}(f_{\alpha})$.
Because 
\begin{equation}
\left. \frac{\partial Q_{\alpha}}{\partial f_{l}}\right|_{f_{\alpha}} = 0,
\ \ \ 
\left. \frac{\partial^{2} Q_{\alpha}}{\partial f_{l}\partial f_{l^{'}}}
\right|_{f_{\alpha}}
=-\left(\frac{\alpha}{f_{l}}\delta_{ll^{'}}+
\frac{\partial^{2} L}{\partial f_{l}\partial f_{l^{'}}}\right)_{f_{\alpha}},
\end{equation}
we can write
\begin{equation}
Q_{\alpha}(f)\approx
Q_{\alpha}(f_{\alpha})-\frac{1}{2}\sum_{l,l^{'}=1}^{N_{\omega}}\delta X_{l}
\left(\alpha\delta_{ll^{'}}+\Lambda_{ll^{'}}
\right)\delta X_{l^{'}},
\label{eq:coco}
\end{equation}
where $\Lambda_{ll^{'}}$ is a real symmetric
$N_{\omega}\times N_{\omega}$ matrix defined as,
\begin{equation}
\Lambda_{ll^{'}} =
\left.
\sqrt{f_{l}}\frac{\partial^{2} L}{\partial f_{l}\partial f_{l^{'}}}
\sqrt{f_{l^{'}}}\right|_{f_{\alpha}}.
\end{equation}
We then obtain
\begin{eqnarray}
{\mathrm P}[\alpha|DHm]
&\approx&
\frac{{\mathrm P}[\alpha|Hm]}{Z_{L}Z_{S}(\alpha)}\!\!
\int \!\prod_{l=1}^{N_{\omega}}dX_{l}
\, exp\!\left[Q_{\alpha}(f_{\alpha})-\frac{1}{2}
\sum_{l,l^{'}}\delta X_{l}
\left(\alpha\delta_{ll^{'}}+\Lambda_{ll^{'}}
\right)\delta X_{l^{'}}
\right]\\
&\propto&
{\mathrm P}[\alpha|Hm]\,e^{Q_{\alpha}(f_{\alpha})}\prod_{l=1}^{N_{\omega}}
\sqrt{\frac{\alpha}{\alpha + \lambda_{l}}},
\label{eq:10}
\end{eqnarray}
here $\lambda_{l}$'s are the eigenvalues of $\Lambda$.

\setcounter{equation}{0}
\subsection{Estimation of uncertainty in MEM}
\label{sec:err}

In MEM, 
it is possible to estimate the uncertainty of 
spectral function averaged 
over a certain region ${\mathrm I}$ of $\omega$,
\begin{equation}
\langle f_{\alpha} \rangle_{{\mathrm I}} = 
\frac{\int_{{\mathrm I}} d\omega \,\langle f(\omega) \rangle}
{\int_{{\mathrm I}} d\omega} \approx 
\frac{\int_{{\mathrm I}} d\omega \, f_{\alpha}(\omega) }
{\int_{{\mathrm I}} d\omega}, 
\end{equation}
where
$\langle \Theta \rangle = \int {\mathcal D}F \,
\Theta \,{\mathrm P}[F|DHm\alpha]
$.
Using the gaussian approximation and the variable $X_{l}(f)$
in Appendix \ref{sec:pag},
the covariance of the spectral function can be calculated as,
\begin{eqnarray}
\langle\delta f(\omega) \delta f(\omega^{'})\rangle
&=&
 \sqrt{f_{\alpha}(\omega)}\,
\langle\delta X(\omega) \delta X(\omega^{'})\rangle
\,\sqrt{f_{\alpha}(\omega^{'})}\\ &\approx&
\sqrt{f_{\alpha}(\omega)}\Gamma^{-1}_{\omega\omega^{'}}
\sqrt{f_{\alpha}(\omega^{'})}\\
&=&
-\left(\frac{\delta^{2}Q_{\alpha}}
{\delta f(\omega) \delta f(\omega^{'})}
\right)^{-1}_{f_{\alpha}}\label{eq:31},
\end{eqnarray}
where $\Gamma = \alpha\delta + \Lambda$.
The form of 
(\ref{eq:31}) is readily available because it is the Hessian of 
the Newton search algorithm\cite{jarr,hats:rev,Bry}
used to find $f_{\alpha}$.
The uncertainty is estimated as,
\begin{equation}
\langle (\delta f_{\alpha})^{2} \rangle_{{\mathrm I}} \approx
\frac{\int_{{\mathrm I}\times {\mathrm I}} d\omega\, d\omega^{'}\,
\sqrt{f_{\alpha}(\omega)}\,\Gamma^{-1}_{\omega\omega^{'}}
\sqrt{f_{\alpha}(\omega^{'})}}
{\int_{{\mathrm I}\times {\mathrm I}} d\omega\, d\omega^{'}}
\label{eq:32}.
\end{equation}
Similar to the spectral function, 
the error of averaged spectral function in a certain region I
is averaged over $\alpha$ with the weight factor ${\mathrm P}[\alpha|DHm]$,
\begin{equation}
\langle \delta \widehat{f} \rangle_{{\mathrm I}} =
\frac{
\int d\alpha\, {\mathrm P}[\alpha|DHm]
\sqrt{\langle (\delta f_{\alpha})^{2} \rangle_{{\mathrm I}}}
}
{
\int d\alpha\, {\mathrm P}[\alpha|DHm]
}.
\end{equation}

\clearpage

\begin{table}[!h]
\caption{Simulation parameters of hadron propagator data
\protect\cite{CPdata,pre}
used in the present MEM analysis.
The numbering of hopping parameter is introduced for convenience.
The smallest number corresponds to the heaviest quark mass, and vice versa.
\label{tab:1}}
\begin{center}
\begin{tabular}{cccc}\hline \hline
$\beta$&lattice size($L^{3}T$)&conf.$\#$&sweep/conf. \\ \hline
5.90   &$32^{3}\ 56$          &800   &200\\
6.10   &$40^{3}\ 70$          &600   &400\\
6.25   &$48^{3}\ 84$          &420   &1000\\
6.47   &$64^{3}\ 112$         &150   &2000\\ \hline
\end{tabular}
\begin{tabular}{cccccc}\hline \hline
\multicolumn{6}{c}{hopping parameter $\kappa$}\\ \hline
$\beta$&1&2&3&4&5\\ \hline
5.90   & 0.1566   & 0.1574  & 0.1583  & 0.1589 &0.1592\\
6.10   & 0.1528   & 0.1534  & 0.1540  & 0.1544 &0.1546\\
6.25   & 0.15075  & 0.15115 & 0.15165 & 0.15200&0.15220\\
6.47   & 0.14855  & 0.14885 & 0.14925 & 0.14945&0.14960 \\ \hline
$m_{\pi}/m_{\rho}$& 0.75    & 0.7     & 0.6    &0.5&0.4\\ \hline
\end{tabular}
\end{center}
\end{table}

\begin{table}[h]
\caption{Parameters used in MEM analysis\label{tab:3}.
The bottom table shows $(N_{D},(\omega a)_{max})$.}
\begin{center}
\begin{tabular}{|c|c|c|c|}\hline
  & $C_{1}$ & $C_{2}$ & $Z$\\ \hline
PS& 3/2     & 11/3    &0.728  \\ \hline
V &  1      & 1       &0.536 \\ \hline
\end{tabular}
\end{center}
\begin{center}
\begin{tabular}{|c|c|c|c|c|}\hline
$\beta$&5.90&6.10&6.25&6.47\\ \hline
PS&(20,4.0)&(32,4.5)&(32,4.5)&(45,4.5)\\ \hline
V&(21,4.2)&(30,4.8)&(30,4.8)&(30,4.8)\\\hline 
\end{tabular}
\end{center}
\end{table}

\begin{table}[p]
\caption{
Comparison of the MEM analysis with 
the double exponential fit,
using the vector meson correlation function at $\beta=5.90$.
The symbol $Kn_{1}n_{2}$ expresses 
the quark mass used in the correlation function,
$n_{1}$ and $n_{2}$ being defined in Table~\ref{tab:1}.
\label{tab:6}}
\begin{center}
\begin{tabular}{|c|ccc|cc|}\hline
     &\multicolumn{3}{c|}{exponential fit}&\multicolumn{2}{c|}{MEM}\\
 \hline
     &   ground   & excited   &$\chi^{2}/d.o.f$    &   ground     & excited\\ \hline
$K11$  & 0.5093(11) &1.08(11)&0.220&0.5094(16)&1.034(30)\\ \hline
$K22$  & 0.4784(12) &1.08(14)&0.359&0.4789(20)&1.018(37)\\ \hline
$K31$  & 0.4772(15) &1.08(14)&0.466&0.4779(20)&1.020(36)\\ \hline
$K32$  & 0.4613(15) &1.07(15)&0.587&0.4623(23)&1.009(40)\\ \hline
$K33$  & 0.4435(22) &1.03(19)&0.687&0.4451(27)&0.997(44)\\ \hline
$K41$  & 0.4668(23) &1.09(17)&0.638&0.4678(23)&1.020(37)\\ \hline
$K42$  & 0.4505(22) &1.06(22)&0.750&0.4519(27)&1.006(44)\\ \hline
$K44$  & 0.4214(43) &1.08(21)&0.890&0.4218(43)&0.969(58)\\ \hline
$K51$  & 0.4622(20) &1.15(21)&0.771&0.4630(25)&1.020(40)\\ \hline
$K52$  & 0.4460(32) &1.11(19)&0.872&0.4469(30)&1.004(46)\\ \hline
$K55$  & 0.4107(37) &1.19(20)&1.191&0.4080(65)&0.929(70)\\ \hline
\end{tabular}
\end{center}
\end{table}

\begin{table}[p]
\begin{center}
\caption{Results obtained from the MEM analysis at each $\beta$. 
Lattice spacings from the standard analysis~\protect\cite{CPdata,pre}
are also listed.
\label{tab:7}}
\begin{tabular}{|c|c|c|c|c|}\hline
$\beta$      &5.90&6.10&6.25&6.47\\ \hline \hline
$a[{\mathrm GeV}^{-1}]$&0.503(6)&0.387(6)&0.321(5)&0.220(25)\\ \hline  
$a^{-1}[{\mathrm GeV}]$&1.986(25)&2.583(40)&3.105(53)&4.52(51)\\ \hline
$a^{-1}[{\mathrm GeV}]$\protect\cite{CPdata,pre}&1.934(16)&2.540(22)&3.071(34)&3.961(79)\\ \hline
$\kappa_{c}$      &0.159881(13)&0.154985(12)&0.152556(9)&0.149809(7)\\ \hline
\multicolumn{5}{|c|}{$\pi$ meson} \\ \hline 
$m_{\pi_{1}}/m_{\rho_{0}}$&2.02(31)&1.30(44)&1.82(62)&1.40(45)\\ \hline
$m_{\pi_{2}}/m_{\rho_{0}}$&\rule[1mm]{1.4cm}{.05mm}&2.61(51)&2.79(23)&3.95(64)\\ \hline
$m_{\pi_{{\mathrm unphys}}}/m_{\rho_{0}}$&4.00(25)&5.86(38)&6.84(29)&10.6(1.2)\\ \hline
$f_{\pi_{0}}/m_{\rho_{0}}$    &0.1157(21)&0.1148(26)&0.1099(28)&0.119(14) \\ \hline
$\Gamma_{\pi_{0}}/m_{\rho_{0}}$    &0.036(16)&0.028(14)&0.029(21)&0.007(4)\\ 
\hline
\multicolumn{5}{|c|}{$\rho$ meson} \\ \hline
$m_{\rho_{1}}/m_{\rho_{0}}$&2.46(19)&2.63(47)&2.48(32)&1.59(67)\\ \hline
$m_{\rho_{2}}/m_{\rho_{0}}$&\rule[1mm]{1.4cm}{.05mm}&3.81(65)&4.02(41)&3.53(71)\\ \hline
$m_{\rho_{3}}/m_{\rho_{0}}$&\rule[1mm]{1.4cm}{.05mm}&\rule[1mm]{1.4cm}{.05mm}&\rule[1mm]{1.4cm}{.05mm}&6.3(1.0)\\ \hline 
$m_{\rho_{{\mathrm unphys}}}/m_{\rho_{0}}$&4.69(14)&6.79(21)&7.76(30)&11.7(1.3)\\ \hline 
$f_{\rho_{0}}$                 &0.2037(20)&0.2088(25)&0.2015(32)&0.178(34)  \\ \hline
$f_{\rho_{1}}$                 &0.1133(46)&0.076(34)&0.102(15)&0.120(40)  \\ \hline
$\Gamma_{\rho_{0}}/m_{\rho_{0}}$    &0.032(19)&0.014(7)&0.008(5)&0.022(15)\\ \hline
\end{tabular}
\end{center}
\end{table}

\begin{table}[h]
\caption{
Masses of excited states normalized by the ground state $\rho$ meson mass
for the $\pi$ and $\rho$ mesons in the continuum limit.
Available experimental values are also given.
\label{tab:fig7}}
\begin{center}
\begin{tabular}{|c|c|c|c|c|}\hline
        &$m_{\pi_{1}}/m_{\rho_{0}}$
        &$m_{\pi_{2}}/m_{\rho_{0}}$
        &$m_{\rho_{1}}/m_{\rho_{0}}$&$m_{\rho_{2}}/m_{\rho_{0}}$\\ \hline
continuum limit&0.86(77)&5.4(1.6)&2.00(74)&3.2(1.8)\\ \hline
$\chi^{2}/d.o.f$&0.514&0.538&0.726&0.240\\ \hline
experimental value&1.68(12)&\rule[1mm]{1.2cm}{.05mm}&\multicolumn{2}{c|}{1.90(3) or 2.20(2)}    \\ \hline
\end{tabular}
\end{center}
\end{table}

\begin{table}[!h]
\caption{
Decay constants for $\pi$ and $\rho$ mesons in the continuum limit
and experimental values.
\label{tab:figdecay}}
\begin{center}
\begin{tabular}{|c|c|c|c|}\hline
        &$f_{\pi_{0}}$
        &$f_{\rho_{0}}$&$f_{\rho_{1}}$\\ \hline
continuum limit&80.3(5.9) MeV&0.2062(84)&0.085(36)\\ \hline
$\chi^{2}/d.o.f$&0.618&2.18&0.555\\ \hline
experimental value&93 MeV&0.198(4)&\rule[1mm]{1.cm}{.05mm}\\ \hline
\end{tabular}
\end{center}
\end{table}

\begin{table}[p]
\caption{
Fit parameters 
and $\chi^{2}/d.o.f.$ of the unphysical state fit for $\pi$ and $\rho$ mesons.
\label{tab:fig8}
}
\begin{center}
\begin{tabular}{|c|c|c|}\hline
&{$\pi$ meson}&{$\rho$ meson}\\ \hline
$C_{1}$&2.57(30)&2.924(25)\\ \hline
$C_{2}$&$-1.05(78)$&$-1.051(58)$\\ \hline
$\chi^{2}/d.o.f$&0.3158&1.476\\ \hline
\end{tabular}
\end{center}
\end{table}

\clearpage

\clearpage

\clearpage

\end{document}